# NovaMoon: A Strategic Lunar Reference Station for Positioning, Timing, and Largely Enhanced Science in the Earth–Moon System

**Serena Molli[1], Agnès Fienga[2], Pascale Defraigne[3], Krzysztof Sośnica[4], Luigi Cacciapuoti[5], Luca Porcelli[6], Lotfi Massarweh[7], Sara Bruni[8], Riccardo Pozzobon[9†], Albert Roura[10], Francesco Vespe[11], Diego Blas[12], Ozgur Karatekin[3], Yoann Audet[13], Floor Melman[5], Richard Swinden[5], Javier Ventura-Traveset[14], Olivier Alibart[2], Marie-Christine Angonin[15], Daniel Arnold[16], Ruth Bamford[17], Emmanuele Battista[6], Marco Belloni[5], J.C. Berton[1], Orfeu Bertolami[18], Mathis Bloßfeld[19], Adrien Bourgoin[15], Giada Bargiacchi[6], Salvatore Buoninfante[20], Nicolò Burzillà[6], Roberto Campagnola[6], Paolo Cappuccio[21], Salvatore Capozziello[22], Giuseppe Cimò[23], Clément Courde[2], Rolf Dach[16], Mario Siciliani de Cumis[11], Simone Dell'Agnello[6], Fabrizio De Marchi[24], Valentina Galluzzi[20], Francesco Gini[1], Philipp Glaeser[25], Klaus Gwinner[10], Alex Guinard[2], Rüdiger Haas[26], Aurélien Hees[15], Hauke Hussmann[10], Luciano Iess[24], Alexander C. Jenkins[27], Siddarth K. Joshi[28], Maria Karbon[37], Sergei Klioner[29], Kaisa Laiho[10], Christophe Le Poncin-Lafitte[15], Marco Lucente[20], David Lucchesi[20], Riccardo March[30], Lucia McCallum[31], Jürgen Müller[32], Weijie Nie[28], Jillian Oduber[5], Roberto Peron[20], Francesco Picciariello[5], Théo Pichavant[33], Michael Plumaris[5], Eleonora Polini[2], Ana-Catalina Plesa[10], Dimitrios Psychas[5], Bernardino Quaranta[5], Nicolas Rambaux[15], Marc Rovira-Navarro[7], Francesco Santoli[20], Matthias Schartner[34], Florian Seitz[19], Ilaria Sesia[35], Yan Seyffert[36], Stefano Speretta[7], Tim Springer[8], Giuseppe Vallone[9], Paolo Villoresi[9], Sebastien Vincent-Bonnieu[5], Ben Wadsworth[5], Pierre Waller[5], Radosław Zajdel[4], Erik Schoenemann[1]**

[1] European Space Operations Centre (ESOC), ESA, Germany
[2] Observatoire de la Côte d'Azur / Université Côte d'Azur, France
[3] Royal Observatory of Belgium (ROB), Belgium
[4] Wrocław University of Environmental and Life Sciences, Poland
[5] European Space Research and Technology Centre (ESTEC), ESA, The Netherlands
[6] Istituto Nazionale di Fisica Nucleare (INFN), Italy
[7] Delft University of Technology (TU Delft), The Netherlands
[8] PosiTim UG for ESA/ESOC, Germany
[9] University of Padova, Italy
[10] Deutsches Zentrum für Luft- und Raumfahrt e.V.(DLR), Germany
[11] Italian Space Agency (ASI), Italy
[12] Institut de Física d'Altes Energies (IFAE) / ICREA, Spain
[13] Telespazio Belgium SRL for ESA/ESTEC, The Netherlands
[14] ESA / Centre Spatial de Toulouse (CST), France
[15] LTE, Observatoire de Paris, Université PSL, CNRS, LNE, Sorbonne Université, Université de Lille
[16] University of Bern, Switzerland
[17] University of Oxford / Rutherford Appleton Laboratory (RAL), United Kingdom
[18] University of Porto, Portugal
[19] Deutsches Geodätisches Forschungsinstitut (DGFI-TUM), Technical University of Munich, Germany
[20] Istituto di Astrofisica e Planetologia Spaziali (INAF-IAPS), Rome, Italy
[21] Aurora Technology B.V. for ESA/ESAC, Villafranca del Castillo, Spain
[22] University of Naples Federico II, Italy
[23] Joint Institute for VLBI ERIC (JIVE), The Netherlands
[24] La Sapienza University of Rome, Italy

[25] Technische Universität Berlin, Institute of Geodesy, Germany
[26] Onsala Space Observatory, Chalmers University of Technology, Sweden
[27] University of Cambridge, United Kingdom
[28] University of Bristol, United Kingdom
[29] Technische Universität Dresden, Germany
[30] National Research Council (CNR), Italy
[31] University of Tasmania, Australia
[32] Leibniz University Hannover, Germany
[33] ISAE-SUPAERO, France
[34] ETH Zürich, Switzerland
[35] Istituto Nazionale di Ricerca Metrologica (INRiM), Turin, Italy
[36] University of Bremen, Germany
[37] University of Alicante, Spain

† Deceased. Dr. Riccardo Pozzobon contributed to this work prior to his passing.

# Abstract

The renewed interest in lunar exploration and the development of future lunar communication and navigation services highlight the need for a precise, stable, and interoperable geodetic and timing infrastructure on the Moon. NovaMoon, proposed as a scientific and navigation payload for ESA's Argonaut lander, is designed as a lunar-based local differential, geodetic, and timing station supporting both the operational needs of the Moon's south polar region and a broad range of scientific investigations. The payload integrates a lunar laser retroreflector, a Very Long Baseline Interferometry transmitter, a receiver for lunar navigation signals compatible with LunaNet standards, high-stability atomic clocks, and direct-to-Earth radio links, making it the first lunar station to co-locate multiple ranging, tracking, and timing techniques.

NovaMoon will enable sub-metre to decimetre positioning in the south polar region, provide local differential corrections for lunar navigation users, and ensure an accurate and stable realisation of position and time for the lander. Through preliminary simulation studies, we show that the resulting multi-technique dataset significantly improves the lunar reference frame, the determination of lunar orientation and ephemerides, and the estimation of interior parameters such as tidal response, core properties, and dissipation.

NovaMoon will also provide the first long-duration physical realisation of a lunar time reference, enabling precise timing for lunar navigation users and contributing to the establishment of a future lunar timescale.

Beyond its primary goals, NovaMoon supports improved cartography, more accurate geolocation of surface features, and higher-resolution topography in the south polar region, contributing to safer and more precise landing and surface operations. Its multi-technique measurements also open new opportunities for fundamental physics, including enhanced tests of the Equivalence Principle, improved constraints on relativistic gravity, and sensitivity to deviations from classical gravitational models or potential variations in fundamental constants.

# 1.Introduction

The renewed interest in space exploration, particularly around the Moon, has prompted significant efforts by space agencies to develop lunar communications and navigation services, such as the European Space Agency's Moonlight program. As a future evolution and complement of the Moonlight Lunar Communications and Navigation System (LCNS) architecture, the European Space Agency (ESA) is proposing the deployment of a Differential and Reference Station on the lunar surface. This initiative, known as NovaMoon, was proposed to and subscribed by ESA Member States at the 2025 Ministerial Council as part of the FutureNAV programme. The implementation phase is planned to start in 2026. Planned as one of the key demonstration payloads of the first ESA Argonaut lunar lander mission (targeted for launch by the end of 2030), NovaMoon is conceived as a local differential, geodetic, and timing station. Its purpose is to enhance the accuracy of Moonlight's PNT services to sub-metre levels across the lunar South Pole region. This increased precision would enable critical capabilities for autonomous robotics, crewed mobility, precision landing, and surface operations, while also supporting high-resolution lunar mapping and improving overall operational safety and efficiency.

In addition to its operational role, **NovaMoon is expected to make a significant contribution to lunar science** through a suite of advanced geodetic instruments. These include a Very Long Baseline Interferometry (VLBI) transmitter, laser retroreflectors, a LunaNet/GNSS receiver, and a set of lunar atomic clocks, which together would provide a stable, long-term lunar-based master clock reference.

This White Paper outlines the scientific objectives of the NovaMoon mission, distinguishing between:

- Primary Objectives, which drive mission design and performance requirements; and
- Secondary Objectives, representing additional scientific opportunities not directly influencing system architecture.

It also introduces potential experimental extensions that could be enabled by future payloads or technological developments. The document details the relevant instruments and measurement techniques, reviews their current state of the art, identifies required technology advancements for NovaMoon implementation, and presents preliminary system-level requirements. While navigation performance and service design are discussed for contextual completeness, these aspects are addressed in full in a dedicated publication (Ventura-Traveset et al., 2025)

This White Paper has been prepared through the collaborative effort of 80 leading scientists representing 40 European scientific institutions, reflecting the strong and coordinated European commitment to advancing lunar navigation and geodesy.

# 2. NovaMoon Concept and Navigation Objectives

NovaMoon's Mission is to provide ***a lunar-based local differential, geodetic and timing station to enhance Moonlight (and other Lunar LCNS Systems) services and support scientific experimentation***.

These general mission objectives can be split into 5 main areas, as follows:

1) **Provision of Local Differential services to Moonlight and other Lunar PNT systems**: Acting as an accurately surveyed local reference station at the lunar South Pole, NovaMoon will augment the Moonlight system by generating and transmitting code (and carrier-phase) measurements and corrections for Moonlight and other LunaNet navigation satellites to lunar users (e.g., rovers, astronauts, landers) across the entire South Pole region. This will enable Moonlight differential code-only (DGNSS) and code-plus-phase (RTK) positioning across the whole South Pole.
2) **Continuous monitoring of Moonlight and other Lunar PNT systems**: NovaMoon will provide continuous signal quality monitoring of Moonlight and other LunaNet navigation satellites using available raw measurements from a well-surveyed and monitored standardised LunaNet receiver. Additionally, NovaMoon will regularly compute the station's position, velocity, and time (PVT) using Moonlight services, assessing the associated errors impacting this signal and the resulting Moonlight service positioning accuracies. This independent monitoring function of NovaMoon will enhance the reliability of lunar navigation services provided to lunar users.
3) **Highly accurate lunar geodetic reference station:** NovaMoon will include four different co-located ranging techniques—a VLBI transmitter, a laser retroreflector, NovaMoon-Moonlight positioning, and Direct-To-Earth ranging—along with communication access via Moonlight. This setup will enable the precise localisation of the NovaMoon station with sub-decimetre accuracy. This capability will support improving lunar reference frame realisations and significantly enhance our understanding of the Moon's composition and interior structure.
4) **Lunar time-reference laboratory**: NovaMoon will include a set of atomic clocks capable of providing a continuous lunar physical time reference supporting the realisation of the LunaNet Reference Timescale.
5) **Scientific support for lunar experimentation**: To support the scientific community, NovaMoon will provide continuous access to individual lunar geodetic sensors and timing data for international research groups.

**NovaMoon is a European scientific and navigation payload** designed for lunar deployment aboard ESA's Argonaut lander. It integrates a compact suite of co-located instruments, making it the first lunar station combining multiple metrological and selenodetic techniques. The baseline payloads include:

- A Very Long Baseline Interferometry transmitter
- A Lunar Laser Ranging (LLR) retroreflector
- A Moonlight navigation signal receiver (LCNS receiver) compliant with the LunaNet standard
- High-stability onboard atomic clocks
- A Direct-to-Earth (DTE) communication terminal

Other potential payloads currently under study are:

- Active laser ranging, by adding a Single Photon Avalanche Detector (SPAD) receiver chain to the LLR retroreflector

- GNSS transmitter
- GNSS receiver, potentially integrated with the LCNS receiver

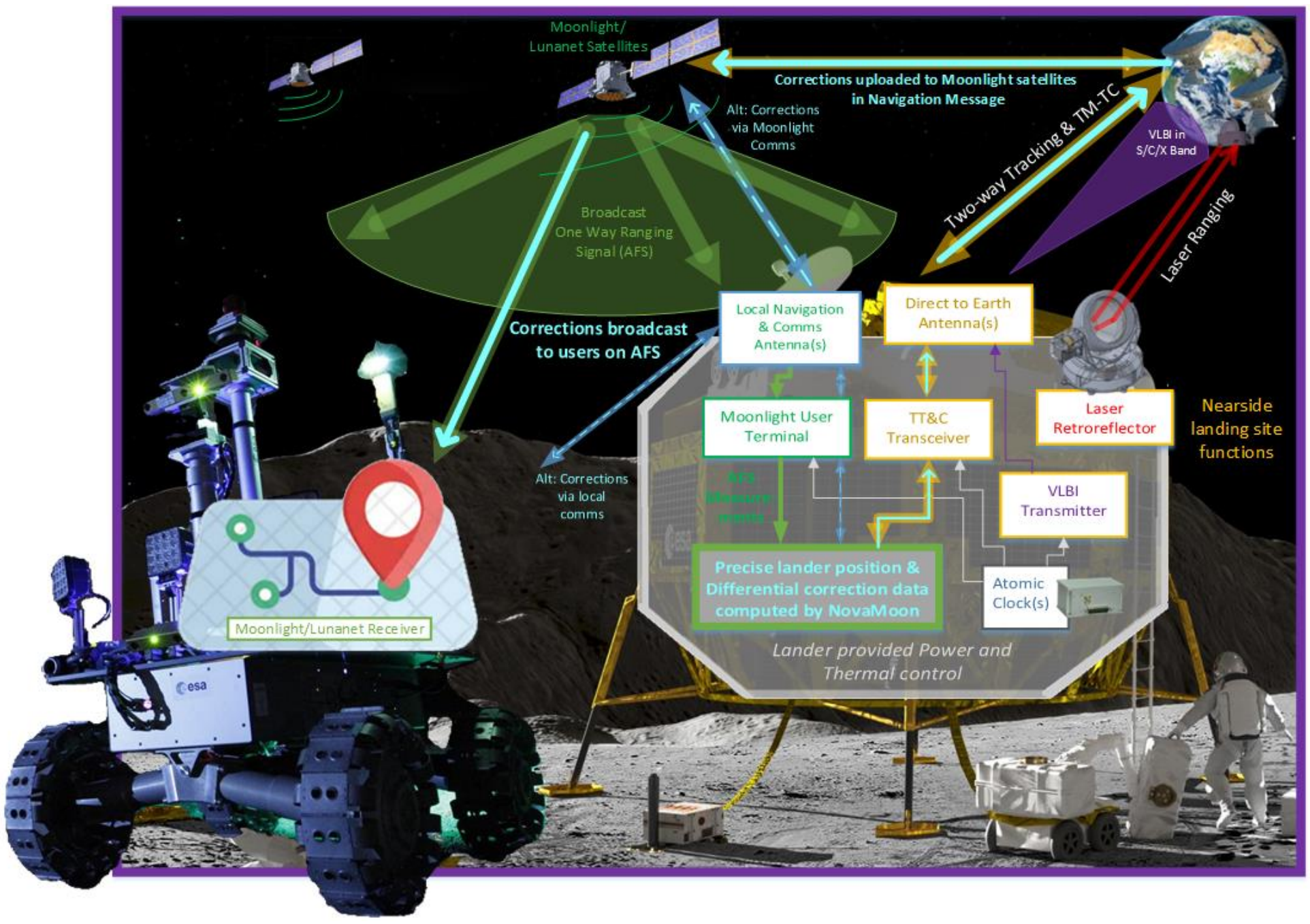

Figure 1. NovaMoon payloads and concept

Together, these instruments establish NovaMoon as a multi-technique geodetic and timing observatory on the lunar surface.

From a scientific standpoint, the station targets:

(i) precise determination of the NovaMoon reference position at the centimetre- to sub-decimetre-level through the combination of co-located ranging techniques, including laser ranging, VLBI, and Direct-to-Earth links;

(ii) the establishment of precise lunar reference frame and time standards;

(iii) the generation of a continuous lunar physical time reference supporting the realisation of the LunaNet Reference Timescale (LRT), computing its offset with respect to UTC for time-transfer and scientific experimentation.

NovaMoon's primary function is to serve as a differential Positioning, Navigation, and Timing (PNT) reference station for the Moon's South Pole region. By broadcasting real-time range corrections, it augments LunaNet Service Provider (LNSP) such as Moonlight, facilitating sub-meter level positioning in an environment currently devoid of local infrastructure. Beyond its navigation role, NovaMoon also operates as a selenodetic and timekeeping observatory. Its co-located LLR and VLBI systems allow multi-technique contributions to the realisation of lunar reference frames and fundamental physics experiments. Meanwhile, the atomic clock supports precision time-tagging, relativistic tests, and foundational steps toward a standardised lunar timescale. In support of community uptake, NovaMoon will enable **open distribution** of geodetic and timing data products and will coordinate **measurement campaigns** with the scientific community.

NovaMoon delivers strategic value in lunar exploration by enabling capabilities critical for the next phase of robotic and human presence on the Moon. This capability underpins a wide spectrum of high-impact foreseen operational applications:

- Reliable autonomous navigation for rovers, landers, and cargo vehicles
- Support for crewed activities through enhanced spatial awareness
- Precision mapping, construction, and in-situ resource utilisation
- Science missions requiring accurate spatial and temporal localisation, including fundamental physics, and selenodesy

Additionally, NovaMoon will serve as the European anchor in the broader lunar reference and timing infrastructure, promoting seamless interoperability within the LunaNet framework. Its integration with Moonlight services will ensure robust access to high-performance lunar communication and PNT capabilities. By contributing real-world data to standardisation bodies such as the International Astronomical Union (IAU) (International Astronomical Union, 2000; 2024) and the International Association of Geodesy (IAG)—including its services like the International Earth Rotation and Reference Systems Service (IERS)—NovaMoon strengthens Europe's role in shaping the future definition and realisation of lunar reference systems and timescales. The mission is currently in early formulation (Phase A/B1), allowing incorporation of scientific requirements into the baseline design.

# 3. NovaMoon Primary Scientific Objectives

## 3.1 Lunar Reference Frames and selenodesy

### *3.1.1 Introduction*

The precise orbit determination of lunar satellites requires a Moon-centred inertial reference frame. In 2024, following the IAU 2000 GCRS concept, the IAU General Assembly adopted the Lunar Celestial Reference System (LCRS), centred at the Moon's centre of mass and defined by a metric tensor suited to the lunar gravitational environment. The LCRS includes the lunar time scale (TCL), analogous to the barycentric coordinate time TCB, with an epoch set to match TCB on 1977-01-01 at 0h 0m 32.184s. For operational use in the cislunar region, TCL—or a scaled version—must be realised to steer lunar clocks. Alongside the LCRS, a lunar body-fixed frame is essential for surface mapping and navigation. The Mean Earth/Rotation Axis (ME) frame, used for cartography, contrasts with the Principal Axis (PA) frame, aligned with the Moon principal axis of inertia and favoured for orbital dynamics. Transformations between these systems and frames (LCRS, PA and ME) currently use time-dependent rotation angles derived from gravity field models and orientation solutions. A Moon-fixed Lunar Reference Frame (LRF), analogous to Earth's ITRF, will be based on the PA frame. Its realisation depends critically on modelling the Moon's internal structure, including core–mantle boundary (CMB) interactions and deformation. For example, the slightly oblate fluid core can cause surface position shifts of ~15 cm over 30 years, while CMB energy dissipation can cause ~100 m shifts over the same period (Fienga et al. 2024). This section focuses on the space-based realisation of the LRF, reviews open questions and limitations in current models and measurements, while it shows how NovaMoon's targeted observations can improve the LRF and understanding of the Moon's interior.

### *3.1.2 Current limitations and open questions*

#### 3.1.2.1 Limitations of the orientation and tidal models

Achieving high-precision lunar laser ranging (LLR) requires adding empirical terms to all lunar and planetary ephemerides. The latest JPL ephemeris, DE440 (Williams and Boggs, 2020), needs six such terms—twice as many as DE430 (Folkner et al. 2014), INPOP21a (Fienga et al. 2021), and EPM2021 (Pavlov et al. 2016)—to account for periodic LLR residuals that vary by reflector location. Williams et al. (2021) attribute reflector position drifts of ~9 mm/year to orientation model errors, not tectonics. Tidal modelling discrepancies persist between LLR (Williams and Boggs., 2020; Pavlov et al., 2016; Viswanathan et al., 2019a) and LRO altimetry (Mazarico et al., 2014; Thor et al., 2021), with no systematic altimetry errors found (Hu et al., 2023). Potential explanations, including viscoelastic effects on orientation modelling and mantle heterogeneities, remain under investigation.

#### 3.1.2.2 Limitation of the lunar centre of mass realisation

The origin of the lunar reference frame should coincide with the Moon's centre of mass (CoM), but current realisation accuracy remains limited to several tens of centimetres due to unevenly distributed retroreflectors and Earth-based measurement constraints (Pavlov, 2020). A fundamental challenge is the strong correlation (r=0.97) between the CoM's Earth-facing X-component and the reference frame's scale, as both rely on similar Earth-Moon distance measurements (Sośnica et al., 2025a). Achieving millimetre-level accuracy requires continuous tracking of lunar orbiters to overcome current geometric limitations and mitigate ephemeris-induced errors (Fienga et al., 2024). Future progress depends on multi-technique tracking to separate CoM and scale parameters, enable precise lunar navigation, and monitor long-term deformations through improved CoM definitions.

#### 3.1.2.3 Limitation of the present Lunar Laser Ranging observations

LLR data analyses currently achieve centimetre-level accuracy (Pavlov et al., 2016; Viswanathan et al., 2019a; Zhang et al., 2024) but are limited by reflector size, atmospheric perturbations, and sparse observation geometry. Modelling errors, especially in tidal deformation, accumulate to decimeter-level over decades. Persistent unexplained motion requires empirical tidal terms in DE/INPOP/EPM ephemerides. Recent infrared tracking (Grasse, Wettzell) has improved coverage, but significant synodic-phase gaps remain, as one can see in *Figure 2*. The NovaMoon mission could address these challenges by enhancing tracking coverage and enabling millimetre-level accuracy in LLR data analysis.

#### 3.1.2.4 Moon Internal structure: open questions

The Moon, after Earth, possesses the most comprehensive geophysical and geochemical dataset, with interior knowledge second only to our planet. Foundational insights stem from the Apollo program seismic, heat-flow, and LLR retroreflector experiments (Weber, 2014). These have been complemented by gravity and altimetry data from various missions, notably enhancing and extending Apollo-era results (Goossens et al., 2011, Konopliv et al., 2013; Steinbrügge et al., 2019; Thor et al., 2021). Despite this, key aspects of the Moon’s internal structure remain unresolved and open questions are still pending, motivating continued investigation through integrated geophysical observations and advanced modelling approaches.

- ***What is the size and physical state of the lunar core?***

Degree-2 gravity and rotational data suggest the Moon has a small liquid metallic core with density stratification. Seismic (Garcia et al., 2019; Nakamura et al., 1974; Weber et al., 2011), rotational (Williams et al., 2001, Viswanathan et al. 2019b), and paleomagnetic data (Hood et al., 1999) indicate an outer core radius of 200–400 km, though the presence of a solid inner core remains uncertain (0–350 km) (Andrews-Hanna et al., 2023; Williams et al., 2014; Matsuyama et al., 2016). Tidal and thermodynamic models support a layered core structure (Briaud et al., 2023; Antonangeli et al., 2015).

- ***What is the thermal state of the lunar mantle?***

Seismic profiles above the lunar core reveal a low-velocity zone, interpreted as a partially molten layer (Nakamura et al., 1974; Weber et al., 2011). Its existence, debated for decades based on tidal observations (Harada et al., 2014; Matsuyama et al., 2016; Walterová et al., 2023; Williams et al., 2014), appears supported by recent tidal dissipation measurements at monthly and yearly periods (Briaud et al. 2023, Goossens et al., 2024). This semi-molten layer constrains the Moon's thermal state and evolution, from early magma ocean crystallisation to today. Deep moonquake data (800–1200 km) suggest brittle-ductile transition temperatures of 1240–1275 K, dependent on strain rate (Kawamura et al., 2017). Combined tidal and seismic constraints indicate a cold, brittle mantle overlying a partially molten mantle.

- ***What is the thickness and composition of the lunar crust?***

Apollo seismic data initially estimated the Moon's average crustal thickness assuming lateral homogeneity (Khan & Mosegaard, 2002). GRAIL gravity data, however, reveal significant crustal variations (Wieczorek et al., 2013). Combined seismic and GRAIL analyses suggest an average thickness of 29–47 km, ranging from 3 km in Mare Crisium to 61 km in Korolev crater (Kim et al., 2025). The far side crust is generally thicker (>50 km) than the nearside, consistent with morphological and volcanic differences, where dense basaltic eruptions are limited on the far side, implying higher intrusive-to-extrusive ratios (Broquet & Andrews-Hanna, 2024).

- ***How did the lunar dichotomy form and what are its consequences for the lunar interior?***

The lunar dichotomy is characterised by significant compositional differences, notably nearside enrichment in radiogenic elements (Jolliff et al., 2000), which is likely a result of magma ocean crystallisation and mantle overturn, with implications for lunar evolution (Gaffney et al., 2023). Some hypotheses suggest the dichotomy extends into the deep interior (Laneuville et al., 2013). High-accuracy GRAIL data reveal a degree-3 tidal response linked to a ~100 K nearside mantle anomaly (Park et al., 2025), possibly explaining discrepancies in degree-2 Love numbers (Thor et al., 2021). NovaMoon, near the South Pole, could help constrain the dichotomy depth and magnitude.

#### 3.1.2.5 Earth Orientation Parameters

Earth Orientation Parameters (EOPs), defining the Earth's absolute orientation relative to celestial frames, are needed for the transformation between the reference frame in which all satellite orbits are integrated and the ground-based observatories with positions given in terrestrial reference frames. Hence, all satellite and interplanetary missions rely on EOPs, which are hard to predict because of the impact of the atmosphere, land hydrology, and oceans that continuously change the Earth's angular momentum. The absolute values of EOPs are primarily based on VLBI and can be supported by LLR (Singh et al. 2022). Pole coordinates can be derived from GNSS and SLR (Zajdel et al., 2020, Mireault et al. 1999, Hefty et al. 2000, Rothacher et al. 2001), but UT1, nutation, and precession require VLBI ( Nothnagel et al., 2016) Satellite techniques can estimate UT1 derivatives (LOD excess) but suffer from long-term instabilities. VLBI uniquely links terrestrial and celestial reference frames, yet it has ~2–3-week latency due to data processing. On the other hand, current UT1 predictions have 0.8 ms error after 10 days, equating to 37 cm at the Earth's equator or 15 m on the Moon (Kur et al., 2022). NovaMoon can improve the latency and accuracy of EOPs, which are needed for proper linking of satellite and interplanetary missions to the terrestrial reference frames. NovaMoon can also provide independent validation of EOPs delivered by VLBI observations.

*3.1.3 NovaMoon contribution*

The NovaMoon objectives for the definition of the Moon reference system and frame and the study of its internal structure are given in Table 1. We will now explain how NovaMoon should meet these objectives.

3.1.3.1 Observation Setups

We simulate the impact of including, for a 5-years lifespan, NovaMoon lunar laser retro-reflector (LLRR), VLBI transmitter, and radio tracking alongside 40 years of LLR data in constructing lunar ephemerides (Williams and Boggs, 2020; Viswanathan et al., 2019a). Key parameters include the Moon's initial positions and velocities (Moon PV) relative to the Earth, initial Euler angles (Moon orientation), and Moon mass, which define the origin, orientation, and scale of the lunar reference frame (LRF). Other influential parameters are the Moon's moment of inertia, degree-3 gravity field, fluid core rotation vector and oblateness, core-mantle tidal dissipation, and LLRR coordinates, which provide an initial LRF realisation. Tidal surface deformation parameters consider the real and imaginary parts of the tidal Love number $h_2$ for surface radial deformation (monthly and yearly) and the modulus of the tidal Love number $l_2$ for monthly surface tangential deformation. We perform covariance matrix analysis starting from 40 years of LLR observations and compute the relative differences between the covariances obtained with the different additional setups described in the following for the parameters of interest described above. The ESA mission analysis GODOT has been used for these simulations, relying on the Moon dynamical modelling from INPOP21a lunar and planetary ephemerides (Fienga et al. 2021).

*3.1.3.1.1 NovaMoon as an additional Lunar Retro-Reflector at the South Pole*

We consider LLR observations to NovaMoon at the South Pole for 5 years with 1 cm (Setup 1) and 1 mm (Setup 2) standard deviation, in addition to 40 years of existing LLR data. The temporal and synodic-angle distribution of observations is important for realistic simulations (*Figure 2*). For the last 18 months, ~1/3 of Apollo 15 observations used medium-to-small LRRs (*Figure 2*), similar to NovaMoon, so we modelled NovaMoon sampling as 1/3 of Apollo 15's data adapted to the South Pole. Setup 1 represents a conservative scenario, while Setup 2 is very optimistic, assuming 1 mm $\sigma$, that accounts for instrumental and post-processing uncertainties. We remind that LRR thermal deformations are also at the mm-level and that Zhang et al. 2024 estimated atmospheric effect of about 5 mm.

3.1.3.1.2 NovaMoon as a new VLBI transmitter

We have considered angular measurements between NovaMoon at the South-Pole and *ICRF3* (Charlot et al., 2020) sources with two standard deviations for the measurements (1 mas and 0.1 mas). The data is supposed to be obtained over 5 years with 2 weekly sessions of 24 hours. A minimum angle of 10° elevation is required, as well as a 5° maximum angle of separation between NovaMoon and the ICRF3 sources. These simulations correspond to Setup 3 and 4 in Table 2. This VLBI concept used in these simulations is similar to Delta-DOR tracking measurements of interplanetary missions used in the planetary ephemeris construction (Folkner and Border 2015).

*3.1.3.1.3 Microwave tracking of NovaMoon from the southern hemisphere*

As it was already identified by Hofman et al. (2013) and Fienga et al. (2014), having observations from the Earth's southern hemisphere will bring new constraints, as most of the LLR stations are located in the northern hemisphere. This is the reason why we assessed the contribution of a 2-way radio link between NovaMoon and the ESA Malargue station in Argentina. We supposed continuous range observations with one normal point of 1 cm standard deviation, each 6 hours for a period of 5 years (Setup 5). This accuracy is technically achievable, but requires an appropriate transponder and, likely, a K-band link. A minimum angle of 10° elevation is required. In order to optimise the access to potential Southern Hemisphere stations, we focus on periods for which the Moon's synodic angles are complementary to those covered by LLR, mainly during new moon and full moon phases. Another possible optimisation can also be to use at least two stations in the Southern Hemisphere, such as Malargue and New Norcia, for example. With this configuration, about 4400 normal points will be acquired by the two stations over 5 years.

3.1.3.2 Impact on the LRF definition and lunar ephemerides

*Figure 3* shows the improvements brought in the covariance matrices of the studied parameters considering the use of LLR to NovaMoon with 1 cm accuracy as described in section 3.1.3.1.1 (yellow curve), the addition of the 2-way microwave link described in section 3.1.3.1.3 (green curve), and finally the addition of VLBI-like measurements (purple curve) proposed in section 3.1.3.1.2. The impact of VLBI is obvious for all parameters, but in particular in the definition of the frame realisation with LLRR coordinates on the Moon's surface. The complementarity of the 2-way microwave link is also visible for the parameters related to the internal structure, such as the spin rotation vector of the fluid core and its flattening, but also the determination of the Moon's mass and moment of inertia.

All these improvements will lead to a more accurate definition of the LRF but will also help identify the possible sources of mismodeling in the present lunar ephemerides as discussed in Section 2.1. The introduction of viscoelastic components in the modelling of the tidal dissipation at the boundary between the fluid core and the mantle could be one of the new directions that can be explored, thanks to the obtained improvements.

3.1.3.3 Impact on the Moon's internal structure

Table 2 also reports improvements in the determination of Moon internal structure parameters, such as the moment of inertia, fluid core properties (rotation vector and oblateness), dissipation mechanisms at the CMB, and tidal surface deformation. Setup 5 (continuous 2-way radio-link with 1 cm standard deviation) provides ~30% improvement for all internal structure parameters, except for the Love number $l_2$. Setup 2 (1 mm LLR) yields the largest improvements, particularly for fluid core characterisation and CMB dissipation mechanisms. These enhancements are critical for refining lunar ephemerides and understanding long-term Moon rotation and thermal evolution. Setup 4 (VLBI) also improves internal structure determination, especially for the degree-3 components of the lunar gravity field. Furthermore, as indicated in (Williams et al. 2012), a better understanding of the Moon libration and its resonances will lead to a better characterisation of its inner core. This will be possible thanks to a better determination of the Moon orientation parameters obtained with NovaMoon and, in particular, with the VLBI (Setup 4). Across all setups, improvements in the Love number $h_2$ are modest. The maximum gain occurs for the real parts of the yearly and monthly tidal excitations, while improvements in the imaginary parts remain below 1%. Also common to all setups, while the $l_2$ uncertainty decreases, it remains too large for definitive detection. As shown by Briaud et al. (2023), Goossens et al. (2024) and others, accurate knowledge of the Love numbers is a key factor in constraining the Moon’s internal structure, alongside precise modelling of its orientation. Parameters related to CMB dissipation mechanisms also provide access to the tidal quality factor Q (Williams et al, 2001, 2014), which is directly correlated to the Moon’s internal structure, as one can see in Figure 4. This Figure also shows the sensitivity of the real part of the Love number $h_2$ to key interior properties such as density, layer thickness, shear modulus, and viscosity.
Both Q and $h_2$ are particularly responsive to the core's size and state (liquid, partially liquid, or solid), offering critical information on the current state and past evolution of the lunar interior. Moreover, these parameters are influenced by the shear modulus of the low-viscosity zone (LVZ), which is closely linked to the thermal state of the lunar mantle. High-precision measurements of Q and $h_2$ could therefore confirm or rule out the presence of a partially molten layer above the core and determine whether this layer extends globally or only regionally.
*Figure 3* also shows the sensitivity of the real parts of $h_2$ and $l_2$ at monthly and yearly periods to the melt fraction within the LVZ.
As shown, achieving determinations of $h_2$ and $l_2$ with better than 6% accuracy - a factor-of-two improvement over current discrepancies between LLR and altimetry measurements - would place stringent constraints on the LVZ melt fraction and provide crucial input for geophysical and thermodynamic models studying the Moon's evolution since its formation.
Finally, preliminary simulations further indicate that a degree-1, order-1 shear modulus anomaly in the lunar mantle, like that described by Park et al. (2025), would produce surface displacements of approximately 1 mm. Such displacements are challenging to detect with present-day LLR accuracy but could be measurable with next-generation LLRR.

#### 3.1.3.4 Contribution of the Moonlight-NovaMoon link

The NovaMoon mission will use the nearly continuous signal of Moonlight satellites to determine the Moon's centre of mass with millimetre-level accuracy, analogous to how Earth's geocenter is determined via ground-based SLR stations tracking geodetic satellites (LAGEOS and LARES, Pearlman et al., 2019; Sośnica et al., 2025b, Chang et al. 2013). By combining ground (lander) and space (satellite) observations, this approach will improve the lunar reference frame, decorrelate the X-component of Moon's center from its scale, and enhance the accuracy of lunar satellite orbits, positioning, and navigation products.

#### 3.1.3.5 Contribution of coherent Doppler tracking from ground

Using the same method as for a 2-way radio Doppler tracking of a spacecraft, it is also possible to acquire Doppler measurements between NovaMoon and directly a station on Earth, (i.e. Marlague), with an accuracy of about $10^{-5}$ Hz in X- or Ka-band and a continuous tracking similar to Setup 5. This technique, analogous to differential lunar laser experiments (Turyshev et al. 2021; Zhang et al. 2024) but without optical limitations, offers superior performance by reducing tropospheric errors that degrade range measurements by ≥5 mm (Zhang et al. 2024). Radio Doppler tracking will enhance studies of lunar dynamics and orientation, improving understanding of the Earth-Moon system (Sections 3.2–3.3), while providing continuous, high-accuracy data less affected by Earth's atmospheric disturbances compared to laser ranging.

#### 3.1.3.6 Impact on EOP

The NovaMoon benefits are not limited to lunar science objectives but could lead to a valuable contribution to Earth's science and/or interplanetary missions. EOPs provided by VLBI are based on two 24h sessions per week and 1h intensive session per day to estimate UT1 only. Moreover, the latency for VLBI solutions is about 2 weeks. NovaMoon can provide nearly continuous observations (70%) with near real-time solutions for EOP determination that are needed for linking the celestial and terrestrial reference frames for all satellite and interplanetary missions. It is expected to improve the latency, continuity, and accuracy of EOPs, offering new possibilities for integrating terrestrial-lunar reference frames. NovaMoon will also allow for studying high-frequency Earth and Moon rotation parameters, i.e., sub-daily polar motions (Zajdel et al., 2021, Hefty et al. 2000, Rothacher et al. 2001), sub-daily LOD, and high-frequency libration parameters, as well as long-term nutation processes.

Geodetic parameters require redundancy to provide reliable and unbiased parameters. Pole coordinates can be derived by GNSS, VLBI, and SLR. Unfortunately, nutation parameters and UT1 are currently based only on VLBI in operational products. NovaMoon can not only improve the latency and accuracy of UT1 and nutation parameters but also provide redundancy that is currently missing for VLBI data.

Today, LLR can be used for EOP determination. However, the number of LLR observations is very limited, especially during the full Moon and new Moon, because of the issues of reflected photons from too-bright lunar surface and the issues with LLR telescope calibration on too-dark craters, respectively. Therefore, LLR observations are typically sparse because the two-way laser measurements at the lunar distances are very challenging for LLR observatories. NovaMoon can provide nearly continuous observations to overcome VLBI and LLR limitations in EOP determination, especially in terms of the solution latency. Table 2 shows a qualitative overview of the potential impact on EOPs' accuracy, continuity, and latency based on VLBI and LLR, as well as what is expected with NovaMoon.

All lunar missions and operations require positioning in the pre-defined reference frames, such as the International Lunar Reference Frame (ILRF). The positions of Earth observatories are expressed in ITRF realisations. The series of transformations is required for linking ITRF positions to ILRF and vice versa. Table 3 summarises the errors for real-time and post-processing lunar operations due to transformation procedures and reference frame realisations at each transformation step.

For real-time and post-processing solutions, the dUT1 component provides the largest error source of 15 m for real-time, when based on predictions, and around 19 cm for post-processing. The total transformation error is 27 cm in post-processing. The error of 15 m can be insufficient for high-precision lunar missions, such as landing or navigation on the lunar surface. Having a permanent station on the lunar surface, such as NovaMoon, and assuming the positioning accuracy of 15 cm would improve the dUT1 predictions by a factor of 100x for (near) real-time users. For the post-processing applications, EOPs can be improved by nearly a factor of two. In this way, NovaMoon can become fully competitive with VLBI in terms of providing all EOPs. High-quality EOPs, including dUT1, are also critical for all interplanetary missions for providing accurate spacecraft positions in real-time.

#### 3.1.3.7 Possible "synergies"

NovaMoon, slated to reach the lunar surface in the 2030s, will join a suite of upcoming missions deploying LLR reflectors and seismometers—including the CLPS FarSide Seismometer (2026; Panning et al. 2023), Chang'e-7 (Zou et al. 2020), and Artemis III LEMS (2027; DellaGiustina et al. 2025)—primarily clustered near the South Pole. As part of the Lunar Geophysical Network (Fuqua Haviland et al. 2022), NovaMoon will distinguish itself as the first *multi-technique selenodetic reference station*, combining LLR, VLBI, radio-link (Moonlight-enabled 1-/2-way ranging), and an atomic clock on a single platform. With a planned 5-year operational lifespan (versus LEMS's 2 years), it will provide long-term, high-precision geodesy to complement other missions' shorter-term deployment

Table 1: NovaMoon objectives for Lunar reference system, frame and internal structure. The colors and indices indicate (from 0-Negligible to 5-Critical ), the impact of the different proposed instrumentation on the topic as obtained in Section 3 and Table 2.

| Objectives | Instrumentation | | | | | | |
|---|---|---|---|---|---|---|---|
| | LLR [cm] | | VLBI [mas] | | 2-way link with Earth SH continuous tracking | 1-way/2-way link to Moonlight | Atomic clock |
| | 1 | 0.1 | 1 | 0.1 | 1 cm | | |
| Consolidate and improve the realisation of the lunar reference frame. | 2 | 4 | 1 | 5 | 3 | 0 | 5 |
| Improve the links (i.e., rotation and transformations) between lunar, Earth terrestrial and celestial reference frames | 2 | 4 | 3 | 5 | 3 | 4 | 5 |
| Contribute to the generation of precise lunar ephemeris. | 2 | 4 | 3 | 5 | 3 | 0 | 0 |
| Improve the precise determination of the lunar rotation parameters. | 2 | 4 | 1 | 5 | 3 | 0 | 0 |
| Improve the precise determination of the lunar tidal deformation parameters | 0 | 4 | 0 | 3 | 3 | 0 | 0 |
| Refine the knowledge of the lunar gravity field | 0 | | 0 | 5 | 1 | 2 | 0 |
| What is the size and physical state of the lunar core? | 0 | 4 | 0 | 4 | 1 | 0 | 0 |
| What is the thermal state of the lunar mantle? | 0 | 1 | 0 | 1 | 1 | 0 | 0 |
| What is the thickness and composition of the lunar crust? | 0 | 0 | 0 | 0 | 0 | 0 | 0 |
| How did the lunar dichotomy form and what are its consequences for the lunar interior? | 0 | 2 | 1 | 2 | 2 | 0 | 0 |

| | VLBI | NovaMoon | LLR |
|---|---|---|---|
| **Accuracy** | **High (lateral)** | **High** | **High (radial)** |
| **Continuity** | No,<br>e.g. 2*24h sessions per week (depending on ground segment availability) | **YES (~70%)** | No LLR observations during<br>full and new Moon |
| **Latency** | Around 2 weeks | **near real-time** | Several days |

Table 2: Impact of VLBI, LLR, NovaMoon continuous tracking for EOP determinations

| | **Budget error - ITRF - ILRF** | **Real-Time** | **Post-processing** |
|---|---|---|---|
| 1 | ITRF realization | 2 cm (real-time accessibility by GNSS) | 0.5 cm |
| 2 | Polar motion (x, y) - ITRF-ICRF | 0.2 mas (0.6 cm Earth, 37 cm at lunar distance) | 0.03 mas (0.1cm Earth, 6 cm at lunar distance) |
| 3 | **dUT1 - ITRF-ICRF** | **0.8 ms (37 cm Earth surf., 15 m at lunar distance)** | **0.01 ms (0.5 cm Earth, 19 cm at lunar distance)** |
| 4 | Sub-daily Earth rotation | 0.007 mas (0.03 cm Earth, 1.3 cm lunar distance) | 0.007 mas (0.03 cm Earth, 1.3 cm lunar distance) |
| 5 | Nutation - ITRF-ICRF | 0.1 mas (0.3 cm Earth, 19 cm at lunar distance) | 0.06 mas (0.2cm Earth, 11 cm lunar distance) |
| 6 | ICRF – LCRS (lunocenter) | 15 cm (surface, ~1.2 m orbiter) | 15 cm (surface, ~1.2 m orbiter) |
| 7 | LCRS – ILRF (Euler angles) | 9 cm (surface, ~0.7 m orbiter) | 9 cm (surface, ~0.7 m orbiter) |
| | **Total: 1+2+3+4+5+6+7** | **~15 m at lunar distance** | **~27 cm at lunar distance** |

Table 3: Error budget for the transformation between terrestrial and lunar reference frames

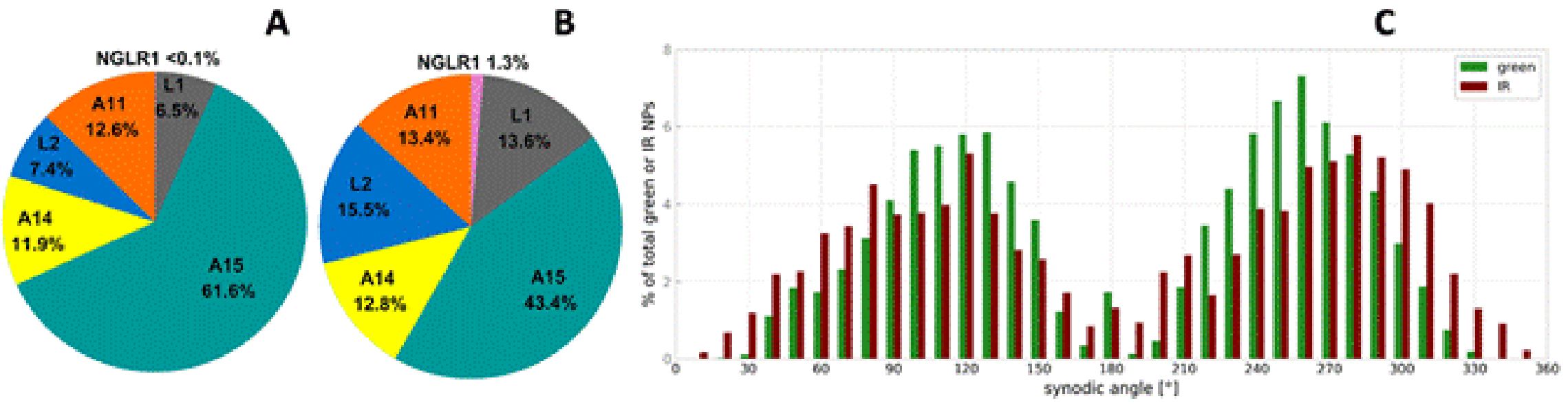

*Figure 2:* Panel A and B: Data distribution over the 6 LLR reflectors for all 34870 normal points from 04.1970 to 05.2025 (A) and only for the past 1564 normal points from 01.2024 to 05.2025 (B). Panel C: Distribution of green and infrared LLR normal points over the synodic angle.

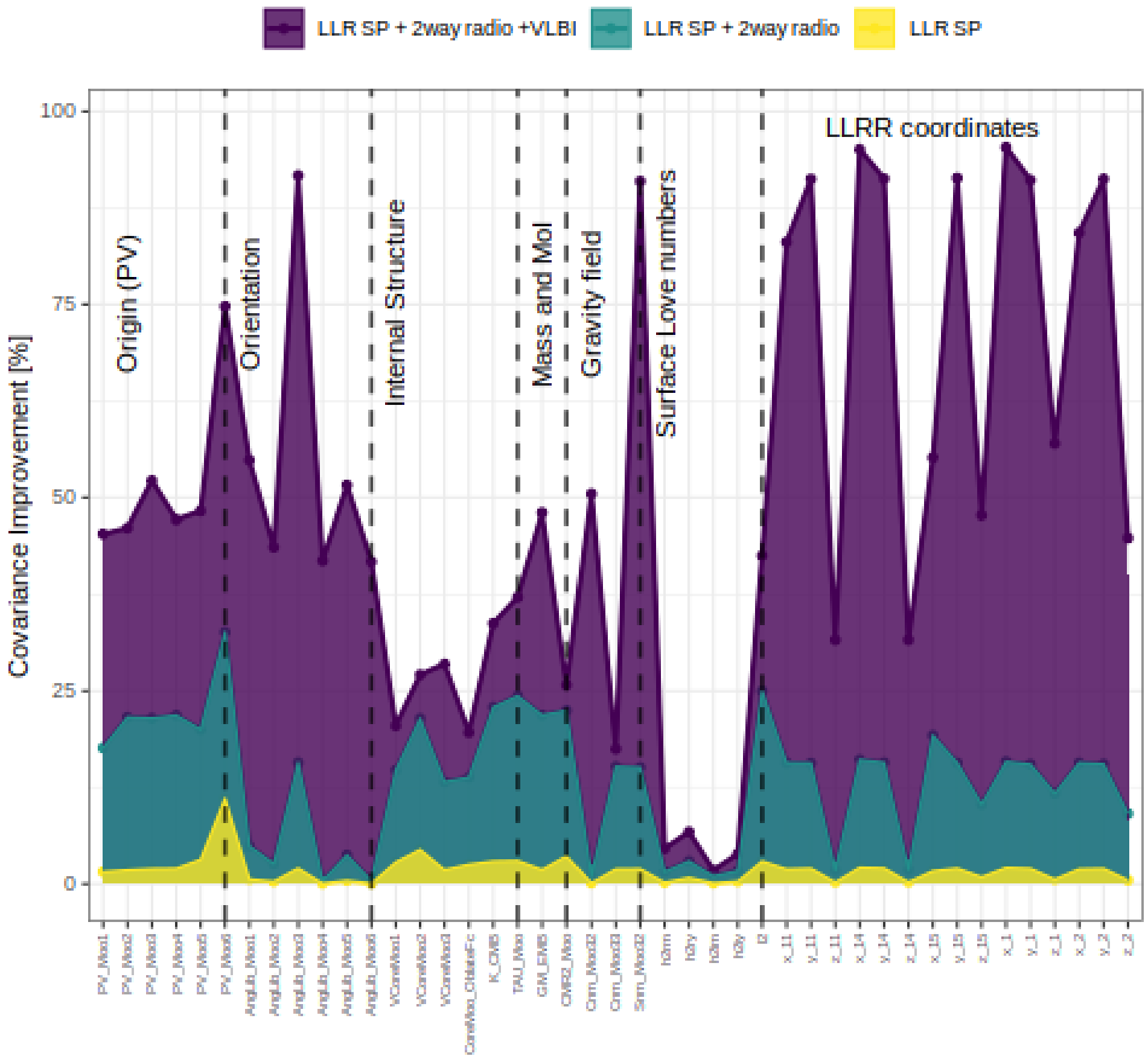

*Figure 3*: Relative improvements (%) in the Lunar Reference Frame (LRF) through covariance matrix analysis of 40 years of LLR data. Categories of parameters are given: the LRF origin (Moon position/velocity), orientation (Euler angles), and scale (mass) the moment of inertia (MoI); LLRR coordinates; and lunar ephemeris/internal structure parameters such as Degree-3 gravity field coefficients ($C_{33}$, $C_{32}$, $S_{32}$), Core dynamics (oblateness, rotation), tidal dissipation at the core-mantle boundary (K_CMB, TAU_Moon). Parameters for the tidal surface deformations (Love numbers) are also indicated.

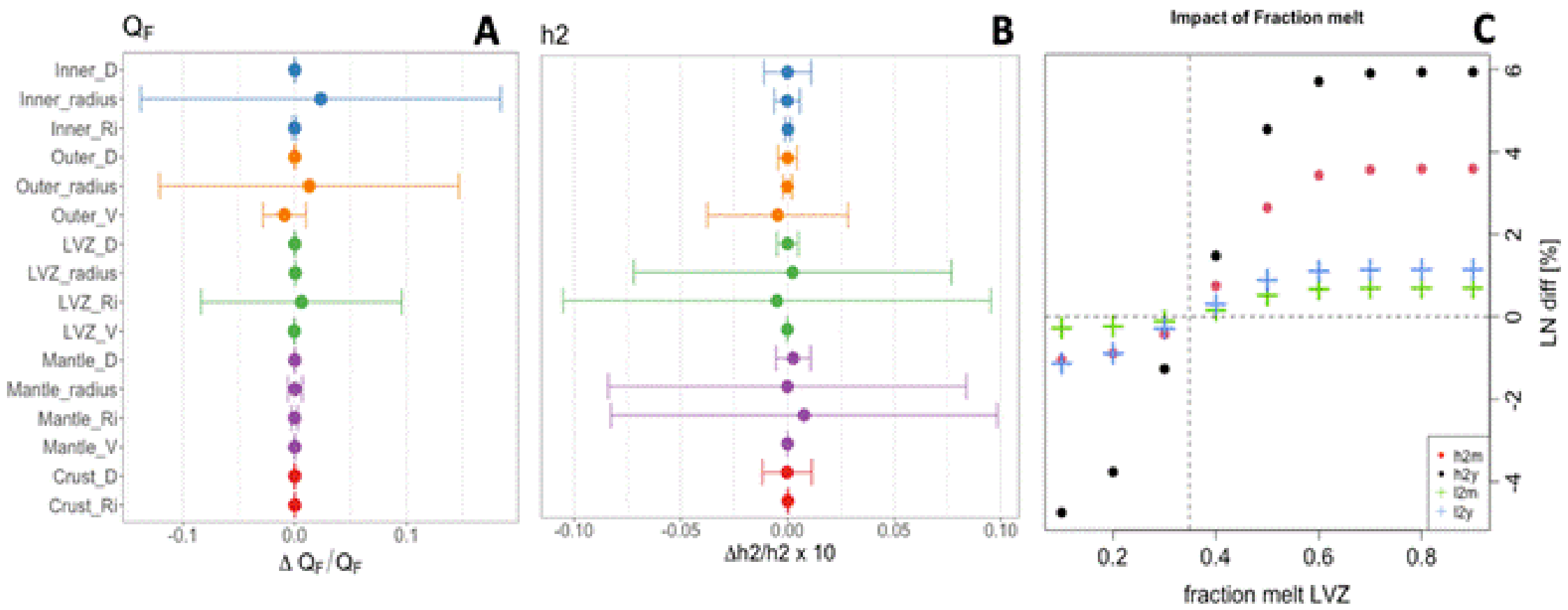

*Figure 4*: Panels A and B: Sensitivity of the monthly Quality factor $Q_F$ and real part of the Love number $h_2$ to the main characteristics of the Moon internal structure layers. These panels are extracted and adapted from (Briaud and al.et al. 2023). Panel C: sensitivity of the real part of the Love number h2m and l2m for the monthly period (resp. yearly period h2y and l2y) to the fraction melt present in the low viscosity zone (LVZ).

## 3.2 Lunar Time scales

### *3.2.1 Introduction*

Based on the concept of the Geocentric Coordinate Time (TCG) as given in the IAU 2000 Resolution B1.3 (Petit and Luzum 2010; Soffel et al., 2003), the IAU Resolution II of the XXXIInd General Assembly recommended in August 2024 to define the Lunar Coordinate Time (TCL) associated with the Lunar Celestial Reference System (LCRS) and the corresponding metric tensor. The transformation between TCL and TCB is given by the relations in IAU Resolution B1.5 of the XXIVth General Assembly (2000), with the substitution of Moon's related quantities for those related to the Earth. The origin of TCL is also defined by the IAU Resolution 2 (2024) so that the reading of TCL be exactly 1977 January 1, 0h 0m 32.184s at the centre of the Moon when the reading of TCB is 1977 January 1, 0h 0m 32.184s at the centre of the Moon.

On Earth, for practical use, a coordinate time scale named Terrestrial Time (TT) has been defined by rescaling the coordinate time TCG, so that the mean rate of TT agrees as much as possible with that of the proper time of a clock located on the geoid. The International Atomic Time (TAI) is then a realisation of TT, built from an ensemble of free-running atomic clocks distributed all over the world. The rate of TAI is obtained from some primary frequency standards corrected for the gravitational redshift associated with their altitude above the geoid.

For practical use in the cislunar region, it is necessary to define a reference coordinate time, and hence its realization. Let's call TL (Lunar Time) the coordinate time that will be either the TCL or a certain linear scaling of it (Bourgoin et al., 2025). If a scaling is applied similarly to how it is done for TT on the Earth, TL would agree as much as possible with that of the proper time of a clock located on the lunar geoid, this scaling would be around $3 \times 10^{-11}$. Associated with this theoretical definition, it will be necessary to determine which is the realisation of the reference time scale on which any lunar clock should be steered.

### *3.2.2 The realisation of TL*

As long as there is no atomic clock on the Moon, there will be no local realisation of TL. The reference used for the lunar timing will necessarily be based on Earth clocks realizing UTC using some time transfer between lunar clocks and the Earth clocks, and applying the relativistic corrections computed from ephemerides as proposed by different works, depending on the coordinate system used for the time transfer expression: (Kopeikin and Kaplan 2024, Ashby and Patla, 2024; Zhang et al., 2025; or Bourgoin et al., 2025).

The first clocks deployed with NovaMoon will be among the candidates listed in Section 6. The frequency accuracy of the selected clocks will determine the level of accuracy of the realisation of TL that will be available on the Lunar surface. If these clocks have to be used for applications requiring TL with higher accuracy, a steering on Earth clocks realising UTC will be needed, see Bourgoin et al. (2025) for more details. We should also mention here that the steering will compensate for both the gravitational redshift due to the altitude, maximum $3 \times 10^{-13}$ (or about 30 ns/day) as seen in Figure 5 or Figure 6, centred on the South Pole, and the clock inaccuracy.

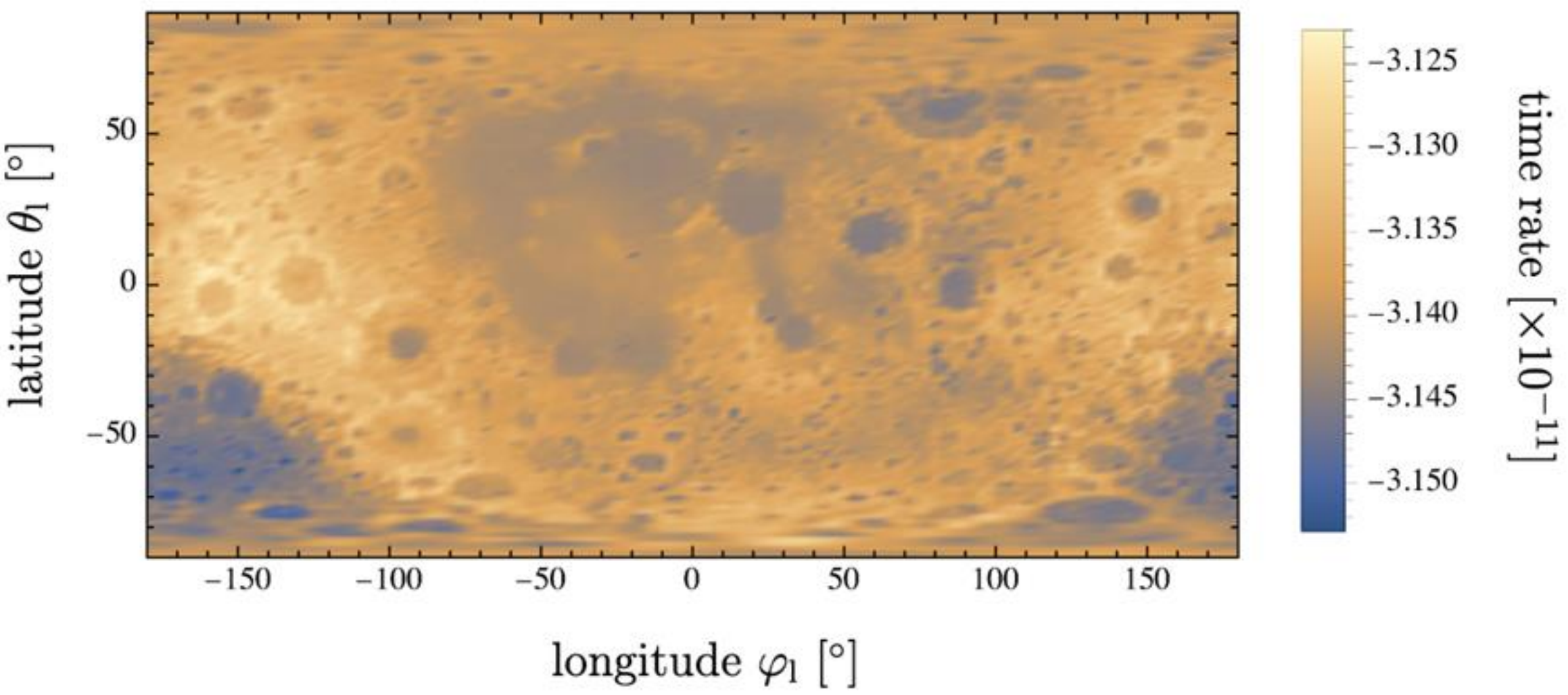

*Figure 5.* Map of the relative frequency difference between the proper time of a clock at rest on the lunar surface and (a) TCL, taken from Bourgoin et al (2025)

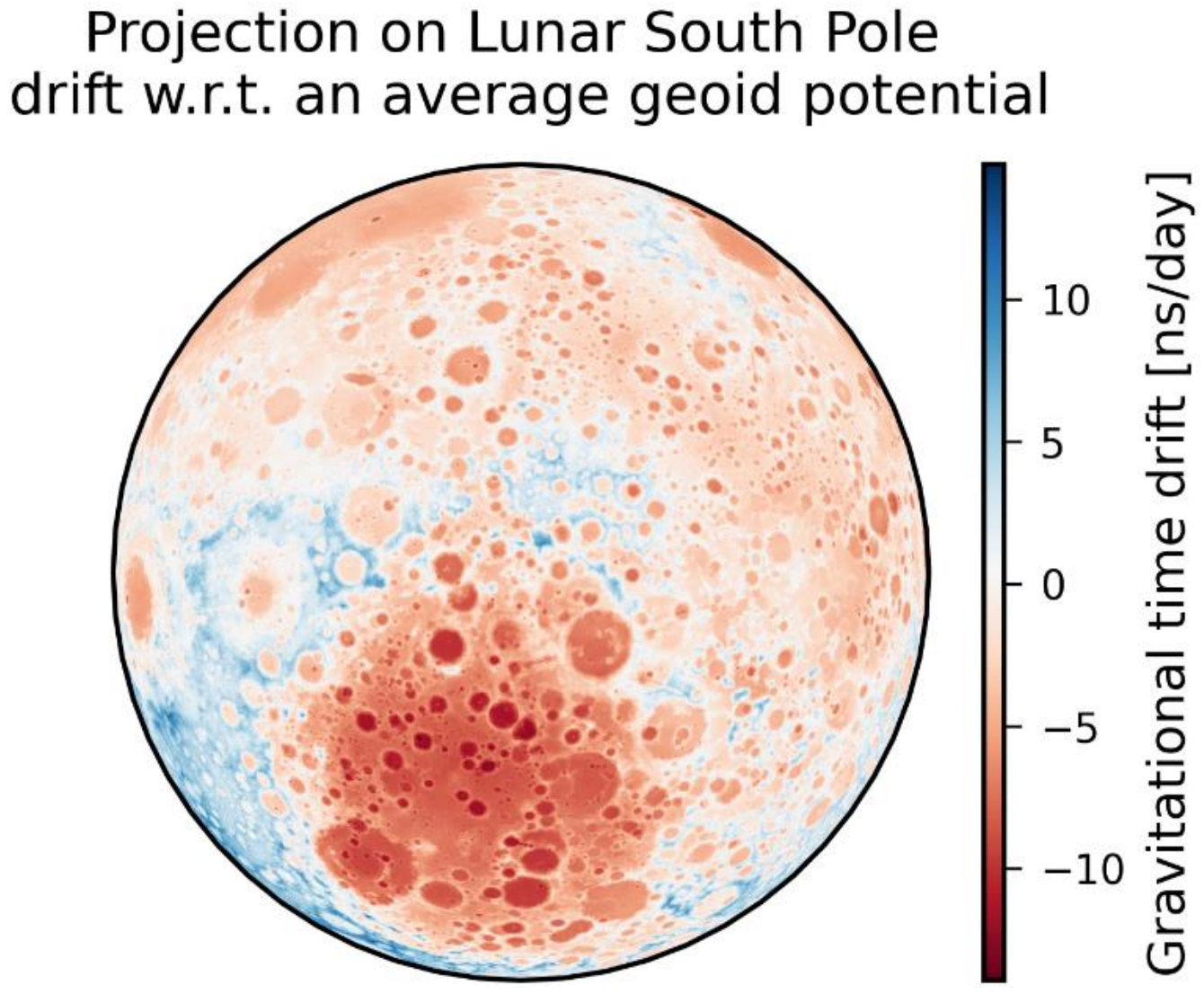

*Figure 6.* Orthographic projection, depicting the Lunar South Pole at the center (near side at the top). The geoid potential (where drift is 0) was arbitrarily chosen as 2.82100e+06 m²/s². (Seyffert, 2025)

### *3.2.3 Validation of Moonlight reference time*

The Lunar Time (TL) will first be made available only from the Moonlight satellite signals, providing the synchronisation error between the satellite clocks and $t_{sys}$, which is the reference time scale of the LCNS system, aligned on TL. From range measurements, the user will then obtain, in the PNT solution, $t_{user} - t_{sys}$ , where $t_{user}$ is the user clock. As for the Earth GNSS, $t_{sat} - t_{sys}$ will be a prediction based on previous measurements of time transfer from the Moonlight satellite with Earth clocks, realising UTC, and taking into account the relativistic corrections. The rapid motion of the satellite and the uncertainties on its position will, however, be a limiting factor for the measurement accuracy, and the stability of the LCNS clocks will be the limiting factor for the quality of the prediction of $t_{sat} - t_{sys}$.

At the NovaMoon station, it will be possible to measure the time difference $t_{NovaMoon} - t_{sys}$ - where $t_{NovaMoon}$ is the NovaMoon clock - from the Moonlight PNT solution. In parallel, a direct time transfer link with some Earth clock realising UTC, named UTC(k), or traced to a UTC(k), will provide a direct solution of the $t_{NovaMoon} - UTC(k)$, and after removing the modelled relativistic terms, provide $t_{NovaMoon} - TL$ (see Figure 7). The difference between these two independent measurements will provide a validation of the $t_{sys}$ obtained through the PNT solution. As the NovaMoon station will be known and stationary, the uncertainty budget of the measurement $t_{NovaMoon} - TL$ should be smaller than the one of $t_{sat} - t_{sys}$ due to the rapid motion of the satellite and the uncertainties on its position. Also, thanks to the fixed position of the NovaMoon station, it will be possible to determine for each Moonlight satellite separately the difference $t_{sat} - t_{sys}$. Adding this to the $t_{NovaMoon} - TL$ determined from time transfer with the Earth will provide a validation of $t_{sat} - t_{sys}$ broadcast by the Moonlight satellites.

If the NovaMoon clock frequency accuracy is better than $1e^{-11}$, which is the case for the PHM in the proposed options (see Table 4), then the NovaMoon clock, as measured from the Earth or from Moonlight satellites, will be a first demonstration of the gravitational redshift of a clock on the lunar surface. This clock frequency accuracy, however, will not be sufficient to distinguish the NovaMoon altitude, which would require a frequency accuracy at the level of $1e^{-13}$ or better. On the other side, the clock frequency stability will allow to measure the frequency variations between the Earth and Moon clocks as predicted by GR coordinate time transformations In general, the difference between the proper time of a clock on the Moon and the proper time of a clock on the Earth will be measured from time transfer between the Earth and the Moon, and will be the sum of three contributions: (i) the difference between clocks' proper time and their local coordinate time (TCL and TCG), (ii) the light-time computed in an intermediate reference system (e.g., LCRS, GCRS, or BCRS), and (iii) the difference between the local coordinate times and the intermediate coordinate time used at step (ii). Considering LCRS as the intermediate reference system and assuming a one-way time/frequency transfer, the contribution from item (iii) (i.e. TCG-TCL) in the proper time difference shows a 1.5 µs/day secular drift (i.e. a $1.710^{-11}$ relative frequency shift) plus monthly and diurnal variations. The monthly term has an amplitude of about 300 ns, and the diurnal terms have an amplitude of about 60 ns (Bourgoin et al., 2025). If the NovaMoon clock is a PHM, RAFS or miniRAFS, it will allow to confirm these theoretical results. Indeed, let's consider the possible clock stabilities as indicated in Table 4, and simulate here for each of them the clock behaviour over several months. Figure 8 provides the so-obtained clock results, first corrected for a deterministic frequency drift, characteristic of this kind of clocks, computed over the whole period (hence not affecting a possible detection of a monthly or daily periodic term). In parallel are displayed the expected periodic variations mentioned above. In addition to these simulations, we also analyse true clock data from three Galileo satellites, two operating with a RAFS master clock, and one with a PHM master clock. From these results, it is confirmed that using several months of time transfer between the NovaMoon clock and an Earth clock, it will be possible to see the monthly periodic variations, whatever the clock used among the possible candidates. Of course, observing diurnal variations will be more challenging due to the non-continuous measurements between the NovaMoon clock and the same Earth clock.

### *3.2.4 Towards an ensemble time scale on the Moon*

Having a clock on the Moon surface, with quasi-continuous possible comparison with Moonlight satellites will allow to start building a local ensemble time scale which would pave the way towards a realisation of TL on the Moon as we are doing for UTC on the Earth. When other clocks will be put on the Moon (on future NovaMoon stations or other), it will furthermore be possible to compare and synchronise their clocks using the Common View of Moonlight satellites; all these clock comparisons being the cornerstone of the development of the ensemble lunar time scale.

*3.2.5 Link with Earth GNSS*

The GNSS receiver of NovaMoon, as connected to the atomic clock, will provide, in the PNT solution, the difference $t_{NovaMoon} - bUTC_GNSS$, where $bUTC_GNSS$ is the realisation of UTC broadcast by the GNSS constellation. This will provide an additional mean of comparison of the NovaMoon clock with UTC, and from there, after correction for the relativistic effects, a possible steering of NovaMoon clock on TL. It must be noted here that only one GNSS satellite is needed to get the clock solution as the NovaMoon station will be fixed with a known position on the Moon surface.

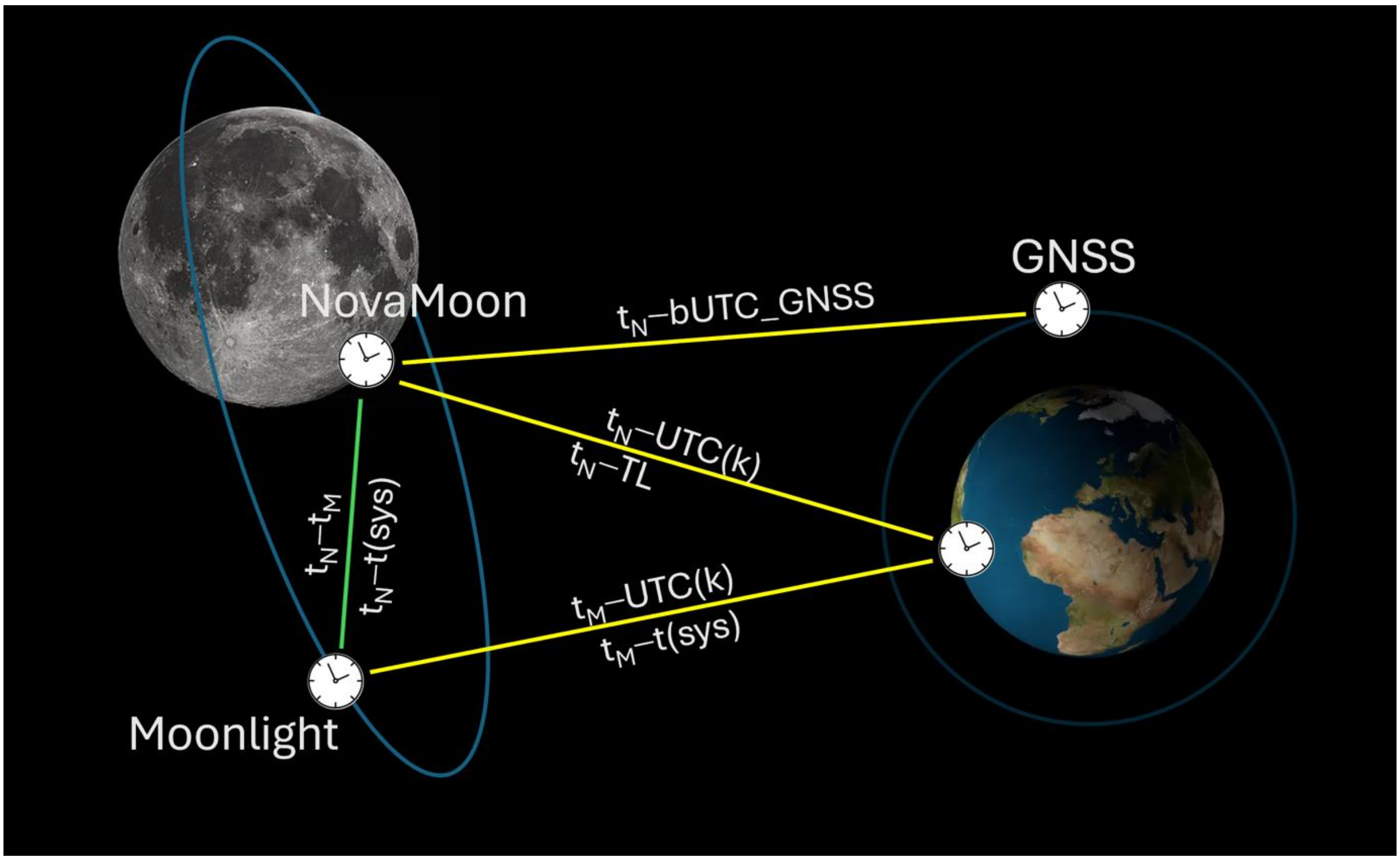

Figure 7. Schematic view of the timing validation architecture. The quantities indicated correspond to the time differences obtained from the time transfer link between the two clocks. The NovaMoon clock is named as $t_N$, and the Moonlight clock as $t_M$.

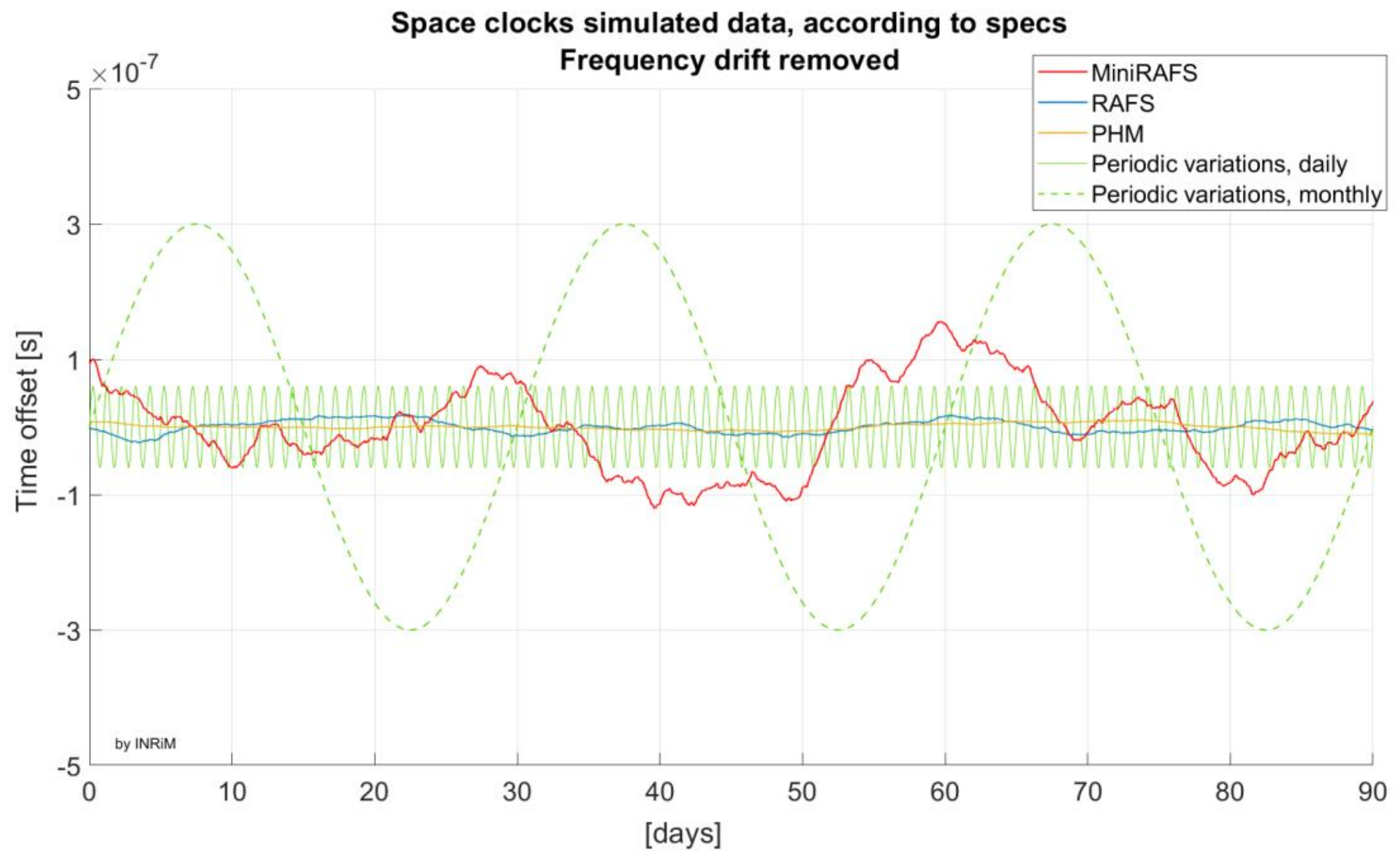

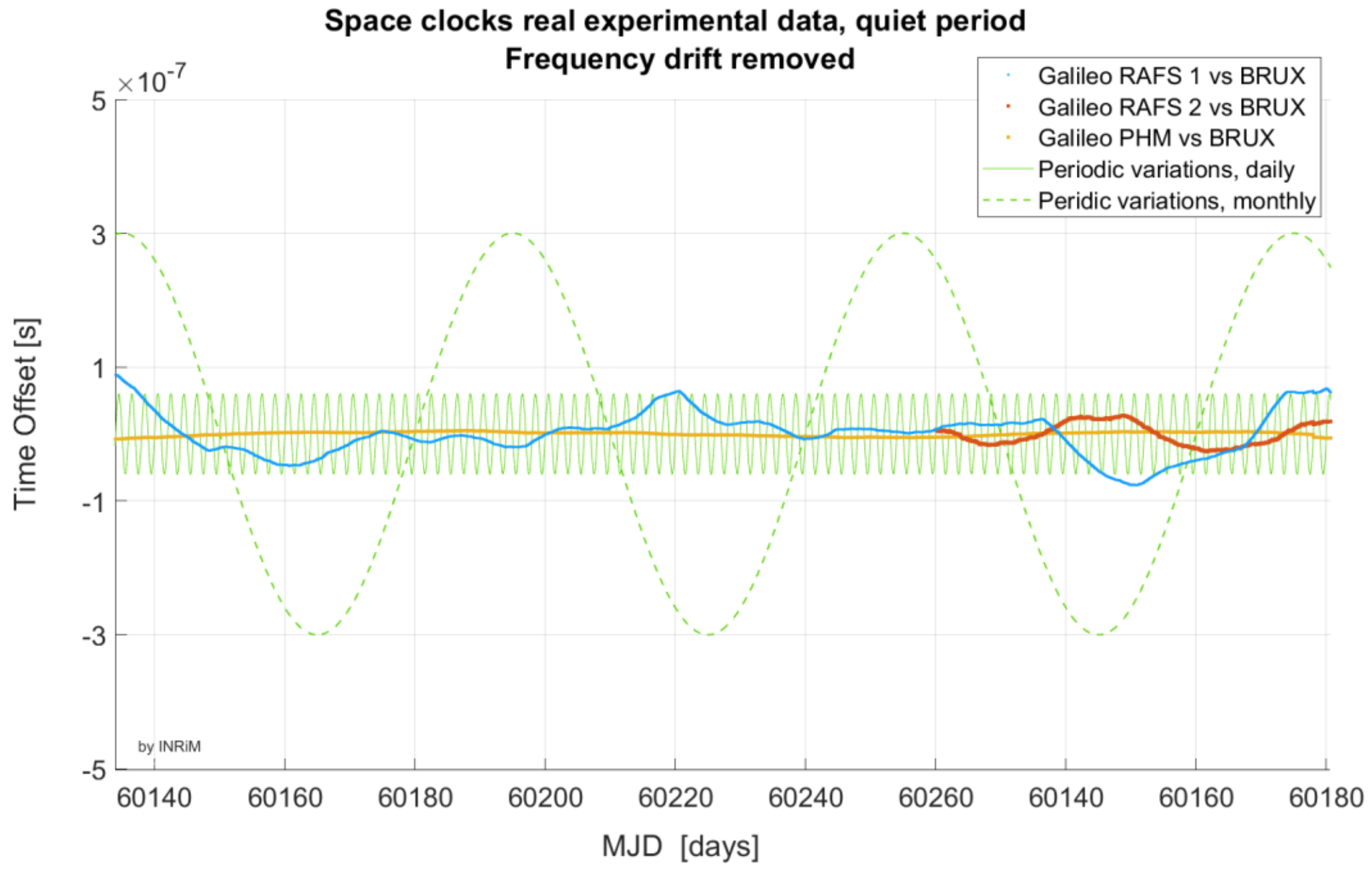

Figure 8. (top) Simulated clock signals compared to the periodic variations in TCG-TCL expected to be observed in the difference between a clock on the Moon and a clock on the Earth. (bottom) True clock signals of some Galileo satellite clocks referred to an active H-maser connected to the GNSS station BRUX.

# 4. NovaMoon Secondary Scientific Objectives (Scientific By-Products)

## 4.1 Lunar Digital Elevation Model (DEM)

The establishment of a new Lunar Celestial Reference System (LCRS) and lunar-body-fixed Lunar Reference Frame (LRF) by NovaMoon, as detailed in this plan, will directly enable more accurate and consistent cartographic mapping of the lunar surface. Precise ties between these modernised lunar reference frames and the terrestrial/celestial reference systems are crucial for rigorously rectifying and controlling lunar imagery and altimetry data into a unified cartographic reference frame. The improved determination of orientation, lunar ephemerides, and centre-of-mass knowledge from NovaMoon's laser ranging, VLBI, and satellite tracking will significantly enhance the transformation between the LCRS/LRF and the cartographic reference frames used for mapping.

Retroreflectors serve as distinct control points physically embodying the two lunar coordinate systems. While the PA coordinate system is valuable for representations such as the gravity field, the ME system is typically employed for surface mapping. Although the ME system is somewhat idealised due to tidal migrations, it remains the primary reference frame for surface cartography, with 0° Lat., 0° Lon representing the point closest to Earth. Retroreflectors in the ME reference frame serve as the foundation for high-resolution cartography, aiding in referencing lunar morphological features or human-made objects (Löcher et al., 2015; Wagner et al., 2016). Improved referencing will be enabled either indirectly through improved geodetic body system parameters, or directly, whenever the retroreflector can be localised in lunar mapping data.

It is acknowledged that lunar coordinate precision decreases rapidly with increasing distance from the reflectors. Consequently, installing new laser targets on the Moon (e.g., near the lunar limb or in polar regions) would extend the applicability of lunar coordinate systems across broader areas, enhancing the accuracy of lunar mapping products.

This will support the generation of globally consistent high-resolution lunar Digital Elevation Models (DEMs), with sub-meter horizontal and vertical accuracy - over an order of magnitude improvement compared to current capabilities. Such highly accurate DEMs are essential for minimising landing hazards, enabling precise localisation for surface operations, and possibly refining surface landform analysis and interior modelling. By providing a solid reference frame foundation tied to an accurately mapped lunar surface, the Moon would become a geodetically controlled body that can enable safer and more effective robotic/human exploration operations and resource utilisation.

### *4.1.1 NovaMoon contribution impact on DEM*

Nowadays, lunar and more generally planetary bodies DEMs are essentially the product of interpolation of laser altimetric data, stereo photogrammetry (for example, LROC WAC and NAC, Adoram-Kershner et al., 2021, Shean et al., 2016, Scholten et al., 2012) or a combination of the two, like for LOLA and SELENE-Kaguya TC (Barker et al., 2016). Vertical and spatial accuracy, in the interpolated lidar-based DEMs and in stereo DEMs, depends on the accuracy of spatial registration of the data to the global coordinate system (in addition to the geometric precision of the observation system). This can involve co-registration to lower resolution, large-scale topographic data (Beyer et al., 2014) to reduce bias and uncertainty in absolute altitude values. Absolute coordinate accuracy affects safety in landing, navigation, object detection and even scientific return.

For this reason, NovaMoon payload will enable the creation of global Digital Elevation Models of sub-metric vertical and horizontal accuracy—an order-of-magnitude leap in both horizontal and vertical fidelity (e.g., Barker et al., 2016)—transforming cartography into mission-critical infrastructure.

Enhanced DEMs will slash landing-site risk and refine hazard maps. An overview of the critical scenarios that will directly benefit both in the short-term and long-term is hereby presented:

#### *4.1.1.1 Impacts of improved DEM accuracy on the interior modelling*

The updated DEM together with the gravity field in spherical harmonics will lead to well-resolved gravity anomalies related to the interior, also profitable for the geological exploration of resources in the subsurface. Also, this improvement will lead to better estimates of crustal and elastic thickness and improved modelling of the regional and global lithosphere.

Tidal deformation (including phase-lags) and both longitudinal and latitudinal librations could be determined from time-dependent measurements. This would have important implications for investigating the deep interior and evolution of the Moon.

#### *4.1.1.2 Impacts of improved DEM accuracy on surface science/geological mapping*

The improved DEM will have a significant impact on the investigation of the surface, such as in:

- the production of high-accuracy controlled-mosaics with control network and refined seamlines among different images;
- the precise location of surface changes (e.g. tied to the future Lunar Meteoroid Impacts Observer, LUMIO, observations; Speretta et al., 2019);
- the possibility to connect small-scale maps to larger-scale maps precisely;
- the creation of interoperable mapping frames;
- an enhanced photometric corrections;
- and the production of a more accurate digital geological cartography.

#### *4.1.1.3 Impacts of improved DEM accuracy on landing site selection and hazard analysis*

A broader consequence of the accurated DEM will be evidenced on the improved reference altitude, to decimetric or centimetric accuracy, which minimises the risk of losing landers mission.

The landing site selection benefits of higher accuracy in locating both major topographic features, but also smaller obstacles detected in satellite images that will be orthorectified onto the higher accuracy DEMs.

#### *4.1.1.4 Impacts of improved DEM accuracy on remote sensing and operations planning*

Real-time location of robotic or humans on the surface of the Moon will tremendously improve the exploration capabilities and the ability to report scientific findings or sampling/contact analysis locations on the lunar surface.

In-Situ Resource Utilisation (ISRU), especially micro-permanently shadowed regions will be resolved much clearly thanks to the combined effect of precise localisation and possible real-time navigation.

Optional imaging cameras onboard of NovaMoon experiment will enhance the science of opportunity in terms of ground-truth lunar surface observations and will also help into constraining surface morphology and features observed from the ground with those observable from satellite (e.g. LRO), creating additional tie-points to correct the pointing and localisation of satellites.

Finally, the illumination conditions for power generation and for identifying PSRs could be more accurately determined. This will be a crucial aspect for landed elements and manned missions).

The updated lunar DEM, together with the LLR and VLBI measurements, will be consistently helpful in defining ESA's Argonaut lander position. The resulting terrain data will power real-time operations, resource prospecting in micro-PSRs, and high-precision geophysics, converting the Moon into a geodetically precisely characterised body, fostering sustainable operations for surface exploration and science.

## 4.2 Tests of General Relativity and Cosmological Models

By combining state-of-the-art lunar laser ranging, VLBI, atomic clocks, and quantum optical techniques, NovaMoon could probe the foundations of General Relativity, search for violations of the Equivalence Principle, constrain possible variations of fundamental constants, and test for new physics beyond the Standard Model. These investigations are aligned with the ESA Fundamental Physics white paper and international scientific priorities.

### *4.2.1 Introduction*

Since the early 1970s, LLR has provided over 56 years of Earth-Moon distance measurements, now achieving centimetre-level accuracy. This legacy is built upon retroreflector arrays placed by the Apollo and Lunokhod missions, tracked by the International Laser Ranging Service (ILRS, Bender et al., 1973). The resulting data have profoundly impacted lunar science, celestial mechanics, and fundamental physics (Müller et al., 2019; Bassi et al., 2022).

In the realm of gravitation, LLR provides some of the tightest constraints on General Relativity (GR) in the weak-field regime of the Solar System (Müller et al., 1996; William et al., 2004; Hofmann and Müller, 2018; Biskupek et al., 2021; Singh et al., 2023). In the three-body Sun-Earth-Moon system (Gutzwiller, 1998), LLR enables stringent tests of the Equivalence Principle (EP), geodetic precession, Yukawa precession and Lorentz symmetry, complementing other experiments like MICROSCOPE and planetary radar ranging. Furthermore, recent studies suggest that three-body systems can serve as probes of low-energy quantum-gravity predictions, through possible shifts in the positions of the Lagrangian points (Battista et al., 2015).

Despite GR's resounding successes—from binary pulsars to gravitational waves and black hole imaging—the motivation to test it with ever-increasing precision remains strong. On the observational front, the mysteries of dark matter and dark energy suggest that our current understanding of gravity and cosmology is incomplete. Many theoretical models proposed to explain these phenomena predict subtle deviations from GR that could be detectable within the Solar System. Improving local tests of gravity, therefore, directly constrains the development of these cosmological theories.

Theoretically, the most profound challenge is the incompatibility between GR and Quantum Mechanics. A unified theory of Quantum Gravity, such as String Theory or Loop Quantum Gravity, is probably required to describe extreme conditions like black hole singularities, where GR breaks down. These candidate theories often predict minute violations of fundamental principles like the Equivalence Principle and Lorentz invariance. Pushing the precision of experiments like LLR is our primary means of finding the experimental clues needed to guide the development of a more fundamental theory through the opportunity it offers to potentially uncover new physics, paving the way for groundbreaking developments (Bassi et al., 2022).

The deployment of a new generation of laser retroreflectors on the Moon (Muccino et al., 2025) promises a significant leap forward. A well-distributed array of new single retroreflectors, like those proposed for NovaMoon and other missions, will not only improve distance measurement accuracy by an order of magnitude or more, but also help decorrelate key parameters, better separating geophysical effects (such as lunar librations) from fundamental physics-related ones. This synergy is crucial: a better understanding of the Moon's interior and rotational dynamics directly enhances the sensitivity of gravitation tests. By integrating data from new, single-element reflectors with the long-standing record from legacy arrays, NovaMoon will enable a new generation of high-precision tests, pushing the frontiers of fundamental physics (Wang et al., 2025).

*4.2.2 Equivalence Principle*

The EP is a cornerstone of all gravitational theories that describe gravity as a manifestation of space-time geometry. It states that gravity couples universally to all forms of matter and energy – a property that fundamentally distinguishes gravitation from the other three known fundamental interactions (electromagnetism, and the weak and strong nuclear forces). As such, the EP stands as one of the most essential foundations of modern physics. However, many theoretical frameworks that extend beyond the Standard Model and General Relativity and also those aiming to explain dark matter and dark energy, predict violations of the EP at some level.

One facet of the EP is the Universality of Free Fall (UFF), which posits that all bodies fall with the same acceleration in a given external gravitational field, regardless of their mass and composition. This principle can be tested in two distinct regimes:

- The Weak Equivalence Principle (WEP), which asserts that the UFF holds for test bodies (i.e. objects with negligible gravitational self-energy).
- The Strong Equivalence Principle (SEP), which extends this universality to bodies with significant gravitational self-interaction.

The UFF is typically parameterised by the Eötvos parameter $\eta$ defined as the relative acceleration difference between two bodies located in the same gravitational field.

Other facets of the EP are the Local Position Invariance (LPI) tested with gravitational redshift measurement, and Local Lorentz invariance (LLI). EP is satisfied by GR and metric theories of gravity, while SEP is satisfied only by GR.

The Earth and the Moon are both falling in the same gravitational field of the Sun, making it an ideal laboratory to test the UFF. A violation of the UFF (SEP regime) between the Earth and the Moon will produce a characteristic signature on the Earth-Moon dynamics, known as the Nordtvedt effect (Nordtvedt, 1968). The Earth, with its massive iron core, and the Moon, a mostly rocky body, possess different non-negligible gravitational binding energies relative to their inertial masses, making this system an ideal probe for the foundations of gravity and, in particular, an ideal probe to test a combination of both the weak and the strong versions of the EP. In addition, both the Moon and the Earth are also sensitive to the gravitational field induced by galactic dark matter, allowing to search for possible interactions between DM and standard matter.

Thus, various tests of the EP are possible using LLR: i) with classical matter (Earth/Moon in the field of the Sun), ii) with respect to dark matter in the galactic centre and iii) the test of the equivalence of active and passive gravitational mass. All of them currently confirm their validity at the 2 x $10^{-14}$ level (Hofmann and Müller, 2018; Singh et al., 2023; Zhang et al., 2022; Biskupek et al., 2021). Viswanathan et al. (2018) obtained a limit of 7 x $10^{-14}$ where they also gave a new interpretation in the frame of dilaton theories of the SEP test.

With the current LLR constraint on the Eötvos parameter of 2 x $10^{-14}$ (Zhang et al., 2022), one of the most stringent tests of both the WEP and the SEP is provided. On the other hand, the current best test of the WEP is provided by the MICROSCOPE space-based experiment, which puts a constraint on the Eötvos parameter at the level of 3 x $10^{-15}$ (Touboul et al., 2022). This is by far better than the best lab test using torsion balances (Schlamminger et al., 2008), which only reached $10^{-13}$. LLR-based EP tests and the other WEP tests are complementary since they test the two distinct regimes of the EP. Taking the MICROSCOPE result to separate the WEP part from the LLR results, the SEP can be tested to 1 x $10^{-4}$. The LLR-based EP test has fallen behind the MICROSCOPE accuracy by about a factor of 5 - 10. NovaMoon can largely help, not only by achieving a higher tracking accuracy, but also by enabling better solutions for lunar motion and rotation, as well as better tracking of the lunar orbit.

As an important caveat, all EP tests are affected by mismodeling of the orientation and tidal deformations (Anderson and Williams, 2001). Therefore, EP test results will depend on how well the dynamical modeling can be improved with NovaMoon (see Section 3.1).

*4.2.3 Variation of Gravitational Constant*

LLR constrains a possible temporal variation of the gravitational constant $\dot{G}/G$ to better than $10^{-14}$ 1/yr as well as a possible accelerated variation $\ddot{G}/G$ at 2 x $10^{-16}$ 1/yr$^2$ (Biskupek et al., 2021). Comparable results are only obtained from ephemeris solutions (Fienga and Minazzoli, 2024) or from the analysis of Messenger data (Genova et al., 2018). But they only reach values in the order of $10^{-13}$ 1/yr and 4 x $10^{-14}$ 1/yr. They are mainly limited by the unknown amount of possible mass change in the Sun.

Furthermore, possible spatial variations of $G$ can be tested with LLR, e.g., related to the distance (5th force, Yukawa potential). For example, the Yukawa coupling constant $\mathbf{a}_{l=400000km}$ (i.e. a test of Newton's inverse square law for the Earth-Moon distance) can be constrained at 4 x $10^{-12}$ (see Hofmann and Müller, 2018).

The next frontier is to use the 56-year+ LLR dataset enhanced by NovaMoon and other next-generation lunar retroreflectors to place tighter constraints on the temporal and spatial variations of $G$.

*4.2.4 Parameterised Post/Newtonian (PPN) tests*

Further relativity tests using LLR include geodetic precession of the lunar orbit (Hofmann and Müller, 2018; Bertotti et al., 1987; Shapiro et al., 1988). Several parameters in the PPN (parameterised post-Newtonian) framework, such as the non-linearity and space curvature parameters $\beta$ and $\gamma$, as well as the two preferred-frame parameters $\alpha_1$ and $\alpha_2$, can also be tested (Biskupek et al., 2021; Hofmann and Müller, 2018). The currently achieved accuracy on some of the relativistic parameters (Biskupek et al., 2021) is comparable with (and often better than) the results from other techniques.

NovaMoon will help by improving the reference frame, the ephemeris, and the rotation of the Moon, also due to a better link to the VLBI system (see Sec. 3.1). Furthermore, with the improvement of the observation network, more observatories will be able to track the Moon.

*4.2.5 Standard Model Extension (SME) and Modified Gravity*

NovaMoon will also improve the search for possible violations of Lorentz Symmetry (LS). LS states that the outcome of any local experiment is independent of both the velocity and the orientation of the frame in which it is performed (Will, 2018). It is a cornerstone of both the Standard Model of particle physics and General Relativity (GR). Given its broad range of applicability, testing the validity of LS provides one of the most powerful probes of fundamental physics. On theoretical grounds, it has been suggested that LS may not be a truly fundamental symmetry of Nature and could be broken at some level. While early motivations for such violations came from string theory, possible LS breaking also arises in loop quantum gravity, non-commutative geometry, multiverse scenarios, brane-world models, ...

To systematically address possible LS violations, an effective field theory has been developed: the Standard Model Extension (SME) (Colladay and Kostelecký 1997, 1998), which covers all sectors of physics. Lunar Laser Ranging already provides the most stringent constraints on several SME parameters—including the minimal gravity sector, the matter-gravity sector, and the next-to-leading order gravity sector. These constraints are typically an order of magnitude tighter than those obtained from pulsar observations (Bourgoin et al. 2016, 2017, 2021). Currently, however, the sensitivity of LS violation searches is limited by systematic errors due to imperfect modelling of the system.

NovaMoon's new retroreflector will significantly advance these searches. First, the new reflector is expected to improve the modelling of the Earth–Moon system, thereby strongly reducing the systematics that currently limit LS tests. In addition, covariance analyses show that the new data—including 5 years of NovaMoon observations—would reduce the statistical uncertainties of SME parameters by about 80% (for the 2027–2032 dataset) and 85% (for 2027–2037), with NovaMoon contributing roughly 11% of the total improvement.

The problem of cosmic acceleration and the related dark energy have motivated the investigation of theories that modify GR on cosmological scales. Such modified gravity theories must satisfy stringent experimental bounds on the Solar System scale, a task which is accomplished by means of a mechanism that screens deviations from GR at this scale. An example of screening mechanism is the so-called chameleon mechanism, which is implemented in theories like f(R) gravity and non-minimally coupled (NMC) gravity.

Particularly, in NMC gravity GR is modified by means of the introduction of a non-minimal coupling between space-time curvature and matter. NMC gravity has an impact on the dynamics of the Sun-Earth-Moon system, which exhibits the above-mentioned screening mechanism. Because of screening, deviations from GR give rise to a fifth force sourced by mass in thin shells at the surfaces of the three bodies. Consequently, the Earth and Moon fall toward the Sun with different accelerations, giving rise to a WEP violation. Constraints on the parameters of the NMC gravity model have been obtained by means of a test of WEP based on lunar laser ranging data.

Computations with a specific NMC gravity model have shown deviations from a cosmological constant, in the effective equation of state of dark energy, ranging from $10^{-2}$ to $10^{-1}$, giving rise to a potentially observable evolving dark energy. It turns out that the LLR bound on WEP violation restricts the admissible values of the parameters of the gravity model that allow the prediction of an observable evolving dark energy. Then the LLR constraints must be intersected with constraints from the recent DESI (Dark Energy Spectroscopic Instrument) experiment, giving rise to cosmological bounds on the NMC gravity model.

Since NovaMoon will contribute to improving the test of EP in the Earth-Moon system, further constraints from LLR will be expected on the parameters of the NMC gravity model that have to be intersected with cosmological predictions.

*4.2.6 Preliminary simulation results and conclusions*

Accurate measurements of the Earth–Moon distance from Lunar Laser Ranging (LLR) data collected over several decades enable the estimation of a wide range of parameters in both geophysics and fundamental physics. We assessed the expected improvements in the coming years, along with the contribution from NovaMoon, using a covariance analysis.

In this simplified analysis, the fundamental physics parameters (FPPs) considered are:

- the Earth–Moon differential acceleration (a test of the Equivalence Principle),
- the PPN parameters ($\beta$, $\gamma$, $\alpha_1$, $\alpha_2$, $\alpha_3$),
- the Yukawa coupling term ($\alpha_Y$),
- possible time variation of Newton's gravitational constant, expressed as $\dot{G}/G$ and $\ddot{G}/G$

The model also accounts for free and forced lunar librations using a rigid-body approximation. To first order, the perturbations introduced by these parameters on the Earth–Moon distance can be derived by solving Hill's equations.

We simulated datasets for the five retroreflector arrays deployed between 1969 and 1973, as well as for Blue Ghost 1 (landed in 2025), IM-3 and NovaMoon (assumed to be deployed in 2026 and 2027, respectively). Between 1970 and 2020, the average acquisition rate was about one normal point every 3.3 days per array (Biskupek et al., 2021). In a conservative scenario, we assume this rate remains constant. For the formal uncertainties of the normal points, we adopted a gradual improvement from 25 cm in 1970 to ~1 cm in 2020, with a lower limit of 1 cm for the older arrays. For Blue Ghost 1, IM-3 and NovaMoon, we assumed 1 mm accuracy after full modelling of geophysical effects (currently limiting LLR residuals to ~1 cm).

The covariance analysis for 1970–2020 yields formal uncertainties for the FPPs consistent with state-of-the-art results ($\sigma(EP) \approx 2.4 \times 10^{-14}$, $\sigma(\beta,\gamma) \approx 10^{-4}$, $\sigma(\alpha_Y) \approx 2 \times 10^{-11}$), validating our simplified approach. Extending the simulation to 2032 and 2037, with 5 and 10 years of NovaMoon data, respectively, shows improvements of 50–75% (2032) and 60–80% (2037) relative to 2025 for all FPP obtained from LLR. NovaMoon's contribution to this improvement is ~9–14% for the FPPs, and as high as 60–70% for lunar librations in latitude and retroreflector positioning in the lunar frame. The NovaMoon contribution was assessed by running simulations with and without the NovaMoon retroreflectors.

The key advantage of LLR in testing gravitational theories lies in the steadily increasing accuracy as more data accumulates. Our preliminary covariance analysis indicates that NovaMoon could make a significant contribution to fundamental physics measurements, even within a relatively short span of 5–10 years.

## 4.4 Radio Occultation

The NovaMoon mission can provide a unique opportunity to perform the atmosphere sounding of Earth from a natural celestial body making use of GNSS Radio Occultation technique.
GNSS radio occultation (RO) is a satellite remote sensing technique that uses radio signals from Global Navigation Satellite Systems (GNSS) to measure physical properties of the Earth's atmosphere. The history of this technology began with planetary exploration (Fjeldbo et al. 1971).
The first significant demonstration for Earth was the GPS/MET (GPS/Meteorology) experiment from 1995 to 1997 (Ware et al. 1996), which was followed by the space missions: CHAMP (2000-2010), COSMIC-1 (2006-2017), GRACE and GRACE-FO (2004-), MetOp (2006-)and COSMIC-2 (2017-**.** Today, GNSS radio occultation is a standard remote sensing method. Its data are used daily in weather forecasting models and climate research because of its high accuracy, stability, and global coverage. The technology is also being used to monitor space weather and study the ionosphere.

The possibility of using GNSS for the navigation of spacecraft with a distance beyond their orbits was deeply investigated till 80th (Jorgensen e al. 1982) to understand if they could serve for navigation of geo-stationary satellites. The tracking of GNSS Navigation signals for Geostationary Satellites, as well as platforms beyond their altitudes, is particularly challenging for their weakness and the poor geometric configuration ( See Fig. 9).
In the framework of the European Lunar Lander Mission, a study was performed to determine the feasibility of using GNSS (GPS/GALILEO) weak-signal technology in future lunar missions to improve the navigation performances (Manzano-Jurado et al. 2014). It was demonstrated the feasibility of a receiver with a signal-to-noise ratio: C/N0 down to 10 dBHz.

**The Moon Radio Occultation Experiment:**
Until now, the RO technique had only been applied to Earth using LEO platform. NovaMoon enables the first-ever implementation of a terrestrial radio occultation experiment leveraging a natural celestial body—specifically, the Moon—as the observational platform (M-RO).
The GNSS signal can be viewed only when the satellite is in opposition to the Moon (see Fig. 1). The beam of GNSS is indeed pointing toward nadir and cannot be viewed by the Moon when it is on the same side. The primary lobes of GPS and GALILEO are wide in turn: 42.6 ° and 41°; while the angular dimension of the Earth viewed at their altitudes is in turn 27.55° and 24.86°. Thus, there is an overshooting of the signal beyond the Earth, which is represented by the yellow hollow cone in Figure 9. The dark blue cone defines the blind zone for which GNSS signal cannot be tracked from the Moon; While the light blue hollow cone defines the zone in which an Earth RO event can be captured from the Moon. The light blue zone has been defined by considering an atmospheric thickness of 600 km (represented by the red dashed circle in Fig. 9), including the ionospheric layers as well. In Figure 10, we have performed a simulation to compute the number of occultations that can be obtained from the Moon in a month.

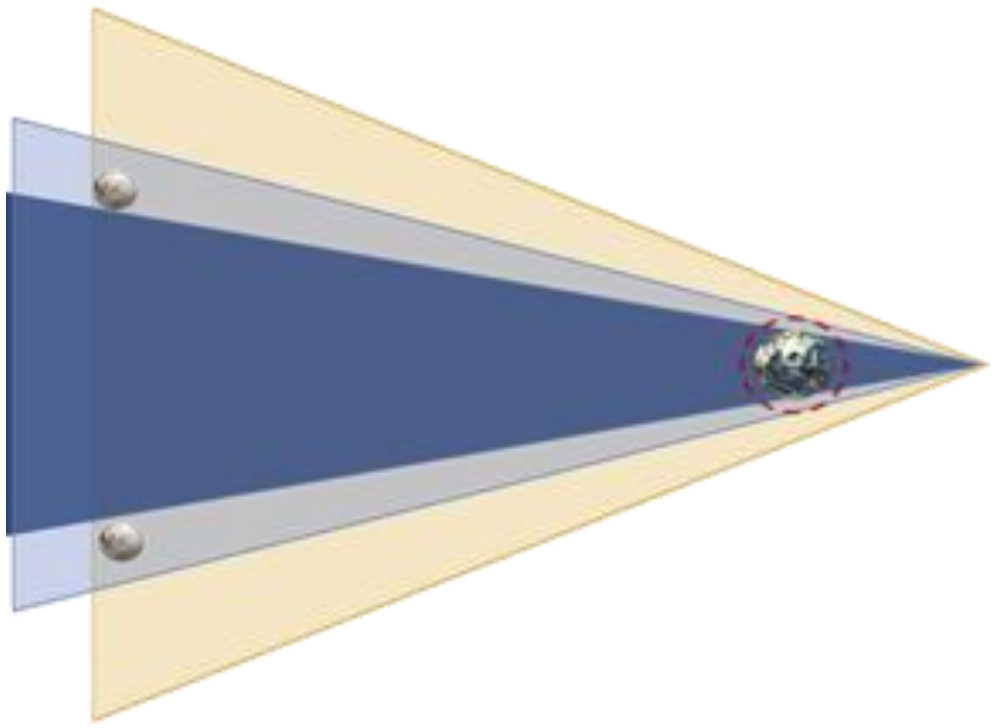

Figure 9. Earth GNSS Radio Occultation Geometry from the Moon.

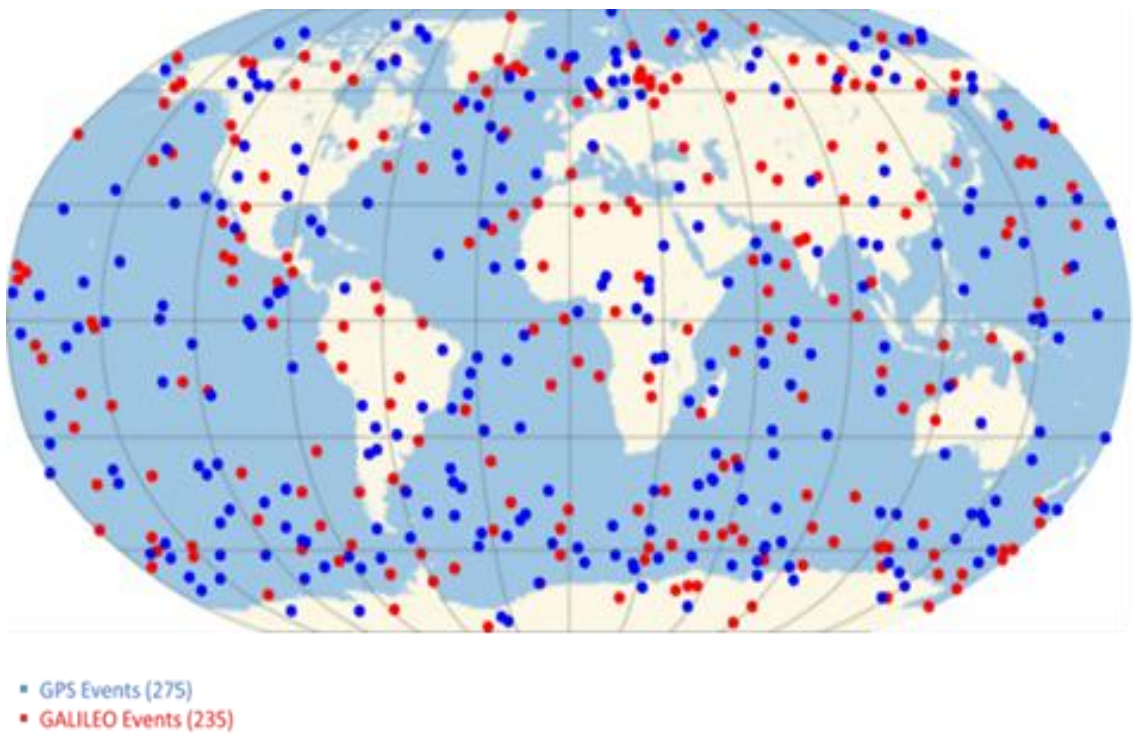

Figure 10. The Moon RO occultation events for GALILEO (red points) and GPS (blue ones) occurred in a Month

The number of GNSS RO events is relatively low compared to those obtained using LEO satellites, with only tens of events per day. This is mainly because the Moon's dynamics are much more relaxed, due to its long orbital period of 27.3 days, compared to the few hundred minutes typical of LEO satellites. The number increases if we add the satellites of other GNSS systems like GLONASS and BDS. Thus, these kinds of observations can be mainly exploited for climate investigations if they provide relevant and added-value information.

The low dynamics of lunar RO events, on the other hand, result in a longer RO duration — typically at least twice as long as usual.

Longer GNSS RO events offer several significant advantages. A prolonged occultation allows the radio signal to traverse a larger vertical extent of the atmosphere and ionosphere, or to spend more time within a specific layer. This results in a greater number of measurements along the signal path, thereby enhancing the quality and scientific value of the derived atmospheric and ionospheric profiles.

While longer GNSS radio occultations offer numerous advantages, as discussed previously, they also come with certain drawbacks and limitations. The most critical is the signal fading and loss of lock due to the big distance from Earth. Thus, it will be necessary to use receivers equipped with high-gain phase array antennas and receivers able to handle weak signals. Today, all the receivers devoted to RO work in open-loop mode. In the last ten years, studies were performed to have more demanding GNSS receivers able to handle the signal at a level of 10 dB-Hz and less around the Moon (Musumeci et al. 2014, Manzano-Jurado et al. 2014).

**Moon exosphere monitoring by Moonlight constellation**

The lunar exosphere is not a dense, self-sustaining blanket of gases, but rather a "surface boundary exosphere" where gas molecules are so spread out that they rarely collide with each other. So, the use of RO technique as applied on Earth or Planets could be unfruitful. It cannot be measurable, indeed, the bending of the radio signal due to the refractivity layers of the atmosphere. Anyway, a small network of Moonlight receivers on S-band could be placed to observe the signals of the Moon navigation system. What can be investigated is the different degrees of extinction suffered by the carrier, in terms of signal-to-noise ratio. What can be performed are measurements of different degrees of extinction between grazing (i.e. very low elevation angles) and zenith directions of the S-band signal coming from Moonlight satellites.

## 4.5 Indirect scientific benefits of a Lunar PNT system

Similar to what is already observed on Earth, a highly accurate Positioning, Navigation, and Timing (PNT) system on the Moon would constitute a transformative capability for lunar science and exploration. By providing precise localisation of orbiters, landers, surface assets, and users, such a system would substantially enhance the scientific return of lunar missions across multiple domains.

### *4.5.1. Enhanced Science from Lunar Orbiters*

The availability of real-time or post-processed orbit determination with sub-meter accuracy for Low Lunar Orbit (LLO) spacecraft would enable a new level of precision in orbital science investigations. Foreseen benefits include:

- Improved lunar gravity mapping, leading to more accurate models of the lunar interior.
- More precise measurements of lunar magnetic anomalies.
- Enhanced detection and mapping of neutron and gamma-ray fluxes.
- Support for future formation flying and interferometric missions.
- Increased safety through accurate collision-risk alerts for spacecraft in LLO.

Then, of course, these improvements have direct implications for other areas of lunar science discussed elsewhere in this paper:

- High-resolution surface topography and Digital Elevation Models (DEMs), as discussed in Section 4.1
- Better characterisation of the lunar plasma and space weather environment (see Section 5.2).

### *4.5.2 Surface Science and Exploration Benefits*

Sub-meter navigation and positioning on the lunar surface would also unlock significant advances in science and exploration:

- **Lunar geological investigations:** Precise geolocation of sampling sites allows direct correlation between in-situ measurements and orbital remote sensing data, providing improved lunar geological context and interpretation.
- **Experiment repeatability:** Scientific instruments requiring repeated measurements at the same site (e.g., seismic stations, volatile detectors) would benefit from precise repositioning capabilities.
- **Distributed and networked science:** Coordinated measurements from distributed assets—such as rover fleets, seismic arrays, or stereo imaging systems—would be enabled through common and precise spatial references.
- **Sensor synchronisation:** Accurate timing synchronisation would support demanding applications such as radio astronomy, interferometry, and distributed sensing networks.
- **Towards a lunar "Internet of Things":** Low-power PNT-enabled devices could form a network of interconnected instruments and sensors on the surface, expanding scientific coverage.

*4.5.3 Precision Landing and Surface Operations*

A lunar PNT system would also enable landers to achieve very accurate precision landings. This capability would directly expand the accessible science return by:

- Allowing safe access to hazardous but scientifically rich regions, such as Permanently Shadowed Regions (PSRs) and Peaks of Eternal Light (PEL).
- Enabling precise targeting for the recovery, inspection, or reuse of past landers and deployed scientific instruments.
- Maximising the payload capability of scientific landers by reducing the need for large landing error ellipses.

In summary, a lunar PNT system represents a cross-cutting enabler for orbital, surface, and human exploration science, with impacts spanning geophysics, geology, space environment monitoring, and astronomy.

# 5. Additional scientific proposals for NovaMoon

This section outlines the future possibilities for upcoming Argonaut missions, focusing on the inclusion of additional NovaMoon payloads and/or their potential to advance scientific research and enable new discoveries.

## 5.1 Gravitational Wave Detection Support

Gravitational waves (GWs) are unique witnesses to all sufficiently energetic processes happening in the Universe. This includes events that are invisible to conventional observatories, such as violent mergers of binary black holes or high-energy phenomena in the primordial Universe. Since 2015, ground-based interferometers have begun detecting GWs with frequencies around 100 Hz, emitted by merging stellar-mass black holes and neutron stars. However, as with electromagnetic astronomy, GWs are expected to be generated across a broad frequency spectrum, including a huge swathe of frequencies that are inaccessible with current GW experiments. As a result, there is currently an extensive effort in the scientific community to widen searches for GWs to cover the broadest possible frequency range. One of the most challenging bands to access is centred around microhertz frequencies. This band contains many expected sources but remains observationally uncharted. The recent detection of a signal consistent with inspiralling supermassive black hole binaries in the nanohertz band by pulsar timing arrays also provides strong motivation for microhertz GW searches, as detecting these sources at higher frequencies would allow us to address fundamental physical questions about the formation and evolution of the largest black holes in the Universe.

GWs perturb the motion of binary systems in ways that are resonantly enhanced when the GW frequency is an integer multiple (harmonic) of the orbital period (Blas & Jenkins 2022). The orbit is then modified by this interaction in a way that accumulates secularly over time if the resonant condition is satisfied. Recent works (Blas & Jenkins2022, Foster et al., 2025) have shown that this effect generates a modification of the Moon's orbit that would result in almost guaranteed sensitivity of current data to the GWs expected from supermassive black hole binaries, provided that these GW-induced perturbations are not strongly degenerate with other parameters describing the Moon's orbit. In reality, however, there will indeed be some degree of degeneracy, particularly with the tidal deformability of the Earth (Foster et al, 2025).

Furthermore, even if perturbations consistent with a GW source are observed, it is not yet clear how to definitively claim a GW detection.

The questions raised above are under active investigation by a community of researchers working in this area. Searches for resonant GW perturbations of the Moon's orbit in existing laser ranging data are in progress, with first results expected by early 2026. As such, it is still too early to determine NovaMoon's potential impact on this program. However, it is fair to say that any advances in the modelling of the Moon's internal and orbital dynamics that may be achieved using NovaMoon would contribute, at zero additional cost, to the effort to detect microhertz GWs. Furthermore, the signal-to-noise ratio for such GW signals will increase with every additional high-quality normal point (Foster et al., 2025). As a result, even if GW detection is not a core science goal of NovaMoon, the mission is likely to make a significant contribution toward detecting (or setting upper limits on) GWs in an entirely new frequency band, with enormous discovery potential for astrophysics and fundamental science.

## 5.2 Space Weather Monitoring

Given NovaMoon's critical role in supporting high-precision Position, Navigation, and Timing (Parker et al, 1958) services on the Moon, the addition of a compact plasma/space weather instrument would provide significant added value. While the Moon lacks a global magnetosphere and dense atmosphere, space weather—especially from solar energetic particle (SEP) events and coronal mass ejections (CMEs)—still plays a key role in the lunar environment.

Including a plasma instrument on NovaMoon would provide:

- **Real-time space weather monitoring** to safeguard navigation accuracy: CMEs and SEPs can introduce disturbances that degrade signal timing, ranging measurements, and tracking performance—especially critical for gravity field mapping, time transfer, and differential positioning (Li et al, 2024)
- **Impact assessment on GNSS-like systems**: Differential GNSS systems near Earth rely on ionospheric corrections. While the Moon lacks an ionosphere, local plasma variations (e.g., transient surface charging, CME-induced wakes) can introduce biases that affect timing and carrier-phase observations. Monitoring these ensures accurate calibration and robust system performance (Kelley et al, 2009).
- **Support for scientific and operational missions**: Lunar orbiters and landers using NovaMoon-based corrections would benefit from co-located plasma data to interpret anomalies in PNT performance or spacecraft dynamics. This is especially relevant for sensitive measurements such as gravimetry or VLBI (Angelopoulos et al, 2008).
- **Contribution to heliophysics**: A plasma/space weather sensor on the Moon's South Pole would fill a key gap in global heliospheric coverage and allow unique comparative studies with data from Earth and L1 spacecraft (e.g., ACE, DSCOVR) (Richardson & Cane, 2010).

Candidate instruments could include low-mass electrostatic analysers, Langmuir probes (Mott-Smith & Langmuir, 1926), or energetic particle detectors (Stone et al., 1998), many of which have been flown on CubeSats or small platforms. Their inclusion would require minimal additional mass or power but offer a high return in operational reliability and science.

## 5.3 *Long-Baseline Quantum Optics*

Experiments over the past five decades demonstrated that quantum systems are sensitive to gravity in measurable ways. From the first observation of a gravitational phase in neutron interferometry [Colella, 1975] to state-of-the-art atom interferometers, precision experiments nowadays test the universality of free fall (Schlippert, 2014; Asenbaum, 2020), measure fundamental constants (Rosi, 2014; Morel, 2020), and even resolve relativistic proper-time differences in delocalized quantum superpositions (Asenbaum, 2017; Overstreet, 2022; Roura, 2022). However, they can be entirely described in terms of Newtonian gravity and non-relativistic quantum mechanics. Quantum-clock interferometry experiments sensitive to gravitational time dilation (Roura, 2020; Roura, 2021), on the other hand, do require a relativistic description, but the velocities of the atoms involved are non-relativistic. In contrast, photons are inherently relativistic and particularly well suited for space-based tests because they preserve coherence over long distances and can benefit from the very long Earth–Moon baseline as well as large differences of gravitational potential. Thus, NovaMoon constitutes a unique platform for **long-baseline quantum optics experiments** probing aspects of **quantum field theory in curved spacetime**, and offers the following opportunities:

1. **Interferometric gravitational-redshift measurements**
   a. *Goal*: Detect tiny relativistic time-dilation effects by comparing optical paths on Earth and the Moon in interferometers spanning the Earth–Moon baseline.
   b. *Payload*: Long, stabilized optical delay line (~100 m), frequency-stabilized laser with accurate frequency reference ($10^{-10}$ level or better), single-photon detectors, polarizing beam splitters and telescope.
   c. *Ground segment*: Analogous optical delay line stabilized with accurate frequency reference, photon source and large-aperture telescope, possibly with adaptive optics.

For 100-m delay lines, the phase shift due to gravitational time dilation amounts to **0.5 rad**, clearly showing the much larger signal that can be achieved with space-based implementations (Terno, 2020; Mohageg, 2022; Wu, 2024), and corresponding to an **enhancement by 4 orders of magnitude** compared to experiments currently pursued on the ground (Hilweg, 2017; Hilweg, 2022; Polini, 2024). A first, simpler version of the experiment would compare the interference of quantum and classical light (with possibly different wavelengths), providing an experimental test of quantum field theory in curved spacetime for quantum states (Birrell, 1994). The more advanced version will require independent calibration and stabilisation of the two delay lines.

2. **Long-baseline Bell tests**
   a. *Goal*: Distribute entangled photons across the Earth–Moon distance to test whether quantum correlations survive under spacetime curvature (Yin, 2017; Mohageg, 2022).
   b. *Payload*: low-dark-count single-photon detectors and large-aperture telescope.
   c. *Ground segment*: large-aperture telescope and adaptive optics to minimise beam losses, as well as a high-rate entangled photon-pair source.

The default configuration involves an uplink. Alternatively, a downlink version would relax the requirements on the adaptive optics and the telescope apertures, but at the expense of a more complex payload.

3. **Bell tests with quantum memories**

a. *Goal*: Exploit the long light travel time to perform "human-decision" Bell tests and tests closing causal loopholes in quantum mechanics.
b. *Payload*: Same as Bell tests above.
c. *Ground segment*: Same as Bell tests above plus quantum memory with high multiplexing capability.

4. **Quantum teleportation**
   a. *Goal*: Extend protocols already demonstrated in orbit (Ren, 2017) to the Earth–Moon distance.

   b. *Payload and Ground segment*: Same as Bell tests, with optional quantum memories for advanced schemes.

NovaMoon's baseline payload does not yet include these components, except for the option of including a SPAD, but the mission architecture could host them as incremental upgrades. A phased approach — from classical interferometry to quantum states of light — would position NovaMoon as a **pathfinder for space-based quantum physics at unprecedented scales** (Rideout, 2012; Mohageg, 2022) and push the boundaries of quantum communication over long distances.

## 5.4 Improving ranging, imaging and gas sensing performance with quantum-inspired correlations

Quantum-inspired optical techniques use classically generated energy-time or spatial-time correlations to significantly improve the signal-to-noise ratio in challenging measurements which can be used for remote metrology, spectroscopic sensing, and imaging (Kaltenbaek et al., 2008, Nie et al., 2025, Dong et al., 2025, Jiang et al., 2019). Quantum-inspired LiDAR offers several advantages compared to traditional classical systems, including the ability to measure remote distance with low light levels under strong background. Due to its advantages of noise reduction, it can also realize sensitive gas detection with wavelength at the vibrational absorption lines of many gases. Furthermore, by combining single-photon imaging techniques, surface reconstruction can be realized via raster scanning or structured illumination (Sun et al., 2016, Li et al., 2021). This class of quantum-inspired technologies currently outperforms fragile entangled-photon systems in bright (including sunlight) and lossy environments due to its enhanced brightness which brings the system to real-world applications (Nie et al., 2025). This robustness makes them ideal for lunar metrology and navigation, where long observation windows, high-sensitivity, and strong noise-resistance are required.

In future NovaMoon missions, such capability could be implemented through a compact optical payload combining an integrated photon source with correlations generated by fast electro-optic modulators and arrayed waveguide grating structures, a compact telescope and detectors for active laser ranging or mapping. This design ensures compatibility with the NovaMoon geodetic suite and integration with Moonlight communication services. Additionally, a photonic integrated chip can be deployed on the lander to realize a compact gas-sensing payload capable of detecting extremely low concentrations of a wide range of organic gas molecules. This can be achieved by measuring the phase shift induced by the target sample within a nonlinear interferometer (Zhou et al. 2024). This would harness the benefits of probing mid-infrared vibrational absorptions while avoiding the complexity and performance limitations associated with mid-infrared detectors.

**Satellite-to-Lunar Ranging**

A quantum-inspired transceiver onboard a lunar orbiter would use the correlated photon source and co-located telescope to perform continuous surface ranging and three-dimensional mapping. The return loss, determined by surface reflectivity, telescope aperture, focal length, source divergence, and orbiter–surface distance, indicates efficient detection achievable within small-spacecraft constraints.

**Earth-to-Lunar Ranging**

The NovaMoon corner cube reflector would operate with large-aperture Earth-based telescopes to perform day-and-night Earth-Moon ranging using quantum-inspired correlations. Compatible with existing LLR infrastructure, this configuration extends operations to daylight conditions by using the energy-time correlations generated in the source.

Together, these configurations demonstrate the feasibility of integrating **quantum-inspired optical payloads** into future NovaMoon missions, enabling continuous, high-precision ranging and timing while bridging classical photonics and quantum metrology to strengthen the lunar reference and navigation framework.

# 6. Scientific payloads and techniques

The NovaMoon mission is conceived as a multi-technique geodetic and timing station on the lunar surface, located near the South Pole. In this chapter, we further discuss payloads and techniques that are foreseen for the mission, based on the primary (Chapter 3) and the secondary (Chapter 4) objectives previously defined. For each payload or technique, we will briefly describe the expected incremental developments, open issues, and main challenges, together with preliminary trade-offs and 1a feasibility analysis. These considerations should be regarded as indicative, e.g., based on

- contribution to scientific objectives (e.g. reference frames, timescales, etc.);
- heritage and maturity (e.g. proven performance in past Earth/Lunar missions, if applicable);
- operational demands (e.g. ground support, calibration, data analysis, etc.);
- synergy and/or complementarity (e.g. maximise the scientific return via a multi-technique approach);

along with other potential engineering constraints (e.g. mass, size, power requirement) to be further addressed in subsequent phases of the NovaMoon mission definition.

At this preliminary stage, several (candidate) instruments have been considered, thus spanning a quite broad range of technology readiness levels (TRLs). Hence, this presents non-trivial challenges in terms of space system engineering. At the same time, this instrumentation diversity represents a unique opportunity for advancing technology development while potentially providing new insights into fundamental physical processes. Some candidates' payload considered are:

- **Very Long Baseline Interferometry (VLBI)**, see Section 6.1;

- **Lunar Laser Ranging (LLR)**, see Section 6.2;
- **Time Laboratory (TimeLab)**, see Section 6.3;
- **GNSS receiver/transmitter**, see Section 6.4;
- **Moonlight LCNS receiver,** see Section 6.5;
- **Single-Beam Interferometry (SBI)**, see Section 6.6;
- **Seismometers**[1], see Section 6.7;

Notice that for each type of payload, different techniques could be considered, nonetheless in some cases this requires substantial modifications or TRL developments, therefore, only few key promising solutions have been discussed in the rest of this chapter.

## 6.1 Very Long Baseline Interferometry (VLBI)

The Very Long Baseline Interferometry (VLBI) is a fundamental technique to establish and maintain geodetic and astrometric reference frames, determining Earth's orientation in space, monitoring Earth's dynamic processes, and it provides a wide range of applications in the domain of astronomy and astrophysics. For instance, VLBI has also been used to track spacecraft and lunar rovers.

Several VLBI-based observing principles exist; each is tailored to specific scientific applications and parameters of interest. In astronomy, the VLBI is commonly utilized for high-resolution imaging of distant radio sources through aperture synthesis, where each baseline provides a Fourier component of the source brightness distribution. With sufficient coverage in time and geometry, it enables image reconstruction with highest angular resolution. For studies of faint sources, phase referencing is frequently applied, where observations alternate between a nearby bright calibrator and the faint target source, and the calibrator's phase solutions are transferred to the target, thereby improving sensitivity and enabling high-precision relative astrometry.

While geodetic VLBI observations typically target extragalactic quasars, near-field targets such as satellites are increasingly studied. VLBI has been shown to be able to position satellites and planetary landers, including demonstrations on Earth-orbiting satellites such as GNSS. These capabilities make geodetic VLBI a versatile technique for both geodesy and navigation.

Geodetic VLBI relies on wideband group-delay measurements of extragalactic quasars, being the difference in signal arrival time between the two telescopes, and provides results in an absolute sense, directly tied to the terrestrial and celestial reference frames. Large bandwidths and high recording rates allow for individual delay precision at millimetre or centimetre level, enabled by observing at frequencies outside the nominal narrow bands allocated to S/C communication. VLBI tracking of the Apollo ALSEP radio beacons on the lunar surface has already demonstrated that lander positions can be determined with accuracies on the order of ten meters (King et al., 1976). More recently, observations of the Chinese Chang'e 3 lunar lander X-band signals with global geodetic networks achieved accuracies of its 2D horizontal position of 8.9 m and 4.5 m in latitude and longitude (Klopotek et al., 2019). These signals were narrow-band S- and X-band DOR tones respectively. Wider observing bandwidths, such as those envisaged for the GENESIS mission's dedicated wideband VLBI transmitter (Delva et al. 2023), are expected to improve delay precision and corresponding position accuracy significantly compared with these earlier narrow-band demonstrations

A related technique, differential or Δ-VLBI, is used for spacecraft tracking. By measuring the differential delay between the spacecraft and a nearby quasar, common-mode errors from clocks and the atmosphere largely cancel out, yielding very accurate relative positioning. VLBI observations to spacecraft differ in their realisation due to the distance to the target, the transmit signal and the ground station capabilities. Current and past implementations are NASA’s ΔDOR technique at the Deep Space Network (DSN), an equivalent system by ESA using their global ground station network (ESTRACK), the Planetary Radio Interferometry and Doppler Experiment (PRIDE) supporting ESA’s JUICE and Mars express mission, the DOR VLBI system by the Chinese VLBI network supporting the Chang'E missions or differential phase delay tracking during the Japanese lunar mission SELENE (Hanada et al., 2008).

[1] This can be expected as an instrument on board the Argonaut lander or as part of a complementary nearby lander mission.

### *6.1.1 Geodetic VLBI*

Geodetic VLBI observations are coordinated by the International VLBI Service for Geodesy and Astrometry (IVS; Nothnagel 2017). Two networks are currently in operation: the legacy S/X system and the next-generation VLBI Global Observing System (VGOS), which will be critical for observing the potential NovaMoon's VLBI transmitter.

Although VGOS telescopes are smaller by design, e.g. with diameters around 12-13 meters, they are optimal for geodetic parameter estimation (Petrachenko et al., 2009). Importantly, the smaller size allows for fast slewing rates of around 12°/s in azimuth and 6°/s in elevation. To compensate for a reduced sensitivity of the small apertures, VGOS employs a broadband observing strategy, spanning four frequency bands between 2 to 14 GHz. Each band currently covers 512 MHz, with potential extension to 1 GHz, centered today near 3.2, 5.5, 6.6, and 10.4 GHz. Alternative setups are being investigated for observations from active VLBI transmitter compliant with ITU regulations. These measurements are currently recorded in two linear polarizations with two-bit quantization, yielding data rates of 8 Gbps. Together, this significantly increases the cadence of VLBI observations that can be carried out, with one scan every 30 to 60 seconds.

VLBI operations require significant supporting infrastructure in addition to the telescope network. The IVS coordinating center organizes VLBI sessions by assigning telescope networks to typically 24-hour long time-windows. Each session is assigned to an operation center, which is tasked to prepare the observation schedules. This task can be seen as solving a complex optimization problem to maximize network efficiency over 24-hour sessions (Schartner et al., 2020). A typical VGOS session with ~15 stations can generate several thousand scans and tens of thousands of delay observations across a few hundred of sources. The recorded data volumes are substantial, with each VGOS telescope currently producing around 25 TB per day, which must be transferred to a correlator for further processing. The correlator determines the group delays, after which datasets are distributed to analysis centers for the derivation of receiver station and source coordinates, EOPs, and related geodetic products.

The VLBI transmitter on board NovaMoon must be compatible with the VGOS observing principle, requiring the emission of broadband signals across four frequency bands within the 2–14 GHz range, ideally using circular polarization. The exact placement of these bands still needs to be coordinated with the ITU and IVS to both minimize interference from local electromagnetic environments at ground stations and to ensure optimal broadband delay resolution. To support different operational modes, the transmitter should be capable of switching between random-noise and pseudo-noise transmissions. The random-noise mode emulates the natural broadband emission of quasars, thus allowing seamless processing with standard VLBI routines. In contrast, the pseudo-noise mode enables single-station delay determination, i.e. providing pseudo-range and carrier-phase measurements analogous to GNSS.

Achievable accuracies for the NovaMoon VLBI observations will depend on the final design of the transmitted signal and frequency channels and their compatibility with the VGOS system. Geodetic VLBI to non-natural, near field targets is not a routine technique yet (McCallum et al., 2025) and findings from the Genesis mission (2028) will be critical for this potential payload.

Simulations of a VLBI transmitter located at the lunar south pole indicate that sub-meter precision in the transverse position components can be achieved under favourable observation geometries. The figure below illustrates results from one month of daily simulations using 12 VGOS stations (Karatekin & Sert, 2025). The simulations adopt identical daily schedules and account for zenith wet delay, station clock variations, and 10 ps thermal noise. The resulting position estimates show a clear dependence on the lunar inclination, reflecting that most VGOS stations are situated in the northern hemisphere, where visibility—and therefore the number of usable observations—is significantly higher. Errors in the line-of-sight direction are approximately an order of magnitude larger, as expected from the weaker VLBI geometric sensitivity along that axis. By combining daily solutions, the precision can be improved to below 10 cm after approximately 10–15 consecutive days of observations. It should be noted that these results depend on the adopted simulation assumptions—such as VT signal properties, number of observations, etc.—and will also vary with the underlying lunar geophysical model, which affects the transformation between terrestrial and lunar reference frames.

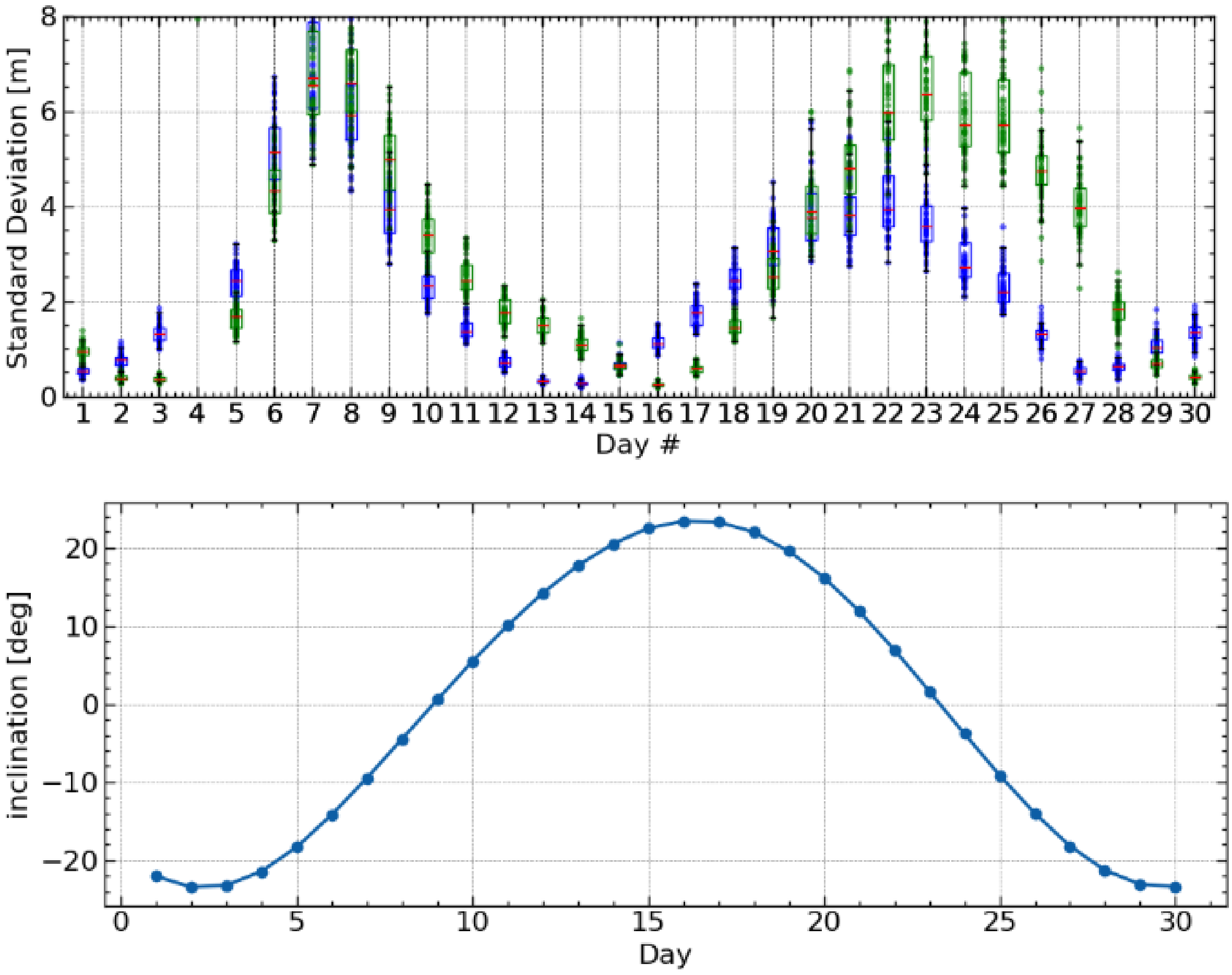

*Figure 11: (Top) Standard deviation of y-(blue) and z-(green) axes for a VT on the lunar south pole during a month. (Bottom) Inclination of the Moon.*

*6.1.2 Delta-DOR*

Delta-Differential One-Way Ranging (Delta-DOR) is an established space navigation technique derived from VLBI that allows the precise angular positioning of spacecraft in deep space. The fundamental observables are the phases of signals emitted from the spacecraft. These signals (tones or spread spectrum) are first received at two geographically well-separated stations, then differentiated and finally compared against similar delay measurements acquired from a quasar of known ICRF coordinates. To allow for an efficient suppression of some error sources (mostly atmospheric delays and, to a lesser extent, uncertainties in station coordinates and EOP values), spacecraft and quasar must be angularly close. In addition, the observations of the two sources must be next to each other to eliminate clock-epoch and clock-rate offsets and to mitigate the effects of the time variability of atmospheric and ionospheric conditions.

In current DDOR measurements, the spacecraft may transmits tones or spread spectrum signals, currently spanning over a bandwidth of about 40 MHz. The frequency spacing of the tones is optimised for phase ambiguity resolution, measurement accuracy, efficient use of spacecraft signal power, efficient use of ground tracking resources, and the frequency allocation for space research (CCSDS, 2019). The use of tones makes the differential measurement sensitive to phase non-linearities of the filters (phase ripples), as the spacecraft and quasar signals are spectrally very different. To suppress this important error source, a wide-band DDOR approach has been proposed and implemented on transponders. The tones are replaced by a PN signals whose spectral characteristics are identical to those of the quasar in the recording bandwidth (typically) spanning up to 8 MHz per channel (Cardarilli et al. 2019). The advantage of wide-band DDOR derives from the cancellation of dispersive phase errors in the differential measurement (quasar minus spacecraft) (Iess et al. 2014). By suppressing one of the two main error sorces, this approach will improve DDOR accuracy, currently at the level of 1-3 nrad level.

The wide-band DDOR would greatly benefit from the use of the VLBI transmitter of Novamoon. Indeed, increasing the recorded bandwidth to 100 MHz or more would reduce the measurement errors as it happens in VLBI.

*6.1.3 Planetary Radio Interferometry and Doppler Experiment (PRIDE)*

The Planetary Radio Interferometry and Doppler Experiment (PRIDE) leverages radio astronomical VLBI technique to determine a spacecraft's position with extreme accuracy (Gurvits et al., 2023). By observing the radio signal from a spacecraft in phase referencing mode (Rioja and Dodson, 2020), a technique related to Delta-DOR, PRIDE provides a direct tie for the lander's position to the quasi-inertial ICRF. Either the TT&C transmitter or the VLBI transmitter on the NovaMoon payload can be key enabler for this experiment, allowing for a synergistic investigation that utilizes existing ground-based infrastructure. A typical PRIDE observation involves tens of telescopes around the world, simultaneously observing for several hours (Duev et al., 2016). This provides a powerful, independent data set to complement the mission's primary navigation and geodetic instrumentation. Hence, for NovaMoon mission, PRIDE might serve as an additional tool for ad hoc measurements aimed at specific science questions. Rather than continuous tracking, PRIDE would be employed for targeted campaigns to provide an accurate position determination on the plane of the sky at critical mission phases. This capability is important for the initial calibration and long-term validation of the other onboard ranging systems. Furthermore, these precise measurements can be complementary to the investigation of specific scientific objectives, such as refining models of lunar libration or tidal response.

*6.1.4 Summary remarks on VLBI concepts*

In an operational scenario, one or more of the potential VLBI concepts may be selectively exploited, depending on the actual mass, power and budget constraints imposed by the hosting platform on the VLBI payload, as well as on potential limitations in the availability and capacity of the VLBI ground network. For instance, equipping NovaMoon with a dedicated transmitter compatible with VGOS frequencies would enable both a geodetic VLBI and DDOR approach, facilitating highly accurate absolute positioning within the ICRF. However, this method demands substantial involvement from the IVS ground network, including a globally distributed and densely populated set of stations, along with sophisticated scheduling support to optimize observation campaigns and accurately model nuisance media propagation effects.

Alternatively (or complementarily), the same payload could support relative angular positioning by referencing the nearest quasar, provided that the angular proximity ensures largely common errors along both lines of sight. This approach could reduce ground network demands and simplify scheduling while potentially enabling more frequent sessions. If platform constraints preclude a VGOS-compatible transmitter, the DDOR exploiting the TT&C link or a dedicated NovaMoon transponder could offer a viable fallback. Though this method may yield reduced performance due to narrower bandwidths and closer observing frequencies, it still provides a means to anchor NovaMoon to an ICRF realization. Ultimately, a careful trade-off between achievable performance and system limitations must be conducted during mission advanced design phases to determine the most effective strategy for precise positioning.

## 6.2 Lunar Laser Ranging (LLR)

### *6.2.1 MoonLIGHT LCNS LRR*

An LRR instrument whose design inherits from the G2G LRR is baselined onboard ESA's Moonlight satellites. The development started in 2023 with an industrial R&D contract in the framework of ESA's LCNS. This R&D led to the design, construction and relatively high (30g RMS) random vibration qualification of an Elegant BreadBoard (of reduced size), which was successful. This LRR LCNS R&D has ended in 2025. The production of the cube corner retroreflectors (CCRs) of about 5 cm diameter initially suffered from optical quality issues, and this was an open point/challenge.

However, as part of a non-ESA/ASI further development, the optical quality of these LCNS CCRs has been improved for a second batch of CCRs. Construction and space qualification of a full-size EM of the LCNS LRR started in 2023 under an ASI contract. This EM will undergo full mechanical qualifications (including pyroshock) and TVT qualification (DRB to ASI by October 2026). This EM is used for mass-volume trade-off and significant risk-reduction to produce PFM1, FM2/3/4 LRRs to the Moonlight constellation. ASI has foreseen the additional production of a PFM Category B LRR for deployment on a commercial lunar orbiter.

Lunar Laser Ranging is a technique of space geodesy developed in the 60's, and it consists of measuring the time of flight of ultra-short optical pulses (one hundred of picoseconds) between optical ground station (OGS) on Earth and laser retroreflector placed on the Moon. Before 2025, five LRRs were continuously tracked by four laser stations of the International Laser Ranging Service (ILRS) network. But on March 2, 2025, a new generation of large mono corner cube (100 mm in diameter) has been successfully deposited on the Moon: the NASA-NGLR-1 LRR (Williams et al., 2022) on board the Firefly Blue Ghost lander paves the way of a new generation of LRR with higher metrological performances. This new target should enable distance measurement residuals to be obtained at the millimeter level. In fact, compared to arrays of small corner cubes, the new design of large LRR prevents the spread of the return pulses due to the inclination of the array with the lunar libration and the reflections of the different corner cubes. However, increasing the size of the corner cube cannot be achieved without limitations: Inhomogeneity of Fused Silica and Thermal gradients in the glass of the corner cubes limit the size to 100 mm in diameter. Moreover, the sensitivity of the far-field diffraction pattern (FFDP) with laser polarisation is stronger.

The challenge for the passive LRR for NovaMoon will be its effective pointing in the direction of the Earth remotely. For that, NovaMoon will benefit from feedback of NGLR-1 and ESA's Moonlight-MPAc deployments. The MPAc (Muccino et al., 2025; Porcelli et al 2021) will be able to perform two continuous perpendicular rotations to accurately point the front face of Moonlight towards the Earth; the device will operate in ultra-high-vacuum space conditions and in a wide temperature range. Two MPAc TRL8 Proto Flight Models have been qualified and delivered to ESA/NASA. PFM2 is for flight in 2026 on the IM-3 CLPS lander. PFM1 is available to ESA for other applications, including NovaMoon preparations.

The deployment of several miniature laser retroreflectors like INRRI (currently 3 are being observed from the laser altimeter on NASA's LRO and about 7 more are foreseen before 2029), especially on the lunar far side and south pole will provide useful geographic lever arm for geometrodynamic measurements (INRRI on Chang'E-6), see Wang et al. (2025). INRRI is TRL9, 25gr, 5x2cm2.

*6.2.2 Active Laser Ranging*

Adding an active part to the passive LRR (A-LRR) of NovaMoon will allow to date on the Moon the arrival of the optical pulse emitted by the OGS. The events date in the NovaMoon-based time scale will allow its comparison with the laser OGS time scale with a level of uncertainty one order of magnitude below the NovaMoon objectives (1ns), like demonstrated before with the CNES-T2L2 instrument (Samain et al., 2015 ; Exertier et al., 2016 ; Leute et al., 2018 ; Samain et al., 2018). On Earth, the construction of a coordinate time scale (TT) requires the comparison of the atomic clocks developed and operated by the National Metrology Laboratory. If fiber network allows the comparison of these clocks on a continental scale, there is no other solution than to implement and share mobile clocks to compare the realization of these time scales between different continents.

An A-LRR on NovaMoon will offer the possibility of doing ground-to-ground time scale comparison over continental distances. It will facilitate the realisation of TT in the context of the Second Redefinition carried out by BIPM with the support of the space agencies. Most of the ILRS stations are not equipped with large telescopes and powerful lasers to perform LLR in round-trip configuration. However, most of them have the link budget to perform one-way LLR, as demonstrated in the past with the LOLA A-LRR of the NASA-LRO orbiter (Bauer et al., 2017; Mao et al., 2017). By increasing the number of ILRS stations participating, the A-LRR on-board NovaMoon will enhance the accuracy of the Lunar reference frame, and it will allow a better link between ITRF & ICRF. LLR already contributes to EOP determination as a validation technique of IVS products (Singh et al., 2022). With more laser stations involved in one-way laser ranging on NovaMoon thanks to the A-LRR, it can improve the contribution of SLR/LLR on EOP determination.

Moreover, contributions from southern hemisphere stations will offer better determination of nutation and precession. For Fundamental Physics, an A-LRR can improve SEP/WEP tests, Lorentz symmetry violation, thanks to the addition of one-way measurements during the new and full Moon periods, two phases during which the noise does not help for the laser echoes detection (Müller et al, 2019). In LLR, there is a large delay between the uplink and the downlink atmospheric crossing. An active-LRR on the Moon can help with the tropospheric delays' determination thanks to the addition of the one-way measurements. Finally, NovaMoon positioning can be done with an A-LRR and laser stations synchronized (e.g. with the distribution of optical reference through the fiber network).

Adding an A-LRR onboard NovaMoon will increase the size, the weight and the power consumption, but could contribute to Precise Orbit Determination (POD). The challenge will be to trade off these aspects and for that, ESA will benefit from the ACES-ELT mission feedback (Schreiber et al., 2009). Open points, will be the potential use of the A-LRR sub-systems for Quantum links (Xu et al., 2019; Spiess et al., 2023) and having potentially two collimation optics for collecting both laser pulses from Earth and from Orbiters like LRO (i.e. Time Transfer and one-way LLR between Moon orbiters & NovaMoon).

## 6.3 Time Laboratory

The “TimeLab” Scientific Payload on board NovaMoon will play a pivotal role in achieving the scientific objectives. First, it will generate, for the first time ever, a physical realisation of the Lunar Reference Time and will allow for its comparison with an Earth-based timescale (UTC) to support tests and experimentations of Lunar Timescales (2$^{nd}$ primary objective). Second, thanks to a well-controlled and calibrated distribution of this reference signal to the various on-board selenodetic instruments (VLBI, A-LRR, GNSS Rx/Tx, LCNS Rx, …), it will guarantee the required time-tie allowing for the consistent realisation of Lunar Reference Frame (1$^{st}$ primary objective).

This payload will be composed of a set of atomic clocks that will guarantee the required level of reliability and robustness in the harsh lunar environment over the specified mission lifetime. The type of atomic clock to be embarked will have to be selected based on a trade-off between the required frequency stability (short, medium, long-term), frequency accuracy as well as the power, thermal, mechanical, radiation constraints of the NovaMoon spacecraft and its operation. The most mature and suitable atomic clock candidates are already identified, and their characteristics are summarised in the table below. Other candidates with lower maturity but better performance are also under consideration.

| | **Accuracy** | **Stability /h** | **Stability /day** | **mass** | **power** | **heritage** | **Qualification status/level** | **FM Availability** |
|---|---|---|---|---|---|---|---|---|
| PHM | few e-12 | 0.06 ns | 0.001us | 18kg | ~60W | Galileo | rad-hard | available |
| RAFS | < 1e-10 | 0.5 ns | 0.1us | 3.2kg | ~30W | Galileo | rad-hard | available |
| miniRAFS | < 1e-10 | 0.7 ns | 0.1us | 500g | <10W | Space dev | rad-hard | 2028 |
| CSAC | few e-10 | 5 ns | 1us | 300g | < 5W | Ground COTS | COTS-screened | 2026 |
| Low Power-CSAC | few e-10 | 5 ns | 1us | 150g | < 0.5W | Ground dev. | COTS/rad-hard | >2028 |

Table 4: candidate atomic frequency standards for the TimeLab payload on NovaMoon spacecraft

To continuously monitor the behaviour and performance of the selected on-board atomic clocks, and to be able to compare them with external signals coming from the other instruments, a dedicated monitoring function will have to be implemented in the NovaMoon TimeLab payload. In addition, to compensate for the intrinsic frequency inaccuracy and drift of the selected atomic clocks, the TimeLab payload will implement a high-resolution frequency steering function, so that the local realisation of Lunar Reference Time is aligned with its definition. Specific algorithms will have to be developed to estimate this steering command.

These steering and monitoring functions, together with the distribution of the reference signal to the various selenodetic instruments will be implemented in a dedicated device called the synchronisation and distribution unit. Such unit will have to be developed and qualified but will take heritage from similar units already embarked on other missions, particularly Galileo.

Overall, the preliminary functional concept of the NovaMoon TimeLab payload is well defined and the associated key enabling technologies are well identified. For successful implementation, dedicated pre-development activities shall be initiated, to consolidate the overall concept and associated requirements (in terms of stability, accuracy, power, mass…) as well as to confirm the suitability of the selected atomic clock technology to meet NovaMoon requirements.

## 6.4 GNSS receiver and transmitter

### *6.4.1 GNSS Receiver*

The availability of GNSS tracking receivers on the lunar surface has been successfully demonstrated in the recent Lunar GNSS Receiver Experiment (LuGRE) mission (Parker et al. 2022), joint effort between NASA and the Italian Space Agency (ASI) to demonstrate the viability of using existing GNSS signals for PNT on the Moon. The first results are presented in (Parker et al. 2025), with several hours of data from GPS and Galileo continuously tracked from the transmitter main lobe. In (ibid), the acquisition of side lobes was potentially limited by sensitivity constraints. NovaMoon mission is expected to reach 15 dB-Hz sensitivity in tracking and 18 dB-Hz in acquisition (Giordano et al. 2022), enabling a more substantial tracking of secondary lobes of the GNSS signals. The demodulation of ephemerides from the navigation message is expected to be challenging (Delépaut et al. 2020), but it could be mitigated by sending these data directly through telecommand. Moreover, a combination of GNSS and LCNS receiver capabilities can be considered to lower the mass of the overall payload.

Several other challenges shall be considered, for instance related to lunar librations causing monthly visibility cycles near the Lunar South Pole, thus resulting in limited visibility half of the time (i.e., two weeks each month). At low elevation angles, multipath effects might play a role in the reception of signals, even if these effects were not observed in LuGRE preliminary data (Parker et al. 2025). With a high gain antenna, up to 16 satellites could be visible when considering GPS+Galileo constellations, whereas limited to around 6 for a low gain antenna. The former will most likely require an antenna pointing mechanism, complicating the lander's design and operations on lunar surface. Lastly, the GNSS receiver is also highly dependent upon the clock technology considered on board, where a few different possibilities are being traded off, including OCXO, MiniRAFS, etc., see Section 6.3.

### *6.4.2 GNSS Transmitter*

The possibility of potentially reusing existing technology from Galileo payload core, like Navigation Signal Generation Unit (NSGU), is surely an interesting candidate solution for NovaMoon. This would allow for the generation of L-band signals from the lunar surface, then tracked from Earth or satellites in low Earth orbit, assuming sufficient visibility from the Lunar South Pole. The need for a more directive antenna should also be considered in the potential design of this specific payload; still, the major challenge lies in the strict ITU regulations that currently prevent transmission below 2 MHz from the shielded zone of the Moon (Inside GNSS, 2024). Hence, S-band could be considered as synergy between Lunar and Terrestrial GNSS signals, whereas C-band was investigated in (ibid) and ruled out due to the protection of radio astronomer observations.

As for the GNSS receiver, the onboard clock technology plays a fundamental role in the transmission of GNSS-like signals, as well as considerations about the antenna gain and required power. These aspects are not yet addressed and traded off in the current preliminary design stage, but surely, they should consider inter-system coordination to ensure no degradation of existing PNT services. In this regard, solutions like those adopted for Galileo E14/E18, wrongly placed in eccentric orbits, might be adopted for science studies (Giorgi et al. 2016), thus setting unhealthy flags even if this would lead to satellites unusable for navigation.

Lastly, for a joint receiver+transmitter design, potential self-interference effects shall be studied, similarly to what is currently under investigation for future LEO-PNT constellations. A key difference will nonetheless be the geometry and directionality of such instrumentation for NovaMoon, as well as additional considerations on mass, size, and power requirements.

## 6.5 Moonlight LCNS receiver

The availability of a receiver tracking Moonlight's Lunar Communications and Navigation Services is an important candidate payload for NovaMoon. A prototype receiver is being developed as part of the In-Orbit Validation Test (IOV-T) for Moonlight mission. This opens to the possibility of processing LCNS satellites' data, as well as considering other 'LunaNet' nodes that can contribute to the Lunar Augmented Navigation Service (LANS). As previously mentioned for the GNSS receiver payload, the synergy between GNSS and LCNS capabilities could be exploited, e.g., to lower the overall payload's mass.

As for the GNSS case, the clocks used on board are fundamental since they drive the measurements' stability, also considering that two receivers could be potentially available and might be interfaced to an internal OCXO or MiniRAFS (see Section 6.3 for an overview). Lastly, IQ data downlink budget is somehow a potential bottleneck to be addressed in future trade-offs for NovaMoon specifications, along with payload-specific requirements yet to be defined.

## 6.6 Single-Beam Interferometry (SBI)

Radio interferometry for orbit determination is typically implemented via delta-DOR, where a quasar serves as a reference to calibrate errors in the measured group delay. Delta-DOR achieves angular accuracy of ~1 nrad, routinely supporting deep space navigation. However, since spacecraft and quasar are observed at different times, media-induced path delays are not fully suppressed.

The same-beam interferometry (SBI) addresses this potential limitation by simultaneously tracking two sources (spacecraft, landers, or rovers) within the beamwidth of widely separated ground antennas. Each source transmits one-way signals, whose phases are differenced at each ground station, then differenced again between stations. This provides highly precise angular separation along the projected baseline. Demonstrated on Pioneer Venus and Magellan (Border et al., 1992), SBI achieved relative angular accuracies of 30–50 prad, equivalent to 12–20 mm lateral uncertainty at lunar distance. The simultaneous tracking of Moonlight satellites and Novamoon signals with Moonlight ground antennas would strongly bind constellation orbits to the lunar surface.

However, the classical SBI approach requires both satellites and the lander to be visible by two antennas, increasing ground segment demands. A simpler alternative, proposed by Bender (1994) and later studied for lunar geophysics (Gregnanin et al., 2012), is **single-dish SBI.** Here, a ground antenna measures the differential phase of the signals from any pair of sources in its beam, yielding their differential range. Unlike classical SBI, the latter requires coherent two-way tracking: a ground uplink is coherently retransmitted by both platforms, and the returned signals are combined to form the interferometer output. The differential phase encodes the relative range between platforms.

A key advantage of single-dish SBI is the strong suppression of common-mode errors (station location, Earth orientation, ephemerides, tropospheric/ionospheric delays, clock and antenna mechanical noise). Residual errors arise mainly from differential atmospheric/ionospheric paths (roughly proportional to the source separation) and drifts in transponder delays. The latter can be minimised with spread-spectrum radio links. According to (Iess et al. 2025), the expected differential range precision is ~1.3 mm at K-band and around 5 mm at X-band for Moonlight satellites, with even better accuracy for NovaMoon/Argonaut-1 pairings due to the smaller angular separation.

SBI offers three major contributions to lunar navigation and geodesy:

- Directly ties Moonlight orbits to the lunar reference frame (current ties are indirect via ITRF and lunar ephemerides);
- Provides near-continuous measurements (in single-dish mode), augmenting data for lunar ephemerides and interior modelling;
- Improves Moonlight orbit determination and strengthens the global orbital solution.

Additionally, nearly continuous two-way coherent range and range-rate measurements, obtained with minimal operational effort, would complement the LLR and the delta-DOR observations planned with NovaMoon.

The onboard hardware required is the transponder/transceiver already in development for Moonlight. The constellation currently supports four nodes for CDM-M uplink modulation, with room for one more without SNR loss. Moonlight ground stations could then deliver IQ samples for all five spread-spectrum signals, generating 10 simultaneous SBI observables (four involving the lander). Ideally, NovaMoon shall carry a dedicated SBI unit integrated into its communication system, though a dual-mode transponder is a practical alternative, operating in spread-spectrum mode for SBI and high data rate mode for telemetry via the ESTRACK deep space antennas.

## 6.7 Seismometers

Although NovaMoon payload aims primarily at establishing co-located space and time references, it could also benefit from having a seismometer as a secondary payload of Argonaut or as part of a complementary nearby lander mission, as mentioned in Section 3.1.3.7, serving as the first node of a future scientific network (Weber et al. 2021). Indeed, a station with a seismometer could also carry out in-situ geophysical measurements, aimed to determine the detailed internal structure of the Moon and to understand its seismic state (Lognonné et al. 2015); such data would be complementary to Apollo's, i.e. collected at high latitudes (poles) and on the lunar far-side, and synergically linked to other future networks (Weber et al. 2021).

Geophysical investigation of the Moon started with the Apollo lunar missions (1969-1972) by deploying instruments including seismometers (5, with Apollo 11, 12, 14, 15 and 16) (Latham et al. 1970, Nunn et al. 2020) . Much of what we know today about the Moon's interior comes from analysis of those data sets. During eight years, they revealed unexpected lunar seismic activity (Amato et al. 2020), about 13,000 events (probably more than double have to be added, despite a recent reanalysis of Apollo data - Onodera et al. 2024), identifying several moonquake types: deep quakes (700-1200 km depth), shallow quakes (less than 200 km depth), thermal quakes (night-day) and natural and artificial impacts. Yet, much work must be done to understand the lunar seismic behaviour, such as the origin of shallow moonquakes (the strongest). Moreover, data on the lunar interior refers only to the central nearside, in a geographically limited triangular network, and close to the equatorial region (Apollo 12/14, 15 and 16).

Upcoming missions (either planned or approved) foresee the deployment of seismometers on the Moon within NASA's Artemis program (LEMS/Lunar Environmental Monitoring System seismometer for Artemis III – Smith et al. 2023) but also within other programs (Chang'e-7 from China, Chandrayaan 3 from India – ISRO- Indian Space Research Organisation; refer to section 3.1.3.7 as well).

| Instrument | Mission /Status | Type | Frequency band (Hz) | Sensitivity | Mass (kg) | Power | Operating Lifetime | Ref. |
|---|---|---|---|---|---|---|---|---|
| ***LP & SP Seismometer*** | Apollo 11, 12, 14, 15, 16 | Passive, long period (LP)/short period (SP) | 0.001-20 | Up to ~10−9 m/s² | ~75 (ALSEP–Apollo Lunar Surface Exp Package) | ~7 W (ALSEP) | Up to 7 years | Latham et al. 2020, Nakamura et al. 1979 |
| ***ILSA - Instrument for Lunar Seismic Activity (ISRO)*** | Chandrayaan-3 (ended) | MEMS accelerometer | 0.05 - 100 | till to 7 $10^{-3}$ m/s² | ~2.2 | Low | ~14 Earth days | John et al. 2024 |
| ***LEMS (NASA)*** | Artemis III, 2027) | Passive, long and short period | SP: 0.05 - 20<br>BB: 0.001 - 1 | $10^{-9}$ m/s² | ~10 | ~5 | 2 years | Smith et al. 2023 |
| ***Farside Seismic Suite (NASA)*** | Future robotic mission (2027) | Passive, ultra-broadband. | 0.008 - 50 | Noise: ~$2\times10^{-10}$ m/s²/sqrt{Hz} | ~10 | ~5 | 4.5+ months | Panning et al. 2024 |
| ***Silicon Seismic Package*** | Prototype / FSS | Passive MEMS seismometer | 0.01 - 40 | Noise: ~$10^{-10}$ m/s²/sqrt{Hz} | ~0.635 | ~0.360 | Long-lived | Pike et al. 2016 |
| ***Farside Explorer*** | ESA Proposal | 3 SP sensors and 3 LP sensors | <2 $10^{-2}$ Hz – 40 Hz (LP+SP channels) | 10–20× better than Apollo | 7 | 3 | Up to 4 years | Mimoun et al. 2012 |

Table 5: Main characteristics of Apollo seismometers along with some planned or approved missions involving such sensors

A seismometer is often operated with Corner Cube Retroreflectors (CCRs) as a suite of instruments devoted to geophysical measurements, as witnessed by the retroreflector arrays (5) delivered by Apollo missions (that were and are in use for LLR), and by several missions aiming at deploying geophysical networks (Panning et al. 2024, Weber et al. 2021). The NovaMoon payload can benefit from a seismometer deployed on the surface, also jointly operated with retroreflectors, contributing to different aspects:

- collecting data to understand the Moon's seismic state and internal structure;
- carrying out coordinated and synergic observations with NovaMoon retroreflectors, to help constrain the deep mantle region, along with core/mantle conditions and boundaries;
- performing landing site monitoring (i.e. seismic activity, micrometeorites), to evaluate how it affects the NovaMoon payload/Argonaut lander and their measurements (also in view of the future human return)
- providing a first ESA testbed for a network of lunar geophysical stations.

Taking into account the Apollo heritage and the recent analyses of gathered data at that time, the characteristics of a NovaMoon seismometer should include a broad bandwidth (around milli-to-tenths of Hertz, covering both short and long periods, in addition to higher frequencies not explored by Apollo instruments), low noise floor ($\leq 10^{-10}\ m/s^2\sqrt{Hz}$).

Some main challenges to be faced for this payload include:

- Deployment of the seismometer: a dedicated mechanism should be developed to deploy the payload on the surface (e.g., with an insulating thermal blanket) or the supporting lander should help in the deployment;
- Survival at the harsh lunar environmental conditions: the supporting lander should provide 1) a thermal system to mitigate the expected large thermal variations and 2) a power system to feed the payload.

# 7. Requirements for NovaMoon

As developed in the previous sections, NovaMoon is a multipurpose mission which involves several different units. This implies that an important harmonisation work needs to be done between the different requirements in regards of the current technologies and future developments to align the requirement set to the targeted mission and scientific goals. This section intends to provide a brief overview of the main requirements for the NovaMoon mission.

As the main objectives of the station, both in terms of navigation and science objectives, rely on the precise positioning of the payload, the main requirement is regarding the given final accuracy for the positioning it can perform which shall be at centimetre level or below. A set of techniques was identified, as defined previously, to achieve this positioning accuracy which are then flown down onto NovaMoon requirements. For each of those techniques, producing a different type of measurement, accuracy and periodicity requirements can be defined to comply with the minimum positioning accuracy requirement. Table 6 presents those preliminary requirements.

Additionally, several interface requirements need to be defined for the payload to ensure each measurement campaign (i.e. for each type of measurement) can be carried out. Thus, the payload shall be able to receive signals coming from the Moonlight LCNS constellation. The VLBI transmitter needs to be compatible with the VGOS stations (the next generation VLBI system for geodesy and astrometry). The diversity of the stations across north and south hemispheres shall be maximised to favour a better positioning. Interface requirements with the lander platform also need to be defined to ensure the calibration and ties between measurement units (at mm (TBC) do not have an impact on the final product. Indeed, calibrations of the different payloads are needed to a very precise accuracy to relate each measurement to a common point during the position estimation process. Finally, requirements need to be derived based on the landing site to make sure that all the needed instruments can be pointed towards the Earth to perform the measurement campaigns.

*Table 6 - Requirements regarding the measurement accuracy and periodicity for NovaMoon*

| Payload & Measurements | Measurement accuracy | Periodicity | External interface |
|---|---|---|---|
| VLBI | < 1mas (TBC) | 2 weekly sessions of 24 hours | NovaMoon shall be compatible with the VGOS station network |
| LLR | < 1 cm (TBC) | 1/3 of Apollo 15's data | NovaMoon retroreflector shall be positioned on the payload to be visible from Earth. The selected payload shall |
| | | | allow LLR measurement to be performed regarding the current ground infrastructure |
| AFS | As per LNIS service definition | continuous | NovaMoon shall embark on a LANS receiver |
| GNSS [opt.] | < 10 m (TBC) | continuous | NovaMoon shall embark on a GNSS receiver |
| DWE | 1 cm (TBC) | During LLR gaps | |

# 8. Summary

The NovaMoon Mission, planned as one of the key demonstration payloads of the first ESA Argonaut lunar lander mission, is conceived as a local differential, geodetic, and timing station. NovaMoon's embarked geodetic instruments and lunar atomic clocks will support the establishment of consolidated lunar geodetic and time reference frames. while unlocking numerous scientific research opportunities on the Moon and in the field of fundamental physics.

This document has described a comprehensive set of *primary* and *secondary* scientific objectives identified for the NovaMoon mission. The document has been developed through a series of specialised working groups, each led by renowned domain experts, and through a sequence of dedicated scientific meetings held between March and October 2025. It has been prepared through the collaborative contribution of 84 European scientific researchers, representing 40 European research institutions, and is further formally endorsed by an additional 20 scientists, bringing the total number of European scientific explicit contributors or supporters to 104.

The principal scientific content described here was presented and discussed in detail during the *1st NovaMoon Science Workshop*, held at ESA's European Space Operations Centre (ESOC) on 8-9 July 2025.

# Acronyms

ALSEP — Apollo Lunar Surface Experiments Package

BCRS — Barycentric Celestial Reference System;

BDS — BeiDou Navigation Satellite System;

CLPS — Commercial Lunar Payload Services;

CMB — Cosmic Microwave Background;

DEM — Digital Elevation Model;

DGNSS — Differential Global Navigation Satellite System;

DM — Dark Matter;

DDOR — Delta-Differential One-Way Ranging;

DTE — Direct-to-Earth;

EOP — Earth Orientation Parameters;

EP — Equivalence Principle

ESA — European Space Agency;

FFT — Fast Fourier Transform;

GCRS — Geocentric Celestial Reference System;

GDOP — Geometric Dilution of Precision;

GEO — Geostationary Earth Orbit;

GNSS — Global Navigation Satellite System;

GPS — Global Positioning System;

IAG — International Association of Geodesy;

IAU — International Astronomical Union;

ICRF — International Celestial Reference Frame;

ICRS — International Celestial Reference System;

ILRS — International Laser Ranging Service;

ILRF — International Lunar Reference Frame;

ILSA — Instrument for Lunar Seismic Activity

INS — Inertial Navigation System;

IR — Infrared;

IRNSS — Indian Regional Navigation Satellite System;

ISRO — Indian Space Research OrganisationISS — International Space Station;

ITRF — International Terrestrial Reference Frame;

JPL — Jet Propulsion Laboratory;

LEMS — Lunar Environment Monitoring Station

LCNS — Lunar Communication and Navigation Services;

LCRS — Lunar Communication and Ranging Service;

LEO — Low Earth Orbit;

LLR — Lunar Laser RangingLNSP – LunaNet Service Provider

LRA — Laser Retroreflector Array;

LRO — Lunar Reconnaissance Orbiter;

LRR — Laser Ranging Retroreflector;

LT — Lunar Time;

LRT — Laser Ranging Transponder;

ME — Mean Earth;

MEMS — Micro Electro-Mechanical Systems;

MEO — Medium Earth Orbit;

MOLA — Mars Orbiter Laser Altimeter;

MRO — Mars Reconnaissance Orbiter;

MSGR — MErcury Surface, GEochemistry, and Ranging (MESSENGER);

NASA — National Aeronautics and Space Administration;

NMC — Non-Minimally Coupled

OD — Orbit Determination;

PA – Principal Axis;

PNT — Positioning, Navigation, and Timing;

PRIDE – Planetary Radio Interferometry and Doppler Experiment

PRN — Pseudo-Random Noise;

PSR — Permanently Shadowed Region;

PVT — Position, Velocity, Time solution;

QZSS — Quasi-Zenith Satellite System;

RA — Right Ascension;

RF — Radio-Frequency;

RO — Radio Occultation;

RTK — Real-Time Kinematic;

SAR — Synthetic Aperture Radar;

SLR — Satellite Laser Ranging;

SME — Standard Model Extension;

SNR — Signal-to-Noise Ratio;

SPAD — Single-Photon Avalanche Diode;

TAI — Temps Atomique International;

TCB — Barycentric Coordinate Time;

TCG — Geocentric Coordinate Time;

TCL — Lunar Coordinate Time;

TL — Lunar Time;

TT — Terrestrial Time;

UFF — Universality of Free Fall;

UT1 — Universal Time 1;

UTC — Coordinated Universal Time;

VLBI — Very Long Baseline Interferometry;

WEP — Weak Equivalence Principle.

# Ethics declaration

## Compliance with Ethical Standards

The authors have complied with the ethical standards of Space Science Reviews in preparing this manuscript and conducting any analysis described therein.

## Conflict of Interest

The authors declare that they have no conflict of interest.

## Ethical approval

The authors approve the ethical standards declaration above.

## Informed consent

All authors have been informed of and agreed to the submission of this manuscript.

## References

A. Bassi et al. (2022). A way forward for fundamental physics in space. *npj Microgravity*, 8, 49. https://doi.org/10.1038/s41526-022-00229-0.

Adoram-Kershner, L. A., Wheeler, B. H., Laura, J. R., Fergason, R. L., & Mayer, D. P. (2021). Automated Kaguya TC and MRO CTX Stereo DEM Generation. *5th Planetary Data and PSIDA 2021*. https://doi.org/10.1029/2007

Amato, M. et al. (2020). Lunar Geophysical Network (LGN), Planetary Science Decadal Survey, Mission Concept Study Final Report, 2020, LGN Decadal Study Final Report

Anderson, J. D., & Williams, J. G. (2001). Long-range tests of the equivalence principle. *Classical and Quantum Gravity*, 18, 2447–2456. https://doi.org/10.1088/0264-9381/18/13/307

Andrews-Hanna, J. C., Weber, R. C., Garrick-Bethell, I., et al. (2023). The structure and evolution of the lunar interior. *Reviews in Mineralogy and Geochemistry*, 89(1), 243–292.

Angelopoulos, V. (2008). The THEMIS mission. Space Science Reviews, 141, 5–34. https://doi.org/10.1007/s11214-008-9336-1

Antonangeli, D., Morard, G., Schmerr, N. C., et al. (2015). Toward a mineral physics reference model for the Moon's core. *Proceedings of the National Academy of Sciences*, 112(13), 3916–3919.

Asenbaum, P., Overstreet, C., Kim, M., Curti, J. & Kasevich, M. A. (2020) 'Atom-interferometric test of the equivalence principle at the $10^{-12}$ level', *Physical Review Letters*, **125**(19), p. 191101. https://doi.org/10.1103/PhysRevLett.125.191101

Asenbaum, P., Overstreet, C., Kovachy, T., Brown, D. D., Hogan, J. M. & Kasevich, M. A. (2017) 'Phase shift in an atom interferometer due to spacetime curvature across its wave function', *Physical Review Letters*, **118**(18), p. 183602. https://doi.org/10.1103/PhysRevLett.118.183602

Ashby, N., & Patla, B. (2024). A relativistic framework to estimate clock rates on the moon. *The Astronomical Journal*, 168, 112. https://doi.org/10.3847/1538-3881/ad643a.

Barker, M. K., Mazarico, E., Neumann, G. A., et al. (2016). A new lunar digital elevation model from the Lunar Orbiter Laser Altimeter and SELENE Terrain Camera. *Icarus*, 273, 346–355. https://doi.org/10.1016/j.icarus.2015.07.039

Bauer, S., Hussmann, H., Oberst, J., et al. (2017). Analysis of one-way laser ranging data to LRO, time transfer and clock characterisation. *Icarus*, 283, 38–54.

Battat, J. B. R., Chandler, J. F., & Stubbs, C. W. (2007). Testing for Lorentz Violation: Constraints on Standard-Model-Extension Parameters via Lunar Laser Ranging. *Physical Review Letters*, 99, 241103.

Bailey, Q. G., & Kostelecky, V. A. (2006). Signals for Lorentz violation in post-Newtonian gravity. *Physical Review D*, 74, 045001.

Bender, P. L. (1994). Proposed microwave transponders for early lunar robotic landers. *Advances in Space Research*, 14(6), 233–242.

Bertotti, B., Ciufolini, I., & Bender, P. (1987). New Test of General Relativity: Measurement of de Sitter Geodetic Precession Rate for Lunar Perigee. *Phys. Rev. Lett.*, 58, 1062.

Beyer, R. A., Alexandrov, O., & Moratto, Z. (2014). Aligning terrain model and laser altimeter point clouds with the Ames Stereo Pipeline. *45th Lunar and Planetary Science Conference*.

Birrell, N. D. & Davies, P. C. W. (1994) *Quantum Fields in Curved Space*. Cambridge: Cambridge University Press.

Biskupek, L., Müller, J., & Torre, J.-M. (2021). Benefit of New High-Precision LLR Data for the Determination of Relativistic Parameters. *Universe*, 7, 34. https://doi.org/10.3390/universe7020034

Blas, D., & Jenkins, A. C. (2022). Bridging the μHz Gap in the Gravitational-Wave Landscape with Binary Resonances. *Phys. Rev. Lett.*, 128(10), 101103.

Böhm, J., Nilsson, T., & Schuh, H. (2010). Prospects for UT1 Measurements from VLBI Intensive Sessions. In *Proceedings of the Sixth General Meeting of the International VLBI Service for Geodesy and Astronomy*, 251–255.

Border, J. S., et al. (1992). Precise Tracking of the Magellan and Pioneer Venus Orbiters by Same-Beam Interferometry, Part I: Data Accuracy Analysis. *TDA Progress Report*, 42-110.

Bourgoin, A., Defraigne, P., & Meynadier, F. (2025). Lunar Reference Timescale. *Submitted to Metrologia*. https://arxiv.org/abs/2507.21597

Bourgoin, A., Hees, A., Bouquillon, S., Le Poncin-Lafitte, C., Francou, G., & Angonin, M.-C. (2016). Testing Lorentz symmetry with Lunar Laser Ranging. *Physical Review Letters*, 117, 241301.

Bourgoin, A., Le Poncin-Lafitte, C., Hees, A., Bouquillon, S., Francou, G., & Angonin, M.-C. (2017). Lorentz symmetry violations from matter-gravity couplings with lunar laser ranging. *Physical Review Letters*, 119, 201102.

Bourgoin, A., Bouquillon, S., Hees, A., Le Poncin-Lafitte, C., Bailey, Q. G., Howard, J. J., ... & Torre, J.-M. (2021). Constraining velocity-dependent Lorentz and CPT violations using lunar laser ranging. *Physical Review D*, 103, 064055.

Briaud, A., Ganino, C., Fienga, A., Mémin, A., & Rambaux, N. (2023). The lunar solid inner core and the mantle overturn. *Nature*, 617, 743–746. https://doi.org/10.1038/s41586-023-05935-7

Broquet, A., & Andrews-Hanna, J. C. (2024). A volcanic inventory of the Moon. *Icarus*, 411, 115954.

Cardarilli, G. C., et al. (2019). Hardware prototyping and validation of a W-ΔDOR digital signal processor. *Applied Sciences*, 9(14), 2909.

Charlot, P., Jacobs, C.S., Gordon, D., Lambert, S., De Witt, A., Böhm, J., Fey, A.L., Heinkelmann, R., Skurikhina, E., Titov, O. and Arias, E.F., 2020. The third realization of the International Celestial Reference Frame by very long baseline interferometry. *Astronomy & Astrophysics*, *644*, p.A159.

Cheng, M. K., Ries, J. C., & Tapley, B. D. (2013). Geocenter Variations from Analysis of SLR Data. In Altamimi, Z., & Collilieux, X. (Eds.), *Reference Frames for Applications in Geosciences*. IAG Symposia, 138. Springer. https://doi.org/10.1007/978-3-642-32998-2_4

Colella, R., Overhauser, A. W. & Werner, S. (1975) 'Observation of gravitationally induced quantum interference', *Physical Review Letters*, **34**(23), pp. 1472–1474. DOI: 10.1103/PhysRevLett.34.1472

Colladay, D., & Kostelecky, V. A. (1997). CPT violation and the standard model. *Physical Review D*, 55, 6760.

Colladay, D., & Kostelecky, V. A. (1998). Lorentz violating extension of the standard model. *Physical Review D*, 58, 116002.

Consultative Committee for Space Data Systems (CCSDS). (2019). *CCSDS Report Concerning Delta-DOR—Technical Characteristics and Performance*.

Delépaut, A., Giordano, P., Ventura-Traveset, J., et al. (2020). Use of GNSS for lunar missions and plans for lunar in-orbit development. *Advances in Space Research*, 66(12), 2739–2756.

DellaGiustina, D. N., Schmerr, N., Benna, M., et al. (2025). LEMS-A3: The Lunar Environmental Monitoring Station — An Artemis 3 Deployed Instrument. *LPICo*, 3090, 2410.

Dong J, Nie W, Cardoso A C, et al. Open-path methane sensing via backscattered light in a nonlinear interferometer. *arXiv preprint arXiv:2506.17107*, (2025).

Duev, D. A., et al. (2016). Planetary Radio Interferometry and Doppler Experiment (PRIDE) technique: a test case of the Mars Express Phobos fly-by. *Astronomy & Astrophysics*, 593, A34. https://doi.org/10.1051/0004-6361/201628869

Exertier, P., Samain, E., Courde, C., Aimar, M., Torre, J. M., Rovera, G. D., ... & Guillemot, P. (2016). Sub-ns time transfer consistency: a direct comparison between GPS CV and T2L2. *Metrologia*, 53(6), 1395.

Fienga, A., Deram, P., Di Ruscio, A., Viswanathan, V., Camargo, J. I. B., Bernus, L., Gastineau, M., & Laskar, J. (2020). INPOP new release: INPOP21a. *Notes scientifique et Techniques de l'IMCCE*. https://www.imcce.fr/content/medias/recherche/equipes/asd/inpop/inpop21a.pdf

Fienga, A., Minazzoli, O. (2024). Testing theories of gravity with planetary ephemerides. *Living Reviews in Relativity*, 27. https://doi.org/10.1007/s41114-023-00047-0

Fienga, A., Rambaux, N., Sośnica, K., & Khatiri, A. (2024). Lunar References Systems, Frames and Time-scales in the context of the ESA Programme Moonlight. *arXiv preprint*, arXiv:2409.10043. https://doi.org/10.48550/arXiv.2409.10043

Fienga, A., Courde, C., Torre, J. M., Manche, H., Murphy, T., Mueller, J., Laskar, J., et al. (2014). Interests of a new lunar laser instrumentation on the ESO NTT Telescope. *arXiv preprint*, arXiv:1405.0473. https://doi.org/10.48550/arXiv.1405.0473

Fjeldbo, G., Kliore, A. J., & Eshleman, V. R. (1971). The neutral atmosphere of Venus as studied with the Mariner V radio occultation experiments. *Astronomical Journal*, 76, 123–140.

Folkner, W. M., Williams, J. G., Boggs, D. H., Park, R. S., & Kuchynka, P. (2014). The planetary and lunar ephemerides DE430 and DE431. *Interplanetary Network Progress Report*, 196(1), 42–196. https://ilrs.cddis.eosdis.nasa.gov/docs/2014/196C.pdf

Folkner, W. M., & Border, J. S. (2015). Linking the planetary ephemeris to the International Celestial Reference Frame. *Highlights of Astronomy*, 16, 219–220. https://doi.org/10.1017/S1743921314005493

Foster, J., Blas, D., Bourgoin, A., Hees, A., Herrero-Valea, M., Jenkins, A. C., & Xue, X. (2025). Discovering μHz gravitational waves and ultra-light dark matter with binary resonances. *arXiv preprint*, arXiv:2504.15334 [astro-ph.CO].

Fuqua Haviland, H., Weber, R. C., Neal, C. R., Lognonné, P., Garcia, R. F., Schmerr, N., Nagihara, S., Grimm, R., Currie, D. G., Dell'Agnello, S., ... (2022). The Lunar Geophysical Network Landing Sites Science Rationale. *The Planetary Science Journal*, 3(2). https://doi.org/10.3847/PSJ/ac0f82

Gaffney, A. M., Gross, J., Borg, L. E., Hanna, K. L. D., Draper, D. S., Dygert, N., ... & van Westrenen, W. (2023). Magmatic evolution I: Initial differentiation of the Moon. *Reviews in Mineralogy and Geochemistry*, 89(1), 103–145.

Garcia, R. F., Khan, A., Drilleau, M., Margerin, L., Kawamura, T., Sun, D., ... & Zhu, P. (2019). Lunar Seismology: An Update on Interior Structure Models. *Space Science Reviews*, 215(8). https://doi.org/10.1007/s11214-019-0613-y

Genova, A., Mazarico, E., Goossens, S., et al. (2018). Solar system expansion and strong equivalence principle as seen by the NASA MESSENGER mission. *Nature Communications*, 9(1), 289. https://doi.org/10.1038/s41467-017-02558-1

Giordano, P., Malman, F., Swinden, R., Zoccarato, P., & Ventura-Traveset, J. (2022). The Lunar Pathfinder PNT experiment and Moonlight navigation service: The future of lunar position, navigation and timing. In *Proceedings of the 2022 International Technical Meeting of The Institute of Navigation*, 632–642.

Goossens, S., Matsumoto, K., Liu, Q., Kikuchi, F., Sato, K., Hanada, H., Ishihara, Y., Noda, H., Kawano, N., Namiki, N., ... (2011). Lunar gravity field determination using SELENE same-beam differential VLBI tracking data. *Journal of Geodesy*, 85(4), 205–228. https://doi.org/10.1007/s00190-010-0430-2

Goossens, S., Matsuyama, I., Cascioli, G., & Mazarico, E. (2024). A Low-Viscosity Lower Lunar Mantle Implied by Measured Monthly and Yearly Tides. *AGU Advances*, 5(5), e2024AV001285. https://doi.org/10.1029/2024AV001285

Gregnanin, E., et al. (2012). Same beam interferometry as a tool for the investigation of the lunar interior. *Planetary and Space Science*, 74, 194–201.

Gurvits, L. I., et al. (2023). Planetary radio interferometry and Doppler experiment (PRIDE) of the JUICE mission. *Space Science Reviews*, 219(8), 79. https://doi.org/10.1007/s11214-023-01026-1

Gutzwiller, M. (1998). Moon–Earth–Sun: The oldest three-body problem. *Reviews of Modern Physics*, 70, 589.

Hanada, H., Iwata, T., Namiki, N., Kawano, N., Asari, K., Ishikawa, T., Kikuchi, F., Liu, Q., Matsumoto, K., Noda, H. and Tsuruta, S., 2008. VLBI for better gravimetry in SELENE. *Advances in Space Research*, *42*(2), pp.341-346.

Harada, Y., Goossens, S., Matsumoto, K., Yan, J., Ping, J., Noda, H., & Haruyama, J. (2014). Strong tidal heating in an ultralow-viscosity zone at the core–mantle boundary of the Moon. *Nature Geoscience*, 7(8), 569–572. https://doi.org/10.1038/ngeo2211

Haruyama, J., Matsunaga, T., Ohtake, M. *et al.* Global lunar-surface mapping experiment using the Lunar Imager/Spectrometer on SELENE. *Earth Planet Sp* **60**, 243–255 (2008). https://doi.org/10.1186/BF03352788

Haviland, H. F., et al. (2022). The Lunar Geophysical Network Landing Sites Science Rationale. *The Planetary Science Journal*, 3(2). https://doi.org/10.3847/PSJ/ac0f82

Hefty, J., Rothacher, M., Springer, T., et al. (2000). Analysis of the first year of Earth rotation parameters with a sub-daily resolution gained at the CODE processing center of the IGS. *Journal of Geodesy*, 74, 479–487. https://doi.org/10.1007/s001900000108

Hilweg, C., Massa, F., Martynov, D., Mavalvala, N., Chruściel, P. T. & Walther, P. (2017) ‘Gravitationally induced phase shift on a single photon’, *New Journal of Physics*, **19**(3), p. 033028. https://doi.org/10.1088/1367-2630/aa638f

Hofmann, F., Müller, J., & Biskupek, L. (2013). Benefit of further ground stations and retro-reflectors for Lunar Laser Ranging analysis. *EGU General Assembly Conference Abstracts*, EGU2013-4330.

Hofmann, F., & Müller, J. (2018). Relativistic tests with lunar laser ranging. *Classical and Quantum Gravity*, 35(3), 035015. https://doi.org/10.1088/1361-6382/aa8f7a

Hood, L. L., Mitchell, D. L., Lin, R. P., Acuña, M. H., & Binder, A. B. (1999). Initial measurements of the lunar induced magnetic dipole moment using Lunar Prospector magnetometer data. *Geophysical Research Letters*, 26(15), 2327–2330. https://doi.org/10.1029/1999GL900487

Hu, X., Stark, A., Dirkx, D., Hussmann, H., Fienga, A., Briaud, A., & Mémin, A. (2023). Sensitivity analysis of polar orbiter motion to lunar viscoelastic tidal deformation. *Celestial Mechanics and Dynamical Astronomy*, 135, 16. https://doi.org/10.1007/s10569-023-10131-w

International Astronomical Union (2000) *Proceedings of the Twenty-Fourth General Assembly, Manchester, 2000*. Paris: International Astronomical Union.

International Astronomical Union (2024) *Proceedings of the Thirty-Second General Assembly, Cape Town, 2024*. Paris: International Astronomical Union.

Inside GNSS. (n.d.). Protecting RA on the shielded side of the Moon. https://insidegnss.com/protecting-ra-on-the-shielded-side-of-the-moon/

Iess, L., Di Benedetto, M., James, N., Mercolino, M., Simone, L., & Tortora, P. (2014). ASTRA: Interdisciplinary study on enhancement of the end-to-end accuracy for spacecraft tracking techniques. *Acta Astronautica*, 94(2), 699–707. https://doi.org/10.1016/j.actaastro.2013.06.011

Iess, L., et al. (2025). A novel orbit determination and time synchronization architecture for a radio navigation satellite constellation in the cislunar environment. *Navigation*, 72(3).

Jenkins, A. C., et al. (2024). Discovering μHz gravitational waves and ultra-light dark matter with binary resonances. *arXiv preprint*, arXiv:2504.15334 [astro-ph.CO]

Jenkins, A. C., et al. (2024). Prospects for gravitational wave and ultra-light dark matter detection with binary resonances beyond the secular approximation. *arXiv preprint*, arXiv:2504.16988 [gr-qc]

Jiang S., et al. (2019). Scan efficiency of structured illumination in iterative single pixel imaging. *Opt. Express, **27**(16),* 22499-22507. https://doi.org/10.1364/OE.27.022499John, J., et al. (2024). Identification and preliminary characterisation of signals recorded by instrument for lunar seismic activity at the Chandrayaan-3 landing site. *Icarus*, 424, 116285. https://doi.org/10.1016/j.icarus.2024.116285

Jolliff, B. L., Gillis, J. J., Haskin, L. A., Korotev, R. L., & Wieczorek, M. A. (2000). Major lunar crustal terranes: Surface expressions and crust–mantle origins. *Journal of Geophysical Research: Planets*, 105(E2), 4197–4216. https://doi.org/10.1029/1999JE001103

Jorgensen, P. (1982). Autonomous navigation of geosynchronous satellites using the NAVSTAR Global Positioning System. In *Proceedings of the National Telesystems Conference*, Galveston, TX, USA, 1–7 December.

Kaltenbaek, R., Lavoie, J., Biggerstaff, D. N., & Resch, K. J. (2008). Quantum-inspired interferometry with chirped laser pulses. *Nature Physics*, 4, 864–868. https://doi.org/10.1038/nphys1093

Karatekin, Ö., & Sert, H. (2025). *VLBI transmitter preliminary performance and link budget estimates for NovaMoon*. Presentation at the 10th International VLBI Technology Workshop (IVTW), 21–25 October 2025, Gothenburg, Sweden.

Kawamura, T., Lognonné, P., Nishikawa, Y., & Tanaka, S. (2017). Evaluation of deep moonquake source parameters: Implication for fault characteristics and thermal state. *Journal of Geophysical Research: Planets*, 122(7), 1487–1504.

Kelley, M. C. (2009). The Earth's Ionosphere: Plasma Physics and Electrodynamics (2nd ed., International Geophysics Series, Vol. 96). Academic Press. https://shop.elsevier.com/books/the-earths-ionosphere/kelley/978-0-12-088425-4

Kim, D., Lekić, V., Wieczorek, M. A., Schmerr, N. C., Collins, G. S., & Panning, M. P. (2025). A new lunar crustal thickness model constrained by converted seismic waves detected beneath the Apollo seismic network. *Geophysical Research Letters*, 52(13), e2024GL114506.

Khan, A., & Mosegaard, K. (2002). An inquiry into the lunar interior: A nonlinear inversion of the Apollo lunar seismic data. *Journal of Geophysical Research (Planets)*, 107. https://doi.org/10.1029/2001JE001658

Kopeikin, S., & Kaplan, G. (2024). Lunar Time in General Relativity. *Physical Review D*, 110, 084047. https://doi.org/10.1103/PhysRevD.110.084047

Konopliv, A. S., Park, R. S., Yuan, D. N., Asmar, S. W., Watkins, M. M., Williams, J. G., ... & Zuber, M. T. (2013). The JPL lunar gravity field to spherical harmonic degree 660 from the GRAIL Primary Mission. *Journal of Geophysical Research: Planets*, 118(7), 1415–1434. https://doi.org/10.1002/jgre.20097

Kur, T., Dobslaw, H., Śliwińska, J., Nastula, J., Wińska, M., & Partyka, A. (2022). Evaluation of selected short-term predictions of UT1–UTC and LOD collected in the second Earth orientation parameters prediction comparison campaign. *Earth, Planets and Space*, 74(1), 191. https://doi.org/10.1186/s40623-022-01753-9

Laneuville, M., Wieczorek, M. A., Breuer, D., & Tosi, N. (2013). Asymmetric thermal evolution of the Moon. *Journal of Geophysical Research: Planets*, 118(7), 1435–1452. https://doi.org/10.1002/jgre.20103

Latham, G. V., et al. (1970). Apollo 11 Passive Seismic Experiment. *Science*, 167(3918).

Leute, J., Petit, G., Exertier, P., Samain, E., Rovera, D., & Uhrich, P. (2018). High accuracy continuous time transfer with GPS IPPP and T2L2. In *2018 European Frequency and Time Forum (EFTF)* (pp. 249–252). IEEE.

Li, Z. P., Ye, J. T., Huang, X., Jiang, P. Y., Cao, Y., Hong, Y., Yu, C., Zhang, J., Zhang, Q., Peng, C. Z., Xu, F. H., & Pan, J. W. Single-photon imaging over 200 km. Optica , 8(3), 344-349 (2021).

Li, W., et al. (2024). Low-latitude ionospheric responses and positioning performance degradation caused by an interplanetary coronal mass ejection. Frontiers in Space and Atmospheric Sciences, 11, 1431611. https://doi.org/10.3389/fspas.2024.1431611

Lognonné, P., Johnson, C. L. (2015), 10.03 - Planetary Seismology, Treatise on Geophysics (Second Edition), Elsevier, Pages 65-120, https://doi.org/10.1016/B978-0-444-53802-4.00167-6.

Lognonné, P., et al. (2019). SEIS: Insight's Seismic Experiment for Internal Structure of Mars. *Space Science Reviews*, 215(12). https://doi.org/10.1007/s11214-018-0574-6

Löcher, A., Hofmann, F., Gläser, P., Haase, I., Muller, J., Kusche, J., & Oberst, J. (2017). Towards Improved Lunar Reference Frames: LRO Orbit Determination. In T. van Dam (Ed.), *REFAG 2014* (pp. 201–207). Springer International Publishing.

Mao, D., McGarry, J. F., Mazarico, E., Neumann, G. A., Sun, X., Torrence, M. H., ... & Zuber, M. T. (2017). The laser ranging experiment of the Lunar Reconnaissance Orbiter: Five years of operations and data analysis. *Icarus*, 283, 55–69.

March, R., Bertolami, O., Muccino, M., & Gomes, C., & Dell'Agnello, S. (2022). Cassini and extra force constraints to nonminimally coupled gravity with a screening mechanism. *Physical Review D*, 105, 044048.

March, R., Bertolami, O., Muccino, M., & Dell'Agnello, S. (2024). Equivalence principle violation in nonminimally coupled gravity and constraints from lunar laser ranging. *Physical Review D*, 109, 124013.

Matsuyama, I., Nimmo, F., Keane, J. T., Chan, N. H., Taylor, G. J., Wieczorek, M. A., ... & Williams, J. G. (2016). GRAIL, LLR, and LOLA constraints on the interior structure of the Moon. *Geophysical Research Letters*, 43(16), 8365–8375. https://doi.org/10.1002/2016GL069952

Mazarico, E., Barker, M. K., Neumann, G. A., Zuber, M. T., & Smith, D. E. (2014). Detection of the lunar body tide by the Lunar Orbiter Laser Altimeter. *Geophysical Research Letters*, 41, 2282–2288. https://doi.org/10.1002/2013GL059085

Mazarico, E., et al. (2016). A new lunar digital elevation model from the Lunar Orbiter Laser Altimeter and SELENE Terrain Camera. *Icarus*, 273, 346–355. https://doi.org/10.1016/j.icarus.2015.07.039

Mazarico, E., et al. (2024). A low-viscosity lower lunar mantle implied by measured monthly and yearly tides. *AGU Advances*, 5(5), e2024AV001285. https://doi.org/10.1029/2024AV001285

Meehan, T., et al. (1996). GPS Sounding of the Atmosphere: Preliminary Results. *Bulletin of the American Meteorological Society*, 77, 19–40.

Mieling, T. B., Hilweg, C. & Walther, P. (2022) 'Measuring space-time curvature using maximally path-entangled quantum states', *Physical Review A*, **106**, p. L031701. https://doi.org/10.1103/PhysRevA.106.L031701

McCallum, L., Schunck, D., McCallum, J. and McCarthy, T., 2025. An experiment to observe GNSS signals with the Australian VGOS array. *Publications of the Astronomical Society of the Pacific*, *137*(4), p.045002.

Mimoun, D., Wieczorek, M.A., Alkalai, L., et al. (2012). Farside Explorer: Unique science from a mission to the farside of the Moon. *Experimental Astronomy*, 33, 529–585. https://doi.org/10.1007/s10686-011-9252-3

Mohageg, M., Mazzarella, L., Anastopoulos, C., Gallicchio, J., Hu, B.-L., Jennewein, T., Johnson, S., Lin, S.-Y., Ling, A., Lohrmann, A., Marquardt, C., Meister, M., Newell, R., Roura, A., Vallone, G., Villoresi, P. & Kwiat, P. (2022) 'The deep space quantum link: prospective fundamental physics experiments using long-baseline quantum optics', *EPJ Quantum Technology*, **9**, article 25. https://doi.org/10.1140/epjqt/s40507-022-00143-0

Morel, L., Yao, Z., Cladé, P. & Guellati-Khélifa, S. (2020) 'Determination of the fine-structure constant with an accuracy of 81 parts per trillion', *Nature*, **588**(7836), pp. 61–65. https://doi.org/10.1038/s41586-020-2964-7

Mott-Smith, H. M., & Langmuir, I. (1926). The theory of collectors in gaseous discharges. Physical Review, 28(4), 727–763. https://doi.org/10.1103/PhysRev.28.727

Müller, J., Murphy, T., Schreiber, U., Shelus, P., Torre, J., Williams, J., ... & Hofmann, F. (2019). Lunar Laser Ranging – A Tool for General Relativity, Lunar Geophysics and Earth Science. *Journal of Geodesy*, 93, 2195–2210. https://doi.org/10.1007/s00190-019-01296-0

Muccino, M. (2025). Moonlight and MPAc: The European Space Agency's Next-Generation Lunar Laser Retroreflector for NASA's CLPS/PRISM1A (CP-11) Mission. *Remote Sensing*, 17(5), 813.

Nakamura, Y., Latham, G., Lammlein, D., Ewing, M., Duennebier, F., & Dorman, J. (1974). Deep lunar interior inferred from recent seismic data. *Geophysical Research Letters*, 1(3), 137–140. https://doi.org/10.1029/GL001i003p00137

Nakamura, Y., et al. (1979). Shallow moonquakes, moonquakes, and lunar tides. *Physics of the Earth and Planetary Interiors*, 20(3).

Nie, W., Zhang, P., McMillan, A., Clark, A. S., & Rarity, J. G. (2025). Entanglement-inspired frequency-agile rangefinding. *arXiv preprint* arXiv:2506.11980.

Nordtvedt, K. (1968). Testing Relativity with Laser Ranging to the Moon. *Physical Review D*, 170, 1186.

Nothnagel, A., Artz, T., Behrend, D. and Malkin, Z., 2017. International VLBI Service for Geodesy and Astrometry: Delivering high-quality products and embarking on observations of the next generation. *Journal of Geodesy*, *91*(7), pp.711-721.

Nunn, C., Garcia, R.F., Nakamura, Y. *et al.* Lunar Seismology: A Data and Instrumentation Review. *Space Sci Rev* **216**, 89 (2020). https://doi.org/10.1007/s11214-020-00709-3

Onodera, K. (2023). New Views of Lunar Seismicity Brought by Analysis of Newly Discovered Moonquakes in Apollo Short-Period Seismic Data. https://doi.org/10.1029/2023JE008153

Overstreet, C., Asenbaum, P., Kim, M., Curti, J. & Kasevich, M. A. (2022) 'Observation of a gravitational Aharonov–Bohm effect', *Science*, **375**, pp. 226–229. https://doi.org/10.1126/science.abf4557

Panning, M. P., Avenson, B., Bailey, S. H., Bremner, P. M., Bugby, D., De Raucourt, S., Garcia, R., et al. (2023). Farside Seismic Suite: A Long-Lived Geophysical Package and a Model for a Potential Lunar Seismic Network. *54th Lunar and Planetary Science Conference*, 2806.

Panning, M. P., et al. (2024). The Farside Seismic Suite: A Novel Approach for Long-term Lunar Seismology. *Annual Review of Earth and Planetary Sciences*.

Parker, E. N. (1958). Dynamics of the interplanetary gas and magnetic fields. Astrophysical Journal, 128, 664–676. https://doi.org/10.1086/146579

Park, R. S., Berne, A., Konopliv, A. S., Keane, J. T., Matsuyama, I., Nimmo, F., Rovira-Navarro, M., et al. (2025). Thermal asymmetry in the Moon's mantle inferred from monthly tidal response. *Nature*, 641, 1188–1192. https://doi.org/10.1038/s41586-025-08949-5

Parker, J. J., Dovis, F., Anderson, B., Ansalone, L., Ashman, B., Bauer, F. H., ... & Valencia, L. (2022). The Lunar GNSS Receiver Experiment (LuGRE). In *Proceedings of the 2022 International Technical Meeting of The Institute of Navigation* (pp. 420–437).

Pavlov, D. A., Williams, J. G., & Suvorkin, V. V. (2016). Determining parameters of Moon's orbital and rotational motion from LLR observations using GRAIL and IERS-recommended models. *Celestial Mechanics and Dynamical Astronomy*, 126, 61–88. https://doi.org/10.1007/s10569-016-9712-1

Pavlov, D. (2020). Role of lunar laser ranging in realization of terrestrial, lunar, and ephemeris reference frames. *Journal of Geodesy*, 94(1), 5.

Pearlman, M. R., Arnold, D., Davis, M., Barlier, F., Biancale, R., Vasiliev, V., ... & Bloßfeld, M. (2019). Laser geodetic satellites: a high-accuracy scientific tool. *Journal of Geodesy*, 93(11), 2181–2194.

Pearlman, M. R., Noll, C. E., Pavlis, E. C., Lemoine, F. G., Combrink, L., Degnan, J. J., Kirchner, G., & Schreiber, U. (2019). The ILRS: approaching 20 years and planning for the future. *Journal of Geodesy*, 93, 2161.

Petit, G., & Luzum, B. (2010). IERS Conventions (2010). *IERS Technical Note* 36:1.

Petrachenko, B., Niell, A., Behrend, D., Corey, B., Böhm, J., Chralot, P., ... & Wresnik, J. (2009). *Design aspects of the VLBI2010 system-Progress report of the IVS VLBI2010 committee* (No. 200901964).

Pike, W. T., et al. (2016). A Silicon Seismic Package (SSP) for Planetary Geophysics. *47th Lunar and Planetary Science Conference (LPSC)*.

Pike, W. T., et al. (2024). [Add full citation if available — title not fully included].

Porcelli, et al. (2021). Next Generation Lunar Laser Retroreflectors for Fundamental Physics and Lunar Science. White Paper Submitted to The National Academies of Sciences of the USA. https://science.nasa.gov/biological-physical/resources/whitepapers/

Richardson, I. G., & Cane, H. V. (2010). Near-Earth interplanetary coronal mass ejections during solar cycle 23 (1996–2009): Catalog and summary of properties. Solar Physics, 264, 189–237. https://doi.org/10.1007/s11207-010-9568-6

Ries, J. C., Cheng, M. K., & Tapley, B. D. (2013). Geocenter Variations from Analysis of SLR Data. In: Altamimi, Z., Collilieux, X. (Eds.) *Reference Frames for Applications in Geosciences*. IAG Symposia, vol 138. Springer. https://doi.org/10.1007/978-3-642-32998-2_4

Rioja, M. J., & Dodson, R. (2020). Precise radio astrometry and new developments for the next generation of instruments. *Astronomy and Astrophysics Review*, 28(1), 6. https://doi.org/10.1007/s00159-020-00126-z

Rosi, G., Sorrentino, F., Cacciapuoti, L., Prevedelli, M. & Tino, G. M. (2014) 'Precision measurement of the Newtonian gravitational constant using cold atoms', *Nature*, **510**(7506), pp. 518–521. DOI: 10.1038/nature13433

Rothacher, M., Beutler, G., Weber, R., & Hefty, J. (2001). High-frequency variations in Earth rotation from Global Positioning System data. *Journal of Geophysical Research*, 106(B7), 13711–13738. https://doi.org/10.1029/2000JB900393

Roura, A. (2020) 'Gravitational redshift in quantum-clock interferometry', *Physical Review X*, **10**, p. 021014. https://doi.org/10.1103/PhysRevX.10.021014

Roura, A., Schubert, C., Schlippert, D. & Rasel, E. M. (2021) 'Measuring gravitational time dilation with delocalized quantum superpositions', *Physical Review D*, **104**(8), p. 084001. https://doi.org/10.1103/PhysRevD.104.084001

Terno, D. R., Vallone, G., Vedovato, F. & Villoresi, P. (2020) 'Large-scale optical interferometry in general spacetimes', *Physical Review D*, **101**(10), p. 104052. https://doi.org/10.1103/PhysRevD.101.104052

Samain, E., Exertier, P., Courde, C., Fridelance, P., Guillemot, P., Laas-Bourez, M., & Torre, J. M. (2015). Time transfer by laser link: a complete analysis of the uncertainty budget. *Metrologia*, 52(2), 423.

Samain, E., Rovera, G. D., Torre, J. M., Courde, C., Belli, A., Exertier, P., ... & Zhang, Z. (2018). Time transfer by laser link (T2L2) in noncommon view between Europe and China. *IEEE Transactions on Ultrasonics, Ferroelectrics, and Frequency Control*, 65(6), 927–933.

Schartner, M., Böhm, J. Optimizing schedules for the VLBI global observing system. *J Geod* **94**, 12 (2020). https://doi.org/10.1007/s00190-019-01340-z

Schlamminger, S., Choi, K. Y., Wagner, T. A., et al. (2008). Test of the equivalence principle using a rotating torsion balance. *Physical Review Letters*, 100(4), 041101. https://doi.org/10.1103/PhysRevLett.100.041101

Schlippert, D., Hartwig, J., Albers, H., Richardson, L.L., Schubert, C., Roura, A., Schleich, W.P., Ertmer, W. & Rasel, E.M. (2014) 'Quantum test of the universality of free fall', *Physical Review Letters*, **112**(20), p. 203002. https://doi.org/10.1103/PhysRevLett.112.203002

Scholten, F. J., Oberst, J., Matz, K.-D., Roatsch, T., Wählisch, M., Speyerer, E. J., & Robinson, M. S. (2012). GLD100: The near-global lunar 100 m raster DTM from LROC WAC stereo image data. *Journal of Geophysical Research*, 117(E12). https://doi.org/10.1029/2011JE003926

Schreiber, U., Prochazka, I., Lauber, P., Hugentobler, U., Schafer, W., Cacciapuoti, L., & Nasca, R. (2009). The European laser timing (ELT) experiment on-board ACES. In *2009 IEEE International Frequency Control Symposium Joint with the 22nd European Frequency and Time Forum* (pp. 594–599). IEEE.

Seyffert, Y. (2025). Relativistic time modeling for lunar positioning navigation and timing. Master's thesis, Université Côte d'Azur, Nice, France. *arXiv preprint*, arXiv:2509.08871. https://doi.org/10.48550/arXiv.2509.08871

Shean, D. E., Alexandrov, O., Moratto, Z. M., Smith, B. E., Joughin, I. R., Porter, C., & Morin, P. (2016). An automated, open-source pipeline for mass production of digital elevation models (DEMs) from very-high-resolution commercial stereo satellite imagery. *ISPRS Journal of Photogrammetry and Remote Sensing*, 116, 101–117. https://doi.org/10.1016/j.isprsjprs.2016.03.012

Shapiro, I. I., Reasenberg, R. D., Chandler, J. F., & Babcock, R. W. (1988). Measurement of the de Sitter Precession of the Moon: A Relativistic Three-Body Effect. *Physical Review Letters*, 61, 2643.

Singh, V. V., Biskupek, L., Müller, J., & Zhang, M. (2022). Earth rotation parameter estimation from LLR. *Advances in Space Research*, 70, 2383. https://doi.org/10.1016/j.asr.2022.07.038

Singh, V. V., Müller, J., Biskupek, L., Hackmann, E., & Lämmerzahl, C. (2023). Equivalence of Active and Passive Gravitational Mass Tested with Lunar Laser Ranging. *Physical Review Letters*, 131, 021401.

Smith, R. K., et al. (2023). The Lunar Environment Monitoring Station (LEMS) for Artemis III.'

Soffel, M., Klioner, S., Petit, G., Wolf, P., Kopeikin, S. Bretagnon, P., Brumberg, V., Capitaine, N., Damour, T., Fukushima, T., Guinot, B., Huang, T. Y., Lindegren, L., Ma, C., Nordtvedt, K., Ries, J.C., Seidelmann, P. K., Vokrouhlický, D., Will, C. M. & C. Xu (2003). The IAU 2000 resolutions for astrometry, celestial mechanics, and metrology in the relativistic framework: Explanatory supplement. Astronomical Journal, 126:2687–2706. http://adsabs.harvard.edu/abs/2003AJ....126.2687S

Sośnica, K., Gałdyn, F., Zajdel, R., Strugarek, D., Najder, J., Nowak, A., ... & Bury, G. (2025b). LARES-2 contribution to global geodetic parameters from the combined LAGEOS-LARES solutions. *Journal of Geodesy*, 99(1), 1–17.

Sośnica, K., Fienga, A., Pavlov, D., Rambaux, N. (2025a). Lunar Reference Frames for the Moonlight - the ESA PNT System. *European Navigation Conference, ENC2025*, Wrocław, Poland, 21–23 May 2025.

Speretta, S., et al. (2019). LUMIO: An Autonomous CubeSat for Lunar Exploration. In: Pasquier, H., Cruzen, C., Schmidhuber, M., Lee, Y. (Eds.), *Space Operations: Inspiring Humankind's Future*. Springer, Cham. https://doi.org/10.1007/978-3-030-11536-4_6

Spiess, C., Töpfer, S., Sharma, S., Kržič, A., Cabrejo-Ponce, M., Chandrashekara, U., ... & Steinlechner, F. (2023). Clock synchronisation with correlated photons. *Physical Review Applied*, 19(5), 054082.

Steinbrügge, G., Steinke, T., Thor, R., Stark, A., & Hussmann, H. (2019). Measuring Ganymede's librations with laser altimetry. *Geosciences*, 9(7). https://doi.org/10.3390/geosciences9070320

Stone, E. C., Frandsen, A. M., Mewaldt, R. A., et al. (1998). The Advanced Composition Explorer. Space Science Reviews, 86, 1–22. https://doi.org/10.1023/A:1005082526237

Sun, M. J., Edgar, M. P., Gibson, G. M., Sun, B., Radwell, N., Lamb, R., & Padgett, M. J., Single-pixel three-dimensional imaging with time-based depth resolution. *Nat. Commun.* **7**, 12010 (2016).

Terno, D. R., Vallone, G., Vedovato, F. & Villoresi, P. (2020) 'Large-scale optical interferometry in general spacetimes', *Physical Review D*, **101**(10), p. 104052. https://doi.org/10.1103/PhysRevD.101.104052

Thor, R. N., Kallenbach, R., Christensen, U. R., Gläser, P., Stark, A., Steinbrügge, G., & Oberst, J. (2021). Determination of the lunar body tide from global laser altimetry data. *Journal of Geodesy*, 95(1). https://doi.org/10.1007/s00190-020-01455-8

Titov, O. (2025). A novel approach for the direct estimation of the instantaneous Earth rotation velocity. *Astronomy & Astrophysics*, 699, A155. https://doi.org/10.1051/0004-6361/202553928

Touboul, P., Métris, G., Rodrigues, M., et al. (2022). MICROSCOPE mission: final results of the test of the equivalence principle. *Physical Review Letters*, 129(12), 121102. https://doi.org/10.1103/PhysRevLett.129.121102

Turyshev, S. G., Shao, M., & Hahn, I. (2021). The Decadal Survey on Biological and Physical Sciences (BPS) Research in Space 2023–2032. *National Academy of Sciences*.

Ventura-Traveset, J., Swinden, R., Melman, F. T., Psychas, D., & Audet, Y. (2025, April 11). *NovaMoon: A new paradigm in lunar exploration*. Inside GNSS. Available: https://insidegnss.com/novamoon-a-new-paradigm-in-lunar-exploration/

Viswanathan, V., Fienga, A., Laskar, J., Gastineau, M., & Manche, H. (2019a). INPOP planetary ephemerides and its applications. *Astronomy & Astrophysics*, 628, A31. https://doi.org/10.1051/0004-6361/201935378

Viswanathan, V., Fienga, A., Minazzoli, O., et al. (2018). The new lunar ephemeris INPOP17a and its application to fundamental physics. *Monthly Notices of the Royal Astronomical Society*, 476(2), 1877–1888. https://doi.org/10.1093/mnras/sty096

Wagner, R. V., Nelson, D. M., Plescia, J. B., Robinson, M. S., Speyerer, E. J., & Mazarico, E. (2017). Coordinates of anthropogenic features on the Moon. *Icarus*, 283, 92–103. https://doi.org/10.1016/j.icarus.2016.05.011

Walterová, M., Běhounková, M., & Efroimsky, M. (2023). Is There a Semi-Molten Layer at the Base of the Lunar Mantle? *Journal of Geophysical Research: Planets*, 128(7). https://doi.org/10.1029/2022JE007652

Wang, Y., Dell'Agnello, S., Di, K., Muccino, M., et al. (2025). The First Laser Retroreflector on the Lunar Far Side onboard China's Chang'e-6 Lander. *Space Science & Technology*, 5, Article 0301.

Ware, R., Exner, M., Feng, D., Gorbunov, M., Hardy, K., Herman, B., ... & Solheim, F. (1996). GPS Sounding of the Atmosphere: Preliminary Results. *Bulletin of the American Meteorological Society*, 77, 19–40.

Weber, R. C. (2014). Chapter 24 – Interior of the Moon. In T. Spohn, D. Breuer, & T. V. Johnson (Eds.), *Encyclopedia of the Solar System* (3rd ed., pp. 539–554). Elsevier. https://doi.org/10.1016/B978-0-12-415845-0.00024-4

Weber, R., Neal, C. R., Grimm, R., Grott, M., Schmerr, N., Wieczorek, M., … Zuber, M. (2021). The scientific rationale for deployment of a long-lived geophysical network on the Moon. Bulletin of the AAS, 53(4). https://doi.org/10.3847/25c2cfeb.674dcfdf

Weber, R. C., Lin, P.-Y., Garnero, E. J., Williams, Q., & Lognonné, P. (2011). Seismic Detection of the Lunar Core. *Science*, 331(6015), 309–312. https://doi.org/10.1126/science.1199375

Wieczorek, M. A., Neumann, G. A., Nimmo, F., Kiefer, W. S., Taylor, G. J., Melosh, H. J., ... & Zuber, M. T. (2013). The Crust of the Moon as Seen by GRAIL. *Science*, 339(6120), 671–675. https://doi.org/10.1126/science.1231530

Williams, J. G., Boggs, D. H., Yoder, C. F., Ratcliff, J. T., & Dickey, J. O. (2001). Lunar rotational dissipation in solid body and molten core. *Journal of Geophysical Research: Planets*, 106(E11), 27933–27968. https://doi.org/10.1029/2000JE001396

Williams, J. G., Boggs, D. H., & Currie, D. G. (2022). Next-generation Laser Ranging at Lunar Geophysical Network and Commercial Lander Payload Service Sites. *The Planetary Science Journal*, 3(6), 136. https://doi.org/10.3847/PSJ/ac6c25

Williams, J. G., Konopliv, A. S., Boggs, D. H., Park, R. S., Yuan, D. N., Lemoine, F. G., ... & Zuber, M. T. (2014). Lunar interior properties from the GRAIL mission. *Journal of Geophysical Research: Planets*, 119(7), 1546–1578.

Williams, J. G., Turyshev, S. G., & Boggs, D. H. (2004). Progress in Lunar Laser Ranging Tests of Relativistic Gravity. *Physical Review Letters*, 93. https://doi.org/10.1103/PhysRevLett.93.261101

Will, C. M. (2018). *Theory and Experiment in Gravitational Physics*. 2nd Edition. Cambridge University Press.

Will, C. M. (2014). The Confrontation between General Relativity and Experiment. *Living Reviews in Relativity*, 17, 4. https://doi.org/10.12942/lrr-2014-4

Wu, H.-N., Li, Y.-H., Li, B., Liu, R.-Z., Ren, J.-G., Yin, J., Lu, C.-Y., Cao, Y., Peng, C.-Z. & Pan, J.-W. (2024) 'Single-photon interference over 8.4 km urban atmosphere: toward testing quantum effects in curved spacetime with photons', *Physical Review Letters*, **133**, p. 020201

Yu, H., Macri, D., Morling, T., Polini, E., Mieling, T. B., Barrow, P., Kabagöz, B., Yin, X., Chruściel, P. T., Hilweg, C., Oelker, E., Mavalvala, N. & Walther, P. (2025) '50-km fiber interferometer for testing gravitational signatures in quantum interference', *arXiv*. arXiv:2511.17022

Zajdel, R., Sośnica, K., Bury, G., et al. (2020). System-specific systematic errors in earth rotation parameters derived from GPS, GLONASS, and Galileo. *GPS Solutions*, 24, 74. https://doi.org/10.1007/s10291-020-00989-w

Zajdel, R., Sośnica, K., Bury, G., et al. (2021). Sub-daily polar motion from GPS, GLONASS, and Galileo. *Journal of Geodesy*, 95, 3. https://doi.org/10.1007/s00190-020-01453-w

Zhang, M., Müller, J., Biskupek, L., et al. (2022). Characteristics of differential lunar laser ranging. *Astronomy & Astrophysics*, 659, A148. https://doi.org/10.1051/0004-6361/202142841

Zhang, M., Müller, J., Biskupek, L. (2024). Advantages of combining Lunar Laser Ranging and Differential Lunar Laser Ranging. *Astronomy & Astrophysics*, 681, A5. https://doi.org/10.1051/0004-6361/202347643

Zhang, M., Müller, J., Kopeikin, S. (2025). Frequency Differences between Clocks on the Earth and the Moon. *Submitted to Physical Review Letters*. https://doi.org/10.48550/arXiv.2506.16377

Zhou, H., Cardoso, A., Faruque, I., Rosenfeld, L., Wollmann, S., Nie, W., Dong, J., Zhang, J., Clark, A. S., Rarity, J. G. (2024). In Gas sensing with undetected light using silicon photonic chips, *Frontiers in Optics*, Optica Publishing Group: p JW4A. 68 https://doi.org/10.1364/FIO.2024.JW4A.68

Zou, Y., Liu, Y., & Jia, Y. (2020). Overview of China's Upcoming Chang'e Series and the Scientific Objectives and Payloads for Chang'e-7 Mission. *51st Lunar and Planetary Science Conference*.